\documentclass[12pt]{article}
\usepackage{amsmath}
\usepackage{graphicx}
\usepackage{enumerate}
\usepackage{natbib}
\usepackage{url} 
\usepackage{hyperref}
\usepackage{bm}

\newcommand{\blind}{1}

\newcommand{\bbeta}{ \mbox{\boldmath $ \beta $} }

\newcommand{\bmu}{ \mbox{\boldmath $\mu$} }

\newcommand{\bSigma}{ \mbox{\boldmath $\Sigma$} }

\newcommand{\bzero}{\textbf{0}}
\newcommand{\bone}{\textbf{1}}

\newcommand{\bI}{\textbf{I}}

\newcommand{\bs}{\textbf{s}}

\newcommand{\bv}{\textbf{v}}

\newcommand{\bx}{\textbf{x}}

\DeclareMathOperator*{\cov}{cov}

\DeclareMathOperator{\dist}{dist}

\DeclareMathOperator{\trendONE}{trend1}
\DeclareMathOperator{\trendTWO}{trend2}

\begin{document}

\def\spacingset#1{\renewcommand{\baselinestretch}%
{#1}\small\normalsize} \spacingset{1}


\if1\blind
{
  \title{\bf Spatio-temporal modeling for record-breaking temperature events in Spain}
  \author{Jorge Castillo-Mateo \\
  Department of Statistical Methods, University of Zaragoza \\
  Alan E. Gelfand \\
  Department of Statistical Science, Duke University \\
  Zeus Gracia-Tabuenca \\ 
  Department of Statistical Methods, University of Zaragoza \\
  Jes\'us As\'in \\
  Department of Statistical Methods, University of Zaragoza \\
  Ana C. Cebri\'an \\
  Department of Statistical Methods, University of Zaragoza \\}
  \maketitle
} \fi

\if0\blind
{
  \bigskip
  \bigskip
  \bigskip
  \begin{center}
    {\LARGE\bf Spatio-temporal modeling for record-breaking temperature events in Spain}
\end{center}
  \medskip
} \fi

\bigskip
\begin{abstract} 
Record-breaking temperature events are now very frequently in the news, viewed as evidence of climate change. With this as motivation, we undertake the first substantial spatial modeling investigation of temperature record-breaking across years for any given day within the year. We work with a dataset consisting of over sixty years (1960--2021) of daily maximum temperatures across peninsular Spain. Formal statistical analysis of record-breaking events is an area that has received attention primarily within the probability community, dominated by results for the stationary record-breaking setting with some additional work addressing trends. Such effort is inadequate for analyzing actual record-breaking data. Effective analysis requires rich modeling of the indicator events which define record-breaking sequences. Resulting from novel and detailed exploratory data analysis, we propose hierarchical conditional models for the indicator events. After suitable model selection, we discover explicit trend behavior, necessary autoregression, significance of distance to the coast, useful interactions, helpful spatial random effects, and very strong daily random effects. Illustratively, the model estimates that global warming trends have increased the number of records expected in the past decade almost two-fold, $1.93$ $(1.89,1.98)$, but also estimates highly differentiated climate warming rates in space and by season.
\end{abstract}

\noindent%
{\it Keywords:} Autoregression; Hierarchical Model; Logistic Regression; Random Effects; Trend

\newpage

\section{Introduction}

Spain has experienced some of the most significant temperature increments in Europe over the past few decades, with temperatures rising faster than the global average \citep{lionello2018}. After the heatwave that hit Europe in 2003, which caused at least 6,600 deaths in Spain, the Mortality Monitoring System (MoMo), coordinated by the Spanish Ministry of Health, was implemented as part of the plan of preventive actions against the effects of excessive temperatures \citep{linares2017}. Nevertheless, temperature records have continued to be broken over the past decade due to the increasing frequency of heatwaves \citep{sousa2019,diaz2023}, bringing unprecedented side effects in terms of human lives, environmental damage, and economic disruptions \citep{cardil2015}. In particular, on August 14, 2021, the province of C\'ordoba in Spain set a new record for the highest temperature ever recorded in the country, while other stations also broke their own records \citep{WMO2022}. Furthermore, the breaking of temperature records is not confined to Spain; it is becoming increasingly frequent worldwide \citep{NYT}. 

This work focuses on the study of calendar day records because the daily scale captures the full range of variability, including short-lived events, such as heatwaves, to which society and ecosystems are critically vulnerable and not necessarily adapted. They yield important consequences in agriculture and biological systems, since different types of crops, livestock and soil moisture are critically sensitive to daytime and nighttime temperature records throughout the year \citep{battisti2009}. The frequency of calendar day records in a period of time and area is a useful indicator of the effect of climate change in the upper tail of temperature, and it is a commonly used metric for the detection and attribution of anthropogenic climate change \citep[see, e.g.,][]{elguindi2013,pan2013}. An advantage of this metric is that the number of calendar day records in a year, season, or month allows a fair comparison of the effects across locations or climates and also across periods of the year.  The international meteorological offices also employ calendar day records as a relevant metric to describe climate; see, e.g., the US National Oceanic and Atmospheric Administration \citep{NOAA}, the Copernicus Climate Change Service \citep[][Events]{ESOTC2020}, or the Spanish State Meteorological Agency \citep[][Section~1.1.3]{aemet23}. 

Records occur even in a stationary climate. Therefore, it is crucial to address the question of whether the observed record rates would be expected without the influence of climate change, and to quantify any increase in the rate if it exists.  In this regard, our effort far exceeds establishing that the occurrence of daily temperature records is substantially different to that expected in a stationary climate.  We seek to quantify their occurrence over time and to identify if the occurrence pattern varies across space, within the year, or both. From a statistical point of view, this is a challenging task due to the intrinsic definition of a record, which results in few observations. Additionally, the magnitude of global warming trends is small compared to the natural variability of daily temperature. Modeling the occurrence of records in Spain is even more challenging due to diverse climate and topography, leading to relevant regional variations \citep{sillmann2017}. As far as we know, most significant effort has been devoted to analyzing the occurrence of records using exploratory analyses or simplified probabilistic models.  However, these approaches fall short with regard to capturing the intricate dependence structure of the underlying climate processes which span both space and time.

We propose a space-time modeling for the occurrence of calendar day records to better understand and quantify their occurrence in the context of global warming. We are not attempting a broader study of the enormous challenge of climate change.  Rather, we are studying records as evidence of climate change. In particular, we address these challenges by developing a Bayesian hierarchical model that accounts for large- and small-scale variation---mean behavior as well as spatial and temporal stochastic dependence--- to obtain high-resolution posterior predictive realizations that can be used for needed inference. Here, we present results based on a dataset comprising daily maximum temperatures from 40 locations across peninsular Spain spanning the years 1960--2021. Records are considered for individual calendar days across years, resulting in 365 time series for each location.

We model record-breaking in terms of annual time series of binary indicators that describe whether or not a record was broken in a given year. We could have modeled the daily temperatures and let this induce the record-breaking model to avoid a potential loss of information. However, while there is a loss of information to explain daily temperatures, there is none for explaining records. A model for daily temperatures would be driven by the bulk of the distribution, where most data is observed, and would yield poor fits for the upper tail and the records. In particular, the covariates influencing the mean may not be relevant for the occurrence of records, and the time evolution of the mean temperature is not the same in the tails \citep{castillo2023c}. Furthermore, a mean model for daily temperature will not incorporate the specific annual trend term for records with interactions that we employ to assess departure from stationarity in records. Another benefit of modeling the indicators is that we do not have to specify a distribution for daily temperatures. There are other models in the literature for capturing different aspects of temperature, such as models for exceeding thresholds or for quantiles, but neither would capture the specific record-breaking behavior.

\subsection{Probabilistic approaches for analyzing records}

Initiated by \cite{chandler1952}, probabilistic  properties of record events have been pursued quite extensively \citep[see][which provides a comprehensive treatment of the topic]{arnold1998}. Given a time series of random variables $(Y_1,\ldots,Y_T)^\top$, an observation $Y_t$ is called a \emph{record} if its value exceeds that of all previous observations, i.e., if $Y_t > \max\{Y_1,\ldots,Y_{t-1}\}$. By definition, $Y_1$ is always considered a \emph{trivial} record. The occurrence of records is completely specified by the sequence of record indicator random variables $(I_1,\ldots,I_T)^\top$, where $I_t$ takes the value $1$ if $Y_t$ is a record and $0$ otherwise. Then, the number of records up to time $t$ is given by $N_t = \sum_{j=1}^{t} I_j$.

The classical record model \citep[CRM;][]{arnold1998} characterizes records arising in a series of continuous i.i.d. (c.i.i.d.) random variables and provides the expected behavior in stationary climatic series. The main property of the variables associated with the occurrence of records under the CRM is that they do not depend on the underlying distribution of the c.i.i.d. variables. The record indicators are mutually independent and follow a Bernoulli distribution with probability $p_t = P(I_t = 1) = 1/t$ for $t = 1,\ldots,T$. Then, the expected number of records up to time $t$ grows as the logarithm of the number of variables, $E[N_t] = \sum_{j=1}^{t} 1 / j = \log(t) + \gamma + O(1/t)$, where $\gamma$ is the Euler-Mascheroni constant.

In climate data, stationary evolution over time is an unrealistic assumption and an alternative specification for the temperature series is the linear drift model \citep[LDM;][]{ballerini1985}, $Y_t = c t + \epsilon_t$, where $c > 0$ is a constant and $\epsilon_t$ are c.i.i.d. random variables. Under this setup, the probabilities of record $p_{t}$ depend on the underlying distribution of $\epsilon_t$ and are rarely known analytically \citep[see][for an asymptotic investigation]{franke2010}. \cite{gouet2020} generalized some of the LDM results for $\delta$-records (observations higher than the previous record plus a constant $\delta$) and used them to characterize $\delta$-records in monthly mean temperatures.

The distribution-free properties under the CRM have been used to develop statistical hypothesis tests to detect non-stationary behavior in the occurrence of extreme and record events in temperature \citep{benestad2003,benestad2004,cebrian2022a,castillo2022,castillo2023}. These and other probabilistic results have been widely used to compare the occurrence of records in observed or gridded temperature datasets with the expected behavior under the CRM or the LDM \citep{rahmstorf2011,coumou2013,wergen2014,mcbride2022,castillo2023b}, and also in climate model projections \citep{elguindi2013,pan2013,fischer2021}. Most of this work has found more records than expected under a stationary climate. In general, the Gaussian LDM or mild generalizations capture observed record behavior better than the CRM.  However, many authors highlight that a simple trend or normal errors are not good enough to explain the actual occurrences.

\subsection{New contributions}

Although the probabilistic properties of records have been widely applied, they must be seen as useful exploratory tools which model the probabilities of record \emph{marginally}. They are unable to capture temporal dependence between contiguous series in time or spatial dependence across locations. A further shortcoming of these probabilistic approaches is that---beyond the independent stationary case---they rarely offer uncertainty associated with their point estimates. 

To better understand the latent spatio-temporal processes that drive the climatological dependence of temperature records in Spain, the contribution of this manuscript is to propose space-time \emph{conditional} models for the record indicators and complementary model-based inference tools. The pooling of data with joint modeling is especially important when studying records because these events, though rare, are highly dependent. We work in a Bayesian framework, using data augmentation Markov Chain Monte Carlo (MCMC) for model fitting. As a result, we provide a fully model-based approach that enables full inference including uncertainty quantification regarding regression coefficients and features related to the occurrence of records over years for any day at any location within the region.

Let $(I_{1\ell}(\bs),\ldots,I_{T\ell}(\bs))^{\top}$ denote the sequence of record indicators across years within day $\ell$ at location $\bs$. We model the probability of record $p_{t\ell}(\bs)$ at day $\ell$ within year $t = 2,\ldots,T$ at location $\bs$ using a version of a logistic regression model. With regard to the Bernoulli distribution of the indicators, we introduce suitable fixed effects and daily spatial random effects, specified within a fully Bayesian hierarchical structure on the logit scale. For long-term trends, we consider forms like $p_t = 1 / t^{\alpha}$ which, on the logit scale yields $-\log(t^{\alpha} - 1)$. To simplify to an asymptotically equivalent linear expression, we adopt $- \alpha \log(t - 1)$. With another link function $g$, we could generalize $p_{t} = g^{-1}(\alpha_{0} + \alpha_{1} g(1/t))$ where $\alpha_{0} = 0$ and $\alpha_{1} = 1$ give the probability of record in the CRM. The $\alpha$ coefficients can be interpreted as a rate of deviation from stationarity. The logit model may be preferred over other link functions due to its easier interpretation through odds ratios (OR's). For large datasets the probit model can offer a faster MCMC algorithm \citep{albert1993}. Further, our modeling also captures seasonal behavior through harmonic terms, persistence by conditioning on the occurrence of records for the previous days, the influence of geographical variables such as distance to the coast, and informative interactions. 

The remainder of the manuscript is organized as follows. Section~\ref{sec:data&EDA} presents the daily temperature data and an exploratory analysis of their record indicators. Section~\ref{sec:methods} introduces the proposed spatial logit model along with fully Bayesian model-based inference tools and validation metrics. Section~\ref{sec:results} presents a comprehensive model comparison and results from the selected model. Finally, Section~\ref{sec:summary} concludes the manuscript with a summary and future work. 
The Supplementary material contains additional exploratory analyses, a detailed MCMC algorithm, model assessment, convergence diagnostics, and further results.

\section{Data and exploratory analysis} \label{sec:data&EDA}

\subsection{Study region and weather stations}

\begin{figure}[t]
    \centering
    \includegraphics[width=.8\textwidth]{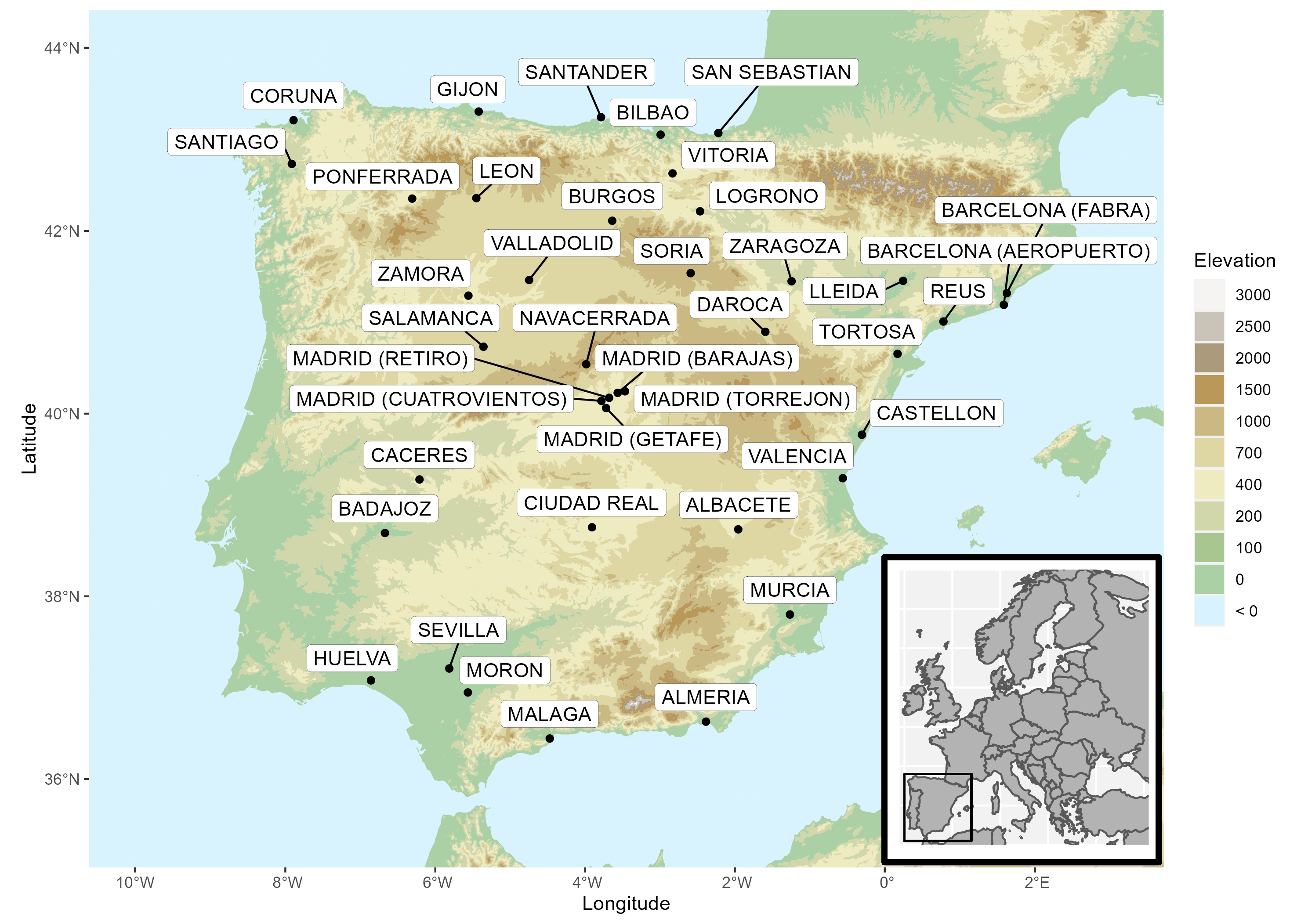}
    \caption{Map of the 40 Spanish stations in southwestern Europe. }
    \label{fig:map}
\end{figure}

We focus on the occurrence of record-breaking temperature events over peninsular Spain, i.e., the area that comprises Spain within the Iberian Peninsula, an area of $492,175$ km$^{2}$. The point-referenced dataset contains daily maximum temperature observational series, from January 1, 1960 to December 31, 2021, obtained from the European Climate Assessment \& Dataset \citep[ECA\&D;][]{tank2002}, at 40 sites. Given that the analysis focuses on the temperature record indicators defined within a day across years, the dataset is organized as 365 binary series (February 29 is removed for convenience) of length 62 for each site, a total of $905,200$ observations. Figure~\ref{fig:map} shows  the stations within the Iberian Peninsula. Spain is a diverse geographic region with several mountain ranges such as the Pyrenees in the northeast, the Central Plateau in the center, and the Sierra Nevada  near the Mediterranean southern coast. The region has a long coastline bordered by the Atlantic Ocean to the north and west, and the Mediterranean Sea to the south and east. The stations are irregularly distributed across the region  representing the diverse climatic zones in Spain, with resulting pairwise site distances ranging from $9$ to $907$ km.  The stations span a wide range of elevations, including five above 800 m, and 16 located on the coast.

We only included stations from the ECA\&D with a minimum of $99.5\%$ reliable data over 1960--2021. There are 53 stations with this requirement over peninsular Spain, but 13 of them are removed from the dataset because they do not meet additional quality criteria as described in Section~1.1 of the Supplementary material.
The retained stations have a small amount of missingness, on average, $0.07\%$ missing values.  To address this, we assigned a value of $-\infty$ to missing data.  Consequently, missing values are only considered records if they appear at $t=1$. A simulation study shows that the impact of this missing data on our results is negligible; see Section~1.2 of the Supplementary material.

\subsection{Data precision and tied records}

Spanish temperature series available in ECA\&D are provided by AEMET and they are measured to the nearest $1/10$th of a $^{\circ}$C.  This rounding/discretization results in some ties when records are identified.  To deal with ties, an observation that is at least as large as any previous observation is called a \emph{weak} record \citep[][Chapter~2]{arnold1998}.  Here, we define a \emph{tied} record in terms \emph{equal} rather than \emph{higher or equal}, i.e., an $r$-tied record ($r \ge 2$) arises when an observation shares the same value with $r - 1$ preceding weak records.  A $3$-tied record has the same value as the two tied previous weak records.  The record before this $3$-tied record was a $2$-tied record, the record before that was a record in the classical sense, and the three of them are weak records.

Among all non-trivial and weak records, the proportion of tied records across stations ranges from $4.2\%$ to $13.5\%$, with a station-wise average of $7.4\%$. The proportion of ties is almost constant across years. Out of a total of $890,600$ observations after the first trivial year, $58,981$ are classified as records, $4,352$ are $2$-tied records, $340$ are $3$-tied records, $30$ are $4$-tied records, and only $4$ are $5$-tied records. 

To accommodate the tied records within the Bayesian framework, we assume that each of the true daily temperatures roundings are i.i.d., following a distribution on the interval $(\text{observed} - 0.05, \text{observed} + 0.05)$.  Therefore, for an $r$-tied record in the rounded data series, the probability of it being a record in the \emph{true} daily temperature series is $1 / r$.  Indicators corresponding to $r$-tied records are sampled from their $\text{Bernoulli}(1 / r)$ distribution at the beginning of each iteration of the MCMC model fitting algorithm.

\subsection{Exploring the occurrence of records} \label{sec:EDA}

An exploratory analysis was conducted to evaluate departures from stationarity in the occurrence of records, identify related regressors, and enable a data-driven variable selection for the model. The covariates studied as fixed effects aim to capture spatio-temporal variability; i.e., annual trend, persistence (dependence on previous days), seasonal behavior, and geographic features. For simplicity, indicators associated with tied records are assigned a value of $0$ in this exploratory analysis. In order to avoid an over-representation of the Madrid region, where there are five stations, only the Retiro series is included in the exploratory analysis, but all of them are used for model fitting in Section~\ref{sec:methods}.

Two types of exploratory tools are used in the analysis: (i) graphical tools, and (ii) exploratory global or local logit models whose response is the vector resulting from stacking the record indicator variables $I_{t\ell}(\bs)$ from all or individual observed sites, respectively. Models with different regressors are compared using the AIC as a goodness of fit measure. This comparison is used to confirm or not the need for the considered temporal and spatial fixed effects; see Section~1.3.4 of the Supplementary material.

\begin{table}[tp]
\centering
\begin{tabular}{lc} 
    \hline
    Linear predictor & AIC \\
    \hline
    $\text{offset}(-\log(t-1)) - 1$   & 343369.2 \\
    $\log(t-1)$                       & 342238.3 \\
    $\text{poly}(\log(t-1), 2)$       & 340112.8 \\
    \ \ $+ \text{ persistence terms}$ & 293101.5 \\
    \ \ $+ \text{ seasonal terms}$    & 292821.8 \\
    \ \ $+ \text{ spatial terms}$     & 292030.7 \\ 
    \hline
\end{tabular}
\caption{Nested fixed effects models and AIC; see Table~3 of the Supplementary material.}
\label{table:EDA:glm:aic}
\end{table}

\begin{figure}[tbp]
    \centering
    \includegraphics[width=0.45\textwidth]{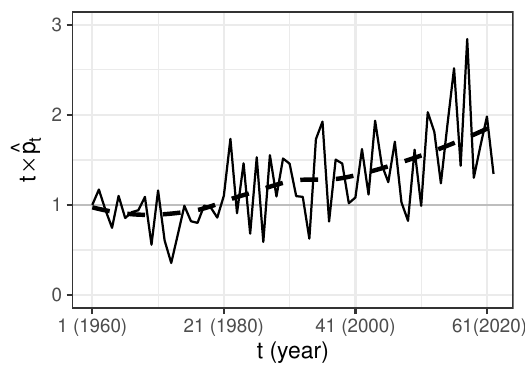}
    \includegraphics[width=0.45\textwidth]{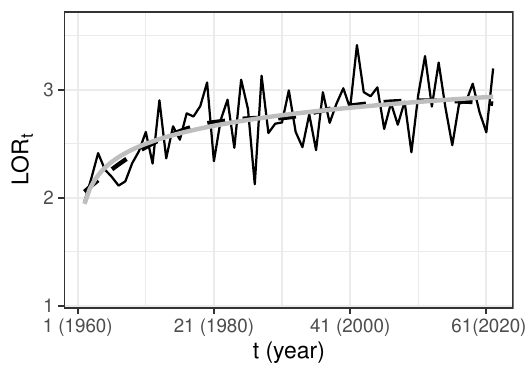} \\
    \includegraphics[width=0.45\textwidth]{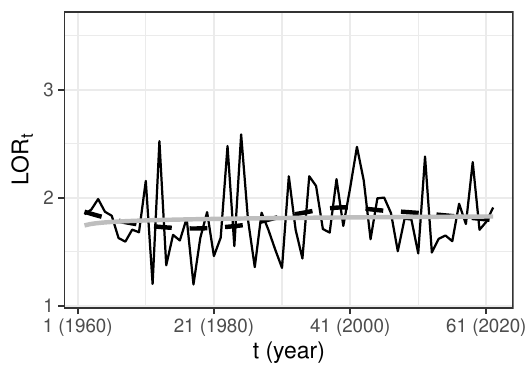}
    \includegraphics[width=0.45\textwidth]{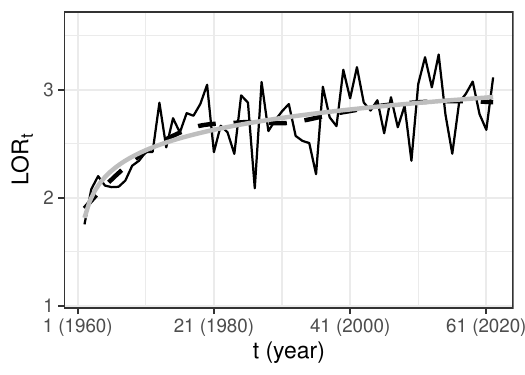}
    \caption{Top--Left: Evolution of $t \times \hat{p}_{t}$ against $t$ with reference value $1$. Top--Right: $LOR_t$ comparing the probability of record given a record the previous day and the probability of record without a record the previous day. Bottom--Left: $LOR_{t}$ comparing the probabilities $p_{t \mid 11}$ and $p_{t \mid 01}$. Bottom--Right: $LOR_{t}$ comparing the probabilities $p_{t \mid 10}$ and $p_{t \mid 00}$. Gray lines are linear models fitted to $\log(t-1)$ (except Top--Left) and dashed lines are LOESS curves.}
    \label{fig:EDA:2nd:LORt}
\end{figure}

\paragraph{Non-stationarity.} The probability of record at time $t$ depends on $t$ (again, $p_t = 1/t$ in the stationary case). Consequently, a model in the logit scale must at least include a trend term $\log(t-1)$, which in the stationary case would have a coefficient roughly equal to $-1$. To explore whether that term is enough to capture temporal evolution across years, the top left plot in Figure~\ref{fig:EDA:2nd:LORt} shows $t \times \hat p_t$ against $t$, whose expected value under stationarity is $1$. An empirical estimate of $p_t$ is obtained by averaging across space and days within year, $\hat{p}_{t} = \sum_{i=1}^{36} \sum_{\ell=1}^{365} I_{t\ell}(\bs_i) / (36 \times 365)$. The plot shows a strong deviation from $1$ starting around $t = 21$ (year 1980) and going up to $2$ in the final years, suggesting that probabilities of record are much higher than in a stationary climate. To allow a flexible modeling of the deviation from stationarity, the inclusion of a polynomial function of $\log(t-1)$ is considered. The first three rows in Table~\ref{table:EDA:glm:aic} show the AIC for three logit models including an offset $-\log(t-1)$, the covariate $\log(t-1)$, and the first and second degree orthogonal polynomials of $\log(t-1)$. According to the AIC, a third order polynomial is unnecessary and the second is preferred; see also Table~3 of the Supplementary material.

In series that are not i.i.d., especially those with an underlying trend like daily temperatures under global warming, the probability of record is a function of time that can vary across series with different distributions, e.g., across series from different locations or measured on different days. The following exploratory analysis assesses which spatio-temporal factors interact with that function, in particular with $\log(t-1)$. Another important feature to study in the exploratory analysis is the existence of temperature persistence, i.e., whether the probability of a record depends on the occurrence of records on previous days.

\paragraph{Persistence.} To study the dependence between the occurrence of records in two consecutive days, we consider the joint distribution $[I_{t\ell}(\bs),I_{t,\ell-1}(\bs)]$ expressed in terms of $2 \times 2$ tables. The evolution of this dependence across years is studied with two-way tables obtained by summing across space and days within year; see Section~1.3.1 of the Supplementary material. The empirical log OR's provide a useful tool for learning about persistence for both records and non-records,
\begin{equation*}
  LOR_t = \log \frac{(n_{t,11}+0.5)(n_{t,00}+0.5)}{(n_{t,01}+0.5)(n_{t,10}+0.5)},
\end{equation*}
where $n_{t,jk} = \sum_{i=1}^{36} \sum_{\ell=1}^{365} \bone(I_{t\ell}(\bs_i)=j, I_{t,\ell-1}(\bs_i)=k)$ for $j,k \in \{0,1\}$ denotes the frequencies in each cell of the table, and $0.5$ is a customary continuity correction. For notation convenience $I_{t0}(\bs) \equiv I_{t-1,365}(\bs)$. The $LOR_{t}$ above compares the probabilities of record given a record or a non-record the previous day. Values close to $0$ express independence while positive values capture persistence. The top right plot in Figure~\ref{fig:EDA:2nd:LORt} shows the $LOR_t$ against $t$: the values are clearly different from $0$, moving from $2$ at the beginning to $3$ at the end, indicating a strong persistence increasing across years. The linear relationship observed between $LOR_t$ and $\log(t-1)$ suggests the inclusion of the autoregressive term $I_{t,\ell-1}(\bs)$ and the interaction $\log(t-1) \times I_{t,\ell-1}(\bs)$ in the model. The interaction allows a different coefficient of $\log(t-1)$ in the model for the probabilities of record depending on whether the previous day was a record or not.

Given the strong persistence of temperature, the introduction of second-order autoregressive terms is also considered. To that end, the joint distribution $[I_{t\ell}(\bs), I_{t,\ell-1}(\bs), I_{t,\ell-2}(\bs)]$, which can be captured by a $2 \times 2 \times 2$ table, is used to extract the conditional distribution $[I_{t\ell}(\bs) \mid I_{t,\ell-1}(\bs), I_{t,\ell-2}(\bs)]$. The evolution of this second-order dependence is studied across years using the four conditional probabilities of record in a day given the occurrence or not of a record one and two days ago; i.e., $p_{t \mid 11}$, $p_{t \mid 01}$, $p_{t \mid 10}$, and $p_{t \mid 00}$. Again, the analysis is based on empirical $LOR_t$'s calculated by summing across days within year and sites; see Section~1.3.1 of the Supplementary material. The bottom plots in Figure~\ref{fig:EDA:2nd:LORt} show the $LOR_{t}$'s that compare  $p_{t \mid 11}$ and $p_{t \mid 01}$, the two conditional probabilities given that two days ago there was a record, and $p_{t \mid 10}$ and $p_{t \mid 00}$, the two conditional probabilities given that two days ago there was not a record. The evolution of the two $LOR_t$'s is very different. The former seems constant across years but higher than $0$, and the latter is quite similar to the $LOR_t$ comparing probabilities $p_{t \mid 1}$ and $p_{t \mid 0}$; this similarity is clearly due to the scarce number of records. These plots suggest that different coefficients of $\log(t-1)$ should be allowed depending on the occurrence of records or not one and two days ago; i.e., $\log(t-1) \times I_{t,\ell-2}(\bs)$ and $\log(t-1) \times I_{t,\ell-1}(\bs) \times I_{t,\ell-2}(\bs)$, and $I_{t,\ell-2}(\bs)$ and $I_{t,\ell-1}(\bs) \times I_{t,\ell-2}(\bs)$ should be included in the model. The similarity between the LOESS curve and the fitted values of the linear model including $\log(t-1)$ suggests that interactions with higher order trend polynomials are not necessary. The AIC values in Table~\ref{table:EDA:glm:aic} confirm the need for the persistence terms.

\paragraph{Seasonality.} Many studies have found that global warming and, in particular, the increase in incidence of extreme temperatures is not homogeneous within the year. In Spain, \cite{castillo2023b} found that the increase in the probabilities of record is significant in summer but not in autumn and relevant differences are also found between months. To allow different annual evolution for days within year, a harmonic given by the terms $\sin(2 \pi \ell / 365)$ and $\cos(2 \pi \ell / 365)$, and its interactions with the first and second degree orthogonal polynomials of $\log(t-1)$, are included in the model. The need for these terms is confirmed by the exploratory logit models in Table~\ref{table:EDA:glm:aic}, and supported by the different patterns in $t\times \hat p_t$ lines by season in Figure~1 of the Supplementary material.

\begin{figure}[t]
    \centering
    \includegraphics[width=.45\textwidth]{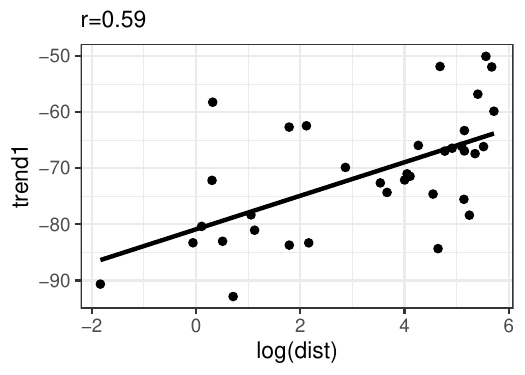}
    \includegraphics[width=.45\textwidth]{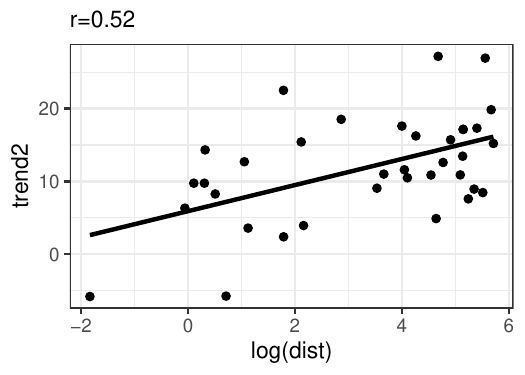}
     \caption{Estimated coefficients of the first (left) and second (right) degree orthogonal polynomials of $\log(t-1)$ in local logit models for each observed site against $\log(\dist(\bs))$.}
    \label{fig:EDA:dist:interac}
\end{figure}

\paragraph{Spatial variability.} The effect of global warming on the occurrence of records is not spatially homogeneous. Previous work \citep{coumou2013} and the characteristics of the study area  suggest the need for including spatial covariates. Latitude, longitude, and elevation were explored but excluded as described in Section~1.3.3 of the Supplementary material. \cite{mcbride2022} and \cite{castillo2023b} found that the increase in the number of records was different between coastal and inland stations both in South Africa and Spain, respectively. The minimum distance to the coast was incorporated into the model as an interaction between the trend terms and a functional form of distance to the coast in km, $\dist(\bs)$. A suitable expression was identified by fitting, for each observed site, local logit models with the covariates above (trend, persistence, and seasonality). The relationship between the estimated coefficients of the trend terms for each site and different functions of $\dist(\bs)$ were explored. Figure~\ref{fig:EDA:dist:interac} shows a linear relationship between those coefficients and $\log(\dist(\bs))$. Analogous results were obtained for the interaction between the persistence terms and $\log(\dist(\bs))$; see Section~1.3.3 of the Supplementary material. The AIC for the logit model with spatial terms in Table~\ref{table:EDA:glm:aic} confirms the need for these terms.

In summary, this exploratory analysis argues, in detail, for the specification of the fixed effects in the full logit model that includes 20 covariates. The resulting fixed effect terms are written precisely in the next section.

\section{Model specifics} \label{sec:methods}

This section proposes a rich spatial logistic regression model across days, for annual temperature records, motivated by the exploratory analysis in Section~\ref{sec:EDA}. It provides details for model fitting and also spatial interpolation at unobserved locations, in a Bayesian framework, to extract desired posterior inference for any spatio-temporal feature of the occurrence of records. The section concludes with proposed metrics for model comparison.

\subsection{Spatial modeling for record breaking} \label{sec:model}

Let $I_{t\ell}(\bs)$ denote the record indicator of the daily maximum temperature for day $\ell$, $\ell = 1,\ldots,365$, within year $t$, $t=1,\ldots,T$, at location $\bs$, $\bs \in D$ with $D$ being peninsular Spain, for the series starting on January 1, 1960; so, $t=1$ corresponds to 1960 and $T = 62$ corresponds to 2021.  We model record indicators beginning with day $\ell=3,\ldots,365$, year $t=2,\ldots,T$, and location $\bs$ according to
\begin{equation} \label{eq:model}
    I_{t\ell}(\bs) \mid I_{t,\ell-1}(\bs), I_{t,\ell-2}(\bs) \sim \text{Bernoulli}(g^{-1}(\eta_{t\ell}(\bs))) \quad \text{with} \quad
    \eta_{t\ell}(\bs) = \bx_{t\ell}(\bs) \bbeta + w_{t\ell}(\bs),
\end{equation}
where $g(p) = \log\{p / (1 - p)\}$ is the logit link function. Here, $p_{t\ell}(\bs) = g^{-1}(\eta_{t\ell}(\bs))$ is the probability of a record for day $\ell$, year $t$, and location $\bs$, with fixed effects $\bx_{t\ell}(\bs) \bbeta$ and random effects $w_{t\ell}(\bs)$. The $\bx_{t\ell}(\bs) = (1, x_{t\ell 1}(\bs), \ldots, x_{t\ell k}(\bs))$ are $k+1$ covariates measured on day $\ell$, year $t$, and location $\bs$ with $\bbeta$ a column vector of length $k+1$ of regression coefficients. The $w_{t\ell}(\bs)$ are space-time correlated errors.

We first supply the fixed effects component.  Apart from the intercept, the entries are: 
\begin{itemize}

\item $\trendONE_{t}$ and $\trendTWO_{t}$, denoting the first and  second degree orthogonal polynomials of $\log(t-1)$; they enrich the trend in the probabilities of record across $t$ beyond the trend under a stationary climate.  

\item persistence terms  to  capture first  and second order autoregressive dependence, $I_{t,\ell-1}(\bs)$, $I_{t,\ell-2}(\bs)$, and $I_{t,\ell-1}(\bs) \times I_{t,\ell-2}(\bs)$, and  their interactions with $\log(t-1)$; this means that different intercepts and first-order time trends are allowed for each of the four possible  persistence situations defined by the occurrence or not of records in each of the two previous days.

\item seasonal terms including one harmonic, $\sin_{\ell} = \sin(2 \pi \ell / 365)$ and $\cos_{\ell} = \cos(2 \pi \ell / 365)$, and their interactions with the trend across years; these terms allow for seasonality in the probability of records within the year and enable it to vary across years. 

\item spatial terms to enrich the effect of $\log(\dist(\bs))$ including interaction with  trend across years and with persistence; the effect of distance from the coast varies with yearly trend and also according to persistence in the previous two days.

\end{itemize}
    
In summary, we have the intercept and the following 20 predictors:
\begin{equation*}
\begin{aligned}
    \bx_{t\ell}(\bs) = (
    &1, \trendONE_{t}, \trendTWO_{t}, I_{t,\ell-1}(\bs), I_{t,\ell-2}(\bs), I_{t,\ell-1}(\bs) \times I_{t,\ell-2}(\bs),\\ 
    &\log(t-1) \times I_{t,\ell-1}(\bs), \log(t-1) \times I_{t,\ell-2}(\bs), \log(t-1) \times I_{t,\ell-1}(\bs) \times I_{t,\ell-2}(\bs),\\     
    &\sin_{\ell}, \cos_{\ell}, \sin_{\ell} \times \trendONE_{t}, \cos_{\ell} \times \trendONE_{t}, \sin_{\ell} \times \trendTWO_{t}, \cos_{\ell} \times \trendTWO_{t} \\ 
    & \log(\dist(\bs)), \log(\dist(\bs)) \times \trendONE_{t}, \log(\dist(\bs)) \times \trendTWO_{t},\\
    &\log(\dist(\bs)) \times I_{t,\ell-1}(\bs), \log(\dist(\bs)) \times I_{t,\ell-2}(\bs), \log(\dist(\bs)) \times I_{t,\ell-1}(\bs) \times I_{t,\ell-2}(\bs)).\\
\end{aligned}
\end{equation*}
While 20 fixed effects terms may seem excessive, in explaining the roughly $900,000$ observations we do find all of them significant using the foregoing fixed effects logistic regression variable selection, with further support from the results in Section~\ref{sec:params}.

We introduce explicit spatial and temporal dependence through random effects. Possibilities for modeling $w_{t\ell}(\bs)$ include: (i) a spatially-varying intercept, $w_{t\ell}(\bs) = w(\bs)$; (ii)  an additive form with annual intercepts, $w_{t\ell}(\bs) = w(\bs) + w_{t}$; (iii) daily intercepts, $w_{t\ell}(\bs) = w(\bs) + w_{t\ell}$; and (iv) daily-spatially-varying intercepts, $w_{t\ell}(\bs)$. From (i) to (iii), spatial dependence is captured by $w(\bs)$ to provide local adjustments to the global intercept. It is supplied as a mean-zero Gaussian process with an exponential covariance function such that $\cov(w(\bs), w(\bs^{\prime})) = \sigma_{0}^{2} \exp\{-\phi_{0} \lvert\lvert \bs - \bs^{\prime} \rvert\rvert\}$, where $\sigma_{0}^{2}$ and $\phi_{0}$ are, respectively, the variance and decay parameters of the Gaussian process, and $\lvert\lvert \bs - \bs^{\prime} \rvert\rvert$ is the distance between $\bs$ and $\bs^{\prime}$.\footnote{More explicitly, $\lvert\lvert \bs - \bs^{\prime} \rvert\rvert$ is the Euclidean distance in km between $\bs$ and $\bs^{\prime}$ using the projected coordinate reference system for peninsular Spain called Madrid 1870 (Madrid) / Spain LCC, EPSG projection 2062 (\url{https://www.ign.es/}).} Temporal dependence is captured by $w_{t}$ or $w_{t\ell}$. These processes are modeled as $w_{t} \sim N(0, \sigma_{1}^{2})$ or $w_{t\ell} \sim N(0, \sigma_{1}^{2})$, respectively. An autoregressive specification is unnecessary for these $w$'s. That is, we implemented autoregression across years and found no significance while daily autoregression is already accounted for in the covariates. The full model (iv) imposes the most demanding implementation, modeling $w_{t\ell}(\bs)$ as Gaussian processes with mean $w_{t\ell}$ and a common exponential covariance function having variance and decay parameters $\sigma_{0}^{2}$ and $\phi_{0}$, as above.

For prediction, the autoregressive model requires an initial condition for $I_{t1}(\bs)$ and $I_{t2}(\bs)$, the first and second values in year $t$. Given the inherent break for each year $t$, we employ a distinct linear specification for modeling these indicators compared to \eqref{eq:model}. We model them as $\eta_{t1}(\bs) = \bx_{t1}(\bs) \bbeta_{1} + w_{t1}(\bs)$ and $\eta_{t2}(\bs) = \bx_{t2}(\bs) \bbeta_{2} + w_{t2}(\bs)$.  The covariate vectors are reduced to $\bx_{t1}(\bs) = (1, \trendONE_{t}, I_{t-1,365}(\bs))$ and $\bx_{t2}(\bs) = (1, \trendONE_{t}, I_{t1}(\bs))$. The $w_{t\ell}(\bs)$ are modeled as above, each a Gaussian process with mean $w_{t\ell} \sim N(0, \sigma_{1,\ell}^{2})$ and exponential covariance function having variance $\sigma_{0,\ell}^{2}$ and the same decay parameter $\phi_{0}$.

\subsection{Prior specification and model fitting}

\paragraph{Prior distributions.} Model inference is implemented in a Bayesian framework. Adopting the data augmentation approach described below, we can use the same conjugate priors as in a standard geostatistical linear model with normal errors. We assign proper but weakly informative priors as follows. Let $\bv \sim N_n(\bmu, \bSigma)$ be an $n$-dimensional random vector that follows a multivariate normal distribution with mean vector $\bmu$ and positive-definite covariance matrix $\bSigma$.  Let $\bone_n$ ($\bzero_n$) be the $n$-dimensional vector of ones (zeros) and $\bI_n$ the $n \times n$-dimensional identity matrix. The regression coefficients are assigned $\bbeta \sim N_{k+1}(\bmu_{\bm{\beta}}, \bSigma_{\bm{\beta}})$ with mean $\bmu_{\bm{\beta}} = \bzero_{k+1}$ and covariance $\bSigma_{\bm{\beta}} = 100^2 \bI_{k+1}$. Equivalently, $\bbeta_{\ell} \sim N_{3}(\bmu_{\bm{\beta}_{\ell}}, \bSigma_{\bm{\beta}_{\ell}})$ with mean $\bmu_{\bm{\beta}_{\ell}} = \bzero_{3}$ and covariance $\bSigma_{\bm{\beta}_{\ell}} = 100^2 \bI_{3}$ for days $\ell = 1,2$. The coefficient associated with $\log(t-1)$ could have a prior mean of $-1$ corresponding to a stationary climate but with the large prior variance there would be no practical difference in posterior inference. The variance parameters are each assigned an inverse gamma distribution, i.e., $\sigma_{0}^{2}, \sigma_{0,\ell}^{2}, \sigma_{1}^{2}, \sigma_{1,\ell}^{2} \sim IG(a_{\sigma}, b_{\sigma})$ with shape parameter $a_{\sigma} = 2$ and scale parameter $b_{\sigma} = 1$. 

Finally, we assign a gamma prior to the decay parameter $\phi_{0} \sim G(a_{\phi}, b_{\phi})$ where $a_{\phi}=2$ and $b_{\phi}=1$ as above. A prior considering the effective range, $3 / \phi_{0} \sim IG(a_{\phi}, 3 b_{\phi})$, the distance beyond which spatial association becomes negligible, could set $a_{\phi} = 2$ and $b_{\phi} = 100$.  This  prior has a mean of $300$ km,  roughly one-third of the maximum pairwise site distances, with infinite variance, and puts $90\%$ of the mass in $(63.2, 844.2)$ km. With this last prior, the posterior distribution placed the majority of its mass on relatively large values for the effective range, so we ultimately opted for the prior with the ``constraint" $3$ km mean and infinite variance. With the large amount of data, we did not find any inference sensitivity to the mean of this prior, or to the hyperparameters of the prior distribution for the regression or variance parameters.

\paragraph{Data augmentation approach.} Specification of the logit model in \eqref{eq:model}, following \cite{held2006} using latent standard logistic variables, leads to the augmented model $I_{t\ell}(\bs) = 1$ if $Y_{t\ell}(\bs) > 0$ and $0$ otherwise. Here, $Y_{t\ell}(\bs) = \bx_{t\ell}(\bs) \bbeta + w_{t\ell}(\bs) + \epsilon_{t\ell}(\bs)$ where we have i.i.d. $\epsilon_{t\ell}(\bs) \sim N(0, \lambda_{t\ell}(\bs))$, $\lambda_{t\ell}(\bs) = (2 K_{t\ell}(\bs))^2$, and $K_{t\ell}(\bs) \sim KS$ follows the asymptotic distribution of the Kolmogorov-Smirnov statistic.  In this case, $\epsilon_{t\ell}(\bs)$ follows a scale mixture of normal form with a marginal logistic distribution, so that the posterior distribution of the model parameters for the augmented model and for model \eqref{eq:model} are equivalent.  The advantage of working with this representation is that we have the usual conjugacy for the model parameters given the latent $Y_{t\ell}(\bs)$’s. Computational details of the MCMC algorithm used to fit the full model are given in Section~2 of the Supplementary material.
An alternative data augmentation approach for fitting Bayesian logistic regression models uses P\'olya-gamma latent variables \citep{polson2013}.

It is worth noting that the specification above implies spatial dependence at the second modeling stage.  That is, the $I_{t\ell}(\bs)$'s are conditionally independent with $P(I_{t\ell}(\bs)=1 \mid \lambda_{t\ell}(\bs)) = \Phi((\bx_{t\ell}(\bs) \bbeta + w_{t\ell}(\bs)) / \lambda_{t\ell}^{1/2}(\bs))$ where $\Phi$ is the cdf of the standard normal distribution, but marginally, they are dependent with direct calculation of the dependence structure available. We are adopting the customary way of introducing spatial dependence in the geostatistical generalized linear model which introduces two sources of error \citep{diggle1998}. As far as the realized surfaces are concerned, $Y_{t\ell}(\bs)$ and $I_{t\ell}(\bs)$ are everywhere discontinuous. 

We could make the $Y_{t\ell}(\bs)$ surface continuous by removing the $\epsilon_{t\ell}(\bs)$.  Then, marginalizing over $w_{t\ell}(\bs)$ after carefully modifying its covariance function to ensure it follows a marginal standard logistic distribution, the indicator surface would be smooth with countable discontinuities.  However, in practice, it is not sensible to argue that the model with only spatial residuals explains the responses perfectly; we almost always envision some pure error which we pass on to the indicator process.  In different words, the conditional independence concern is related to what is typically referred to as ``micro-scale" dependence \citep[see, e.g.,][Chapter~6]{BCG}.  After we condition on the spatial random effects, we may still retain some fine-scale spatial dependence which we are modeling as noise.  Such dependence is known to be difficult to extract and, in the context of our very rich modeling for the indicator functions, will not affect inference.

\subsection{Spatial interpolation and inference} \label{sec:inference}

\paragraph{Spatial interpolation.} Our inference objective is the posterior predictive distribution for the record indicators (or their probabilities) for any location in the study region within the observed time period, i.e., for $I_{t\ell}(\bs_0)$ (or $p_{t\ell}(\bs_0)$) for any $\bs_0 \in D$. From \eqref{eq:model}, a posterior sample from $I_{t\ell}(\bs_0)$ can be obtained by sampling from a Bernoulli distribution with probability $p_{t\ell}(\bs_0)$. A sample of $p_{t\ell}(\bs_0)$ is obtained through the inverse link function taking a sample of the linear predictor $\eta_{t\ell}(\bs_0)$, which is a function of the model parameters and process realizations. Posterior samples for the parameters are available from each iteration of the MCMC algorithm, and posterior samples for the spatial processes are available using posterior samples of the parameters through usual Bayesian kriging \citep[][Chapter~6]{BCG}. Sampling must be done dynamically in $t$ and $\ell$ because the response variable depends on its own previous values. 

\paragraph{Model-based tools.} Employing $G_{D}$, a fine spatial grid for $D$, posterior samples from $I_{t\ell}(\bs)$ can be realized at each location in $ G_{D}$ for every day $\ell$, $\ell = 1,\ldots,365$, within year $t$, $t=2,\ldots,T$.  So, we can make inference about any feature of interest related to the occurrence of records. Let $1 \leq \ell_1 \leq \ell_2 \leq 365$ and $1 \leq t_1 \leq t_2 \leq T$. Then, a general feature of primary interest is the average cumulative number of records across days from $\ell_1$ to $\ell_2$ and across years from $t_1$ to $t_2$.  It is defined as
\begin{equation} \label{eq:N}
  \bar{N}_{t_1:t_2,\ell_1:\ell_2}(\bs) = \frac{1}{\ell_2 - \ell_1 + 1} \sum_{t=t_1}^{t_2} \sum_{\ell = \ell_1}^{\ell_2} I_{t\ell}(\bs).
\end{equation}
The average total number of records $\bar{N}_{T}(\bs)$ arises with $\ell_1 = 1$, $\ell_2 = 365$, $t_1 = 1$ and $t_2 = T$. Also of interest is the number of records in the observed period of the 21st century and its comparison across seasons, leading to the statistics $\bar{N}_{(T - 21):T,\text{DJF}}(\bs)$, $\bar{N}_{(T - 21):T,\text{MAM}}(\bs)$, $\bar{N}_{(T - 21):T,\text{JJA}}(\bs)$, and $\bar{N}_{(T - 21):T,\text{SON}}(\bs)$, where DJF is winter (December, January, and February), MAM is spring, JJA is summer, and SON is autumn, with obvious notation. Other periods of interest could be the months of the past decade.

Comparison between the average number of records predicted by the model and the expected number of records under the stationary case, $E_0[\bar{N}_{t_1:t_2,\ell_1:\ell_2}(\bs)] = \sum_{t=t_1}^{t_2} 1 / t$ may be of interest. The ratio expression 
\begin{equation} \label{eq:ratio}
    R_{t_1:t_2,\ell_1:\ell_2}(\bs) = \frac{\bar{N}_{t_1:t_2,\ell_1:\ell_2}(\bs)}{E_0[\bar{N}_{t_1:t_2,\ell_1:\ell_2}(\bs)]},
\end{equation}
captures records expected by the model compared to a scenario without climate change.

Computing the above quantities for all $\bs \in G_{D}$ we can draw maps of the posterior mean or borders of the $90\%$ credible intervals (CI's) of the quantities of interest. This enables a useful picture of the spatio-temporal characteristics of the occurrence of records and assessment of regions and time periods with higher risk of exceeding temperatures above all of the previous measurements.

In a different direction, consider the extent of record surface (ERS) over a block $B \subseteq D$ of peninsular Spain for a given day $\ell$ within a given year $t$ \citep[see][for more details of the concept of extent]{Cebrian2022b}.  Suppose we compute 
\begin{equation} \label{eq:ERS}
    \widehat{\text{ERS}}_{t\ell}(B) = \frac{1}{\lvert G_{B} \rvert} \sum_{\bs_j \in G_{B}} I_{t\ell}(\bs_j),
\end{equation}
where $G_{B}$ is a fine spatial grid of $B$ and $\lvert G_{B} \rvert$ is the number of grid cells in $G_{B}$.  We define $ERS_{t\ell}(B)$ as the limit of this expression as the size of the grid cells go to $0$.  We interpret the limit as the proportion of $B$ which produced a record on day $\ell$ in year $t$.

We could compute $\widehat{\text{ERS}}_{t\ell}(B)$ for different blocks $B_{1}, B_{2} \subseteq D$ to compare trends in different subregions. Also, we can compute it for all of peninsular Spain, $D$, that would be the proportion of the country which is suffering a record-breaking temperature event on day $\ell$ within year $t$.  Following the above notation, we can go further and average these amounts for all or some $\ell$'s giving rise to a yearly evolution of the average number of records all over $B$, say $\widehat{\overline{\text{ERS}}}_{t}(B)$. Again, we can plot the posterior mean and $90\%$ CI of $t \times \widehat{\overline{\text{ERS}}}_{t}(B)$ against $t=1,\ldots,T$ to assess the nature of deviations from $1$. Alternatively, the cumulative sum of $\widehat{\overline{\text{ERS}}}_{t}(B)$ in $t=1,\ldots,T$ could be drawn as the proportion of the total number of records across the entire region.

\subsection{Model comparison and checking} \label{sec:metrics}

The performance of the models for record-breaking is compared using metrics for cross-validation and for the deviance.  Both approaches globally evaluate the goodness of fit of the models. The adequacy of the selected model is criticized using posterior predictive checks for features of interest. 

\paragraph{Cross-validation.} Model comparison is considered in the context of performance of the interpolation across the 40 locations. In particular, the models are compared using 10-fold cross-validation. This means that the dataset is split at random into 10 groups with data from four different sites for each group; see Section~3.1 of the Supplementary material. Then, each group of four sites is taken as hold-out and the model is fitted with the remaining 36 sites. Finally, posterior samples from the conditional probabilities of record for the hold-out sites are obtained using one-step ahead prediction. This is doable since the $I_{t,\ell-1}(\bs)$ and $I_{t,\ell-2}(\bs)$ to condition on are known (or we use the data rounding strategy for indicators associated to tied records). Indicators associated with tied records are not included in the metrics because we do not know their true value.

Here, we consider two metrics for comparing model performance and a third metric is given in Section~3.1 of the Supplementary material. The first of these is based on the Brier score \citep[BS;][]{brier1950} which, for binary events, is a strictly proper scoring rule \citep{czado2009} given by the squared difference $[I_{t\ell}(\bs) - \hat{p}_{t\ell}(\bs)]^2$, where $\hat{p}_{t\ell}(\bs)$ is the posterior mean of the probability of record on day $\ell$ within year $t$ at a location $\bs$ in a group given data in the remaining nine groups of locations. We average the BS's over all observations, and a lower score indicates better performance. Since the BS penalizes false negatives and false positives equally, it may not be the most suitable metric for record-breaking events because of the unbalanced nature of the data. As a consequence, we also consider the area under the receiver operating characteristic curve (AUC) obtained via the \texttt{R}~package \texttt{pROC} \citep{pROC} for each held-out group one by one. We average the AUC's across all groups; its value ranges from $0$ to $1$ and a higher value indicates better performance.

\paragraph{Deviance.} Model comparison is also considered in terms of deviance. A commonly used measure in this regard for Bayesian models is the deviance information criterion \citep[DIC;][]{spiegelhalter2002} which is the sum of two components, one for quantifying the model fit and the other for penalizing its complexity; see Section~3.2 of the Supplementary material. Models with smaller DIC are better supported by the data. 

\paragraph{Posterior predictive checks.} Model adequacy is criticized in terms of calibration and sharpness of the probabilistic forecasts or predictions. Scatter plots that display the actual observations for particular features versus a summary of their corresponding predictive distributions based on the posterior mean and $90\%$ CI are useful to observe the consistency between the predictions and the observations together with the concentration of the predictive distributions. 

Another diagnostic tool to identify model deficiencies is the probability integral transform (PIT). This is the value that the posterior cdf attains at an observation. If the observation is drawn from the predictive distribution and the distribution is continuous, the PIT has a standard uniform distribution. Here, we adapt the PIT histogram for discrete data proposed by \cite{czado2009}; see Section~3.3 of the Supplementary material.

\section{Results} \label{sec:results}

We first present model comparison with differing inclusion of the foregoing spatial and temporal dependence introduced in Section~\ref{sec:model} and model checking for the selected model. Then, we present the results for the full model over peninsular Spain. 

The models were fitted by scaling the covariates to have zero mean and unit variance to improve the mixing of the MCMC. For each model we ran two chains of the MCMC algorithm, each chain with different initial values, each to $200,000$ iterations for the full model (smaller models require fewer iterations), to obtain samples from the joint posterior distribution. The first half of the samples were discarded as burn-in and the remaining samples were thinned to $500$ samples from each chain. Section~4 of the Supplementary material contains usual convergence diagnostics for the MCMC samples.

\subsection{Model comparison and checking}

\paragraph{Model comparison.} The metrics in Section~\ref{sec:metrics} are used to compare the models in Table~\ref{tab:metricsKFCV} (BS, AUC, and DIC) using 10-fold cross-validation and the deviance. The models include the fixed effects already discussed; they differ from each other in the spatial and temporal random effects described in Section~\ref{sec:model}, with the full model $M_5$ being the richest specification. Comparison with the stationary model $M_0$ is solely for illustration.  With regard to the cross-validation model performance metrics, two periods for record breaking are considered.  The first employs the first 30 years of the series when records are more frequent; the second employs the last 31 years of the series when records are more rare.

\begin{table}
\centering
\begin{tabular}{clccccc} \hline
 Model & Linear predictor & BS 1 & BS 2 & AUC 1 & AUC 2 & DIC \\ \hline
 $M_{0}$ & $\eta_{t\ell}(\bs) = - \log(t - 1)$ & $7.88$ & $3.23$ & $0.733$ & $0.514$ & $393805.4$ \\
 $M_{1}$ & $ \eta_{t\ell}(\bs) = \bx_{t\ell}(\bs) \bbeta$ & $6.49$ & $2.91$ & $0.831$ & $0.686$ & $333193.6$ \\
 $M_{2}$ & $\eta_{t\ell}(\bs) = \bx_{t\ell}(\bs) \bbeta + w(\bs)$ & $6.49$ & $2.91$ & $0.831$ & $0.687$ & $333130.3$ \\
 $M_{3}$ & $\eta_{t\ell}(\bs) = \bx_{t\ell}(\bs) \bbeta + w(\bs) + w_{t}$ & $6.46$ & $2.90$ & $0.837$ & $0.716$ & $331142.9$ \\
 $M_{4}$ & $\eta_{t\ell}(\bs) = \bx_{t\ell}(\bs) \bbeta + w(\bs) + w_{t\ell}$ & $5.20$ & $2.48$ & $0.924$ & $0.904$ & $251653.5$ \\
 $M_{5}$ & $\eta_{t\ell}(\bs) = \bx_{t\ell}(\bs) \bbeta + w_{t\ell}(\bs)$ & $4.32$ & $2.13$ & $0.944$ & $0.924$ & $182680.2$ \\ \hline
\end{tabular}
\caption{The different nested models compared and their corresponding metrics. Metrics: $100 \times \text{BS}$ and AUC 1 in 1961--1990 and 2 in 1991--2021, and DIC.}
\label{tab:metricsKFCV}
\end{table}

Each dependence component included improves the performance of the model, with the full model being the best in all metrics. The most significant improvements are produced by the inclusion of covariates and the inclusion of daily-varying intercepts. However, the daily spatial effects also lead to a substantial gain. The cross-validation AUC values higher than $0.9$ for the full model, even for the period 1991--2021, indicate that the model has excellent ability to explain incidence of record and non-record events. The two parts of the DIC to account for model fit and complexity are shown in Table~4 of the Supplementary material. Additional disaggregated AUC's for the full model by decade, season and location are shown in Figure~8 of the Supplementary material; which is useful to know periods of time or spatial regions that the model reproduces better or worse. In general, performance is homogeneous between decades and seasons, while there is some differences between locations. Most AUC's are greater than $0.9$, except for M\'alaga and Almer\'ia, the two southernmost weather stations, which present some difficulties probably due to their orography.

\paragraph{Model checking.} The posterior predictive checks assess the adequacy of the full model for explaining record-breaking occurrences. Figure~9 of the Supplementary material shows the scatter plot and associated PIT histogram of $R_{53:62,month}(\bs)$ in \eqref{eq:ratio} for each observed location and $month$ going through the $12$ months.  The same is shown for $t \times \widehat{\overline{\text{ERS}}}_{t,season}(D)$ averaging in \eqref{eq:ERS} (defined over the grid of $40$ observed locations) for $t=2,\ldots,T$ going through the $4$ seasons. The discussion and figure in the Supplementary material effectively demonstrate very good model adequacy.

\subsection{Inference} \label{sec:params}

Here, we show results for the full model. First, a summary of the posterior distribution of the regression coefficients and other hyperparameters. Then, the model-based tools from Section~\ref{sec:inference} are used to make inference on the record-breaking events over peninsular Spain.

\begin{figure}[t]
    \centering
    \includegraphics[width=0.8\textwidth]{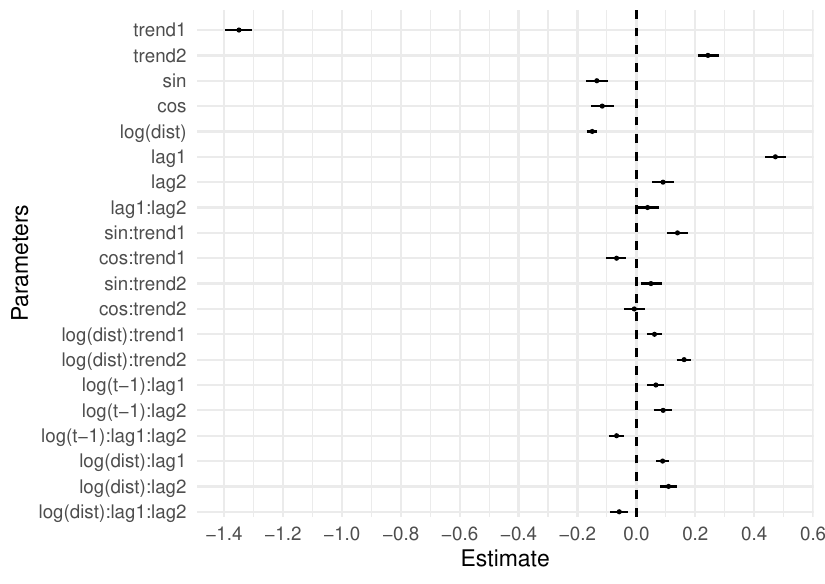}
    \caption{Posterior mean and $90\%$ CI of the  $\bm{\beta}$ coefficients from the full model with scaled covariates. \texttt{lag1}  and \texttt{lag2} denote the covariates $I_{t,\ell-1}(\bs)$ and $I_{t,\ell-2}(\bs)$, respectively.}
    \label{fig:coeffs}
\end{figure}

\paragraph{Posterior distribution of the model parameters.} The posterior mean and $90\%$ CI of the regression coefficients from the full model, with scaled covariates, are shown in Figure~\ref{fig:coeffs}. No CI includes zero except for the interaction between $\cos_{\ell}$ and $\trendTWO_{t}$. For ease of interpretation, Table~5 of the Supplementary material includes the posterior mean and $90\%$ CI for all model parameters, but expressed in terms of non-scaled covariates. Here, we analyze the OR's associated with persistence terms, and the hyperparameters. The trend is further studied below using model-based tools. 

The posterior distribution of parameters associated with $I_{t,\ell-1}(\bs)$, $I_{t,\ell-2}(\bs)$, and their interactions reveals a big increase in the probabilities of record due to short-term dependence. This increase is higher close to the coast or in later years.  For example, the occurrence of records gives an OR up to $20$ when the two previous days were records, $12$ when there was a record only on the previous day, and slightly lower than $3$ when  there was a record two days ago but not the previous day; see Section~5.2 of the Supplementary material.

The posterior mean and $90\%$ CI for the intercept $\beta_0$ are $-6.90$ $(-6.98,-6.80)$. For the variances of the random effects we have $11.19$ $(10.74,11.70)$ for $\sigma_{0}^{2}$, and $2.01$ $(1.57,2.43)$ for $\sigma_{1}^{2}$. These variances give evidence of the strong daily and spatial variability in the occurrence of records. Finally, the decay parameter $\phi_0$ is $0.0021$ $(0.0019,0.0022)$, which gives an effective range between $1390.6 $ and $1550.2$ km, yielding smooth surfaces indicating that record occurrence in even the two most distant observed sites within peninsular Spain have a correlation of $0.16$ $(0.14,0.17)$. This is in agreement with the co-occurrence of records in the region. Table~8 of the Supplementary material summarizes the posterior distribution of all the hyperparameters in the model.

\paragraph{Model-based analysis over peninsular Spain.} This section reports results for three of the large number of model-based tools that can be utilized to draw inferences from the model. Results from additional tools are shown in Section~5.3 of the Supplementary material. First, we need a fine grid $G_{D}$ of points within peninsular Spain, we choose $G_{D}$ with resolution of $10 \text{ km} \times 10 \text{ km}$ yielding $\lvert G_{D} \rvert = 4,937$ grid cells. Altogether, the entire posterior predictive time series results in a very large dataset from which maps and time series are computed. For instance, we have $61$ years by $365$ days by $4,937$ grid centroids by $1,000$ replicates yielding approximately 110 billion points. 

\begin{figure}[tb]
    \centering
    \includegraphics[width=.45\textwidth]{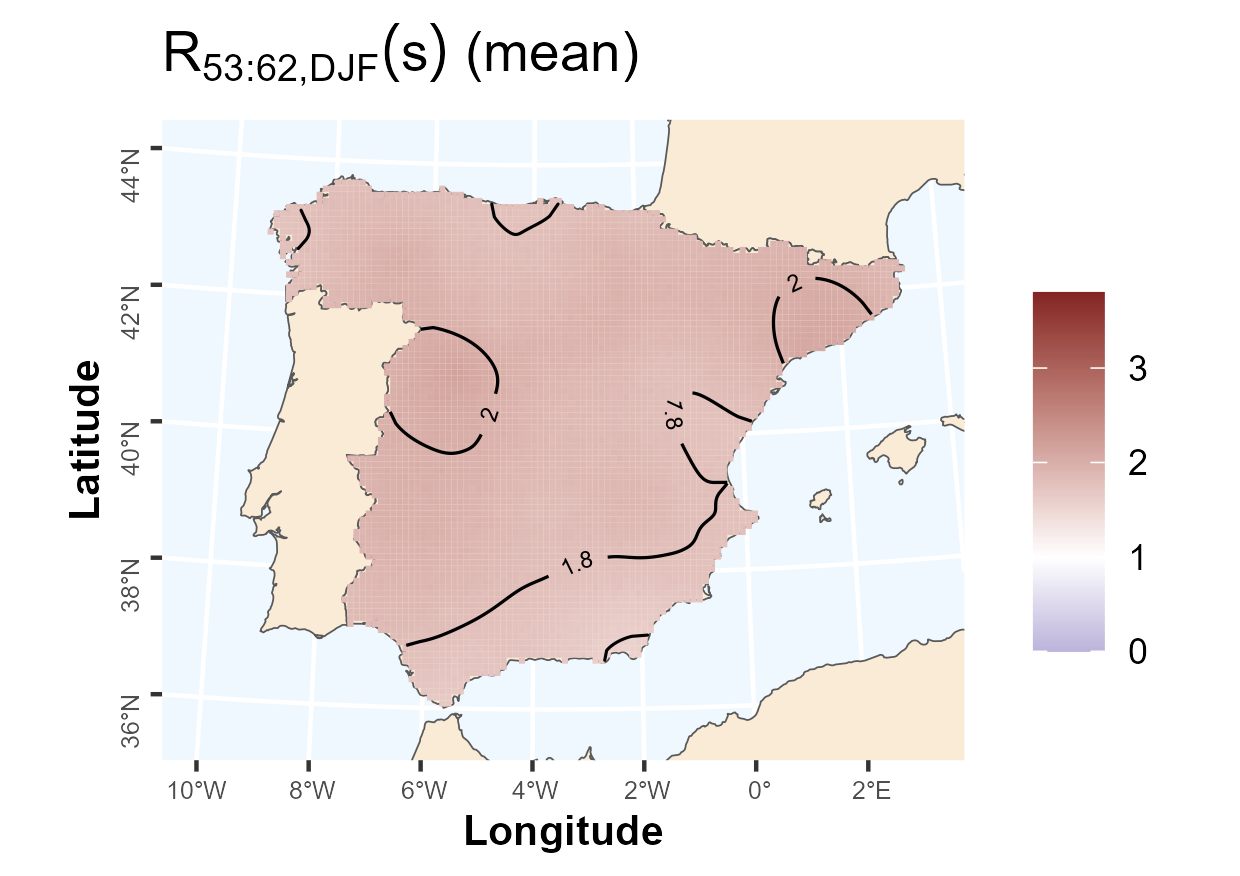}
    \includegraphics[width=.45\textwidth]{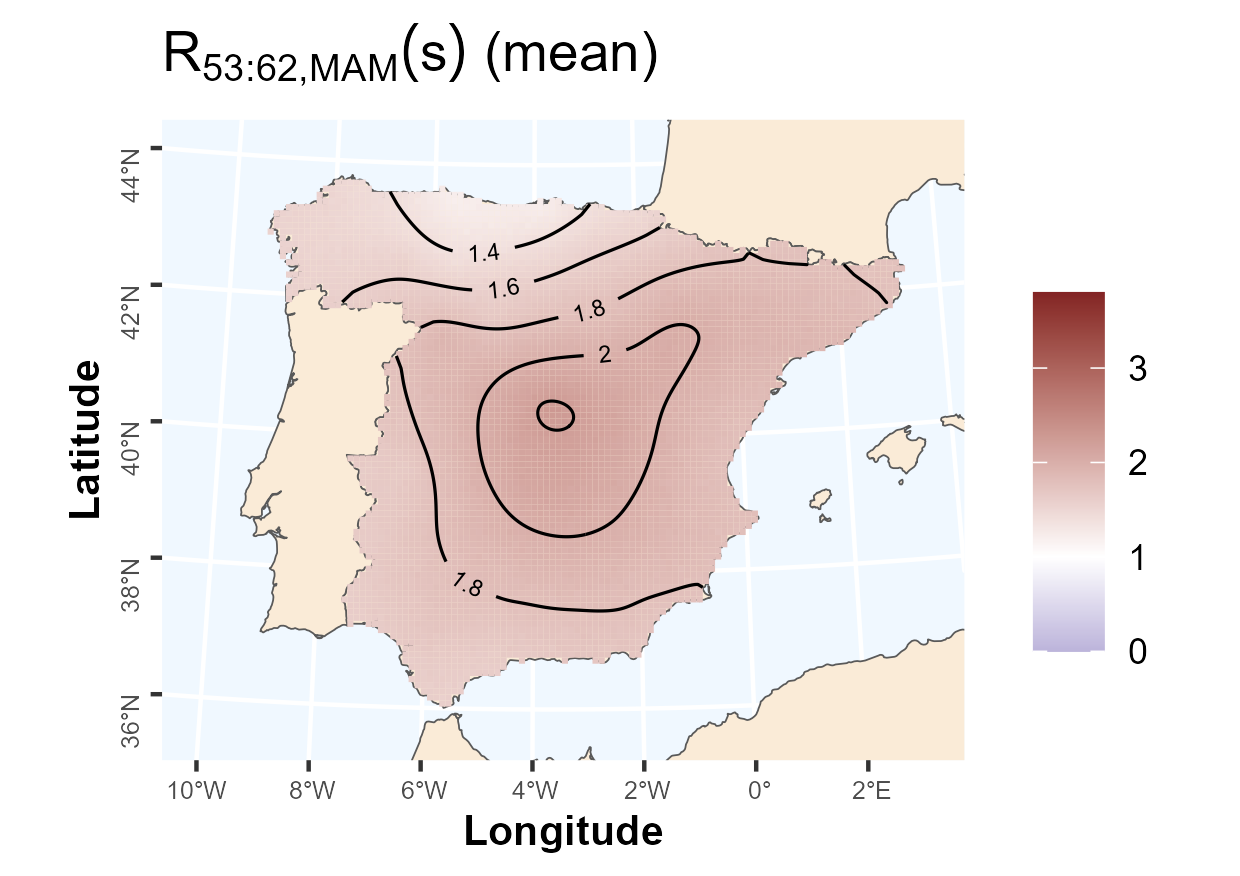}
    \includegraphics[width=.45\textwidth]{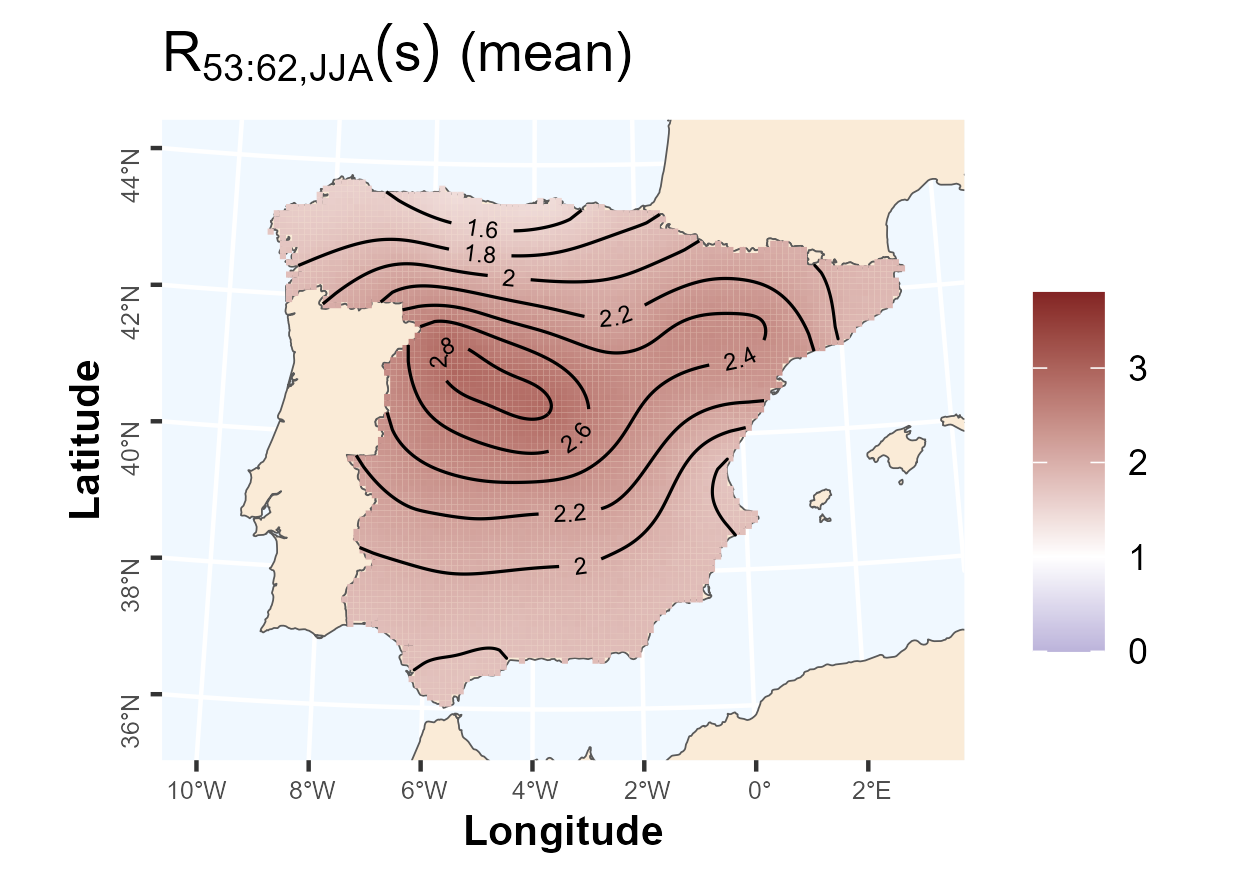}
    \includegraphics[width=.45\textwidth]{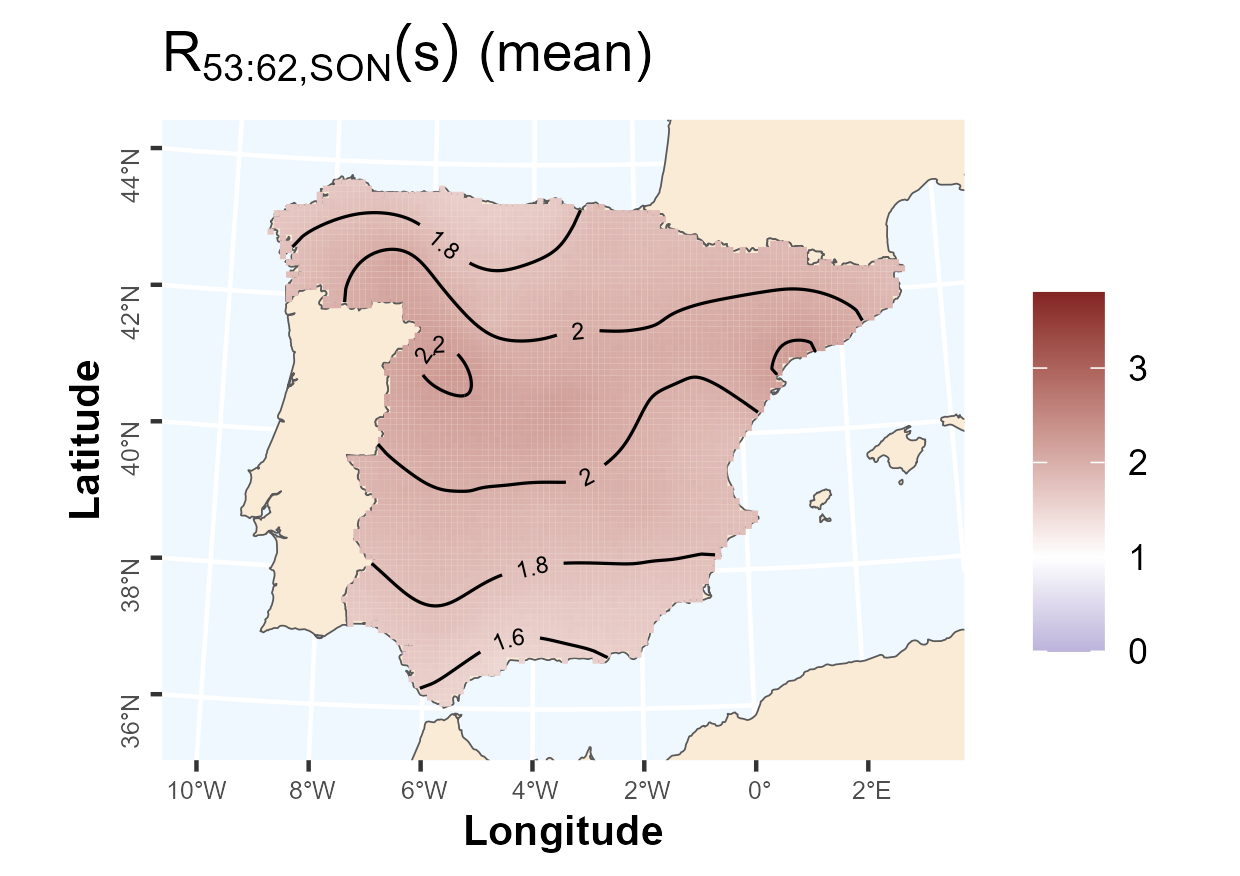}
    \caption{Map of the posterior mean of $R_{53:62,DJF}(\bs)$ (winter, top--left), $R_{53:62,MAM}(\bs)$ (spring, top--right), $R_{53:62,JJA}(\bs)$ (summer, bottom--left), and $R_{53:62,SON}(\bs)$ (autumn, bottom--right). Contour lines indicate ratios between intervals of length $0.2$.}
    \label{fig:lastDecade}
\end{figure}

First, the total number of records $\bar{N}_{62}(\bs)$ in \eqref{eq:N} is analyzed in Figure~11 of the Supplementary material. In this regard, $90.0\%$ of the region shows a significantly higher number of records compared to the stationary case, while no points exhibit a significantly lower number of records. The Atlantic and southern Mediterranean coasts seem to have a number of records compatible with stationarity.
However, we are interested in studying the number of records during the last years of the studied period, as they provide a more current picture of the climate and are not influenced by the high probability of occurrence in the early stages. Analyzing records by season is also crucial because each season has distinct spatial and temporal patterns. To quantify the occurrence of records in the past decade, Figure~\ref{fig:lastDecade} shows the posterior mean of the ratio $R_{53:62,season}(\bs)$ in \eqref{eq:ratio} for years from 2012--2021 by $season$ going through the four seasons. This ratio compares the average number of records predicted by the model and the expected average number under stationarity for the desired period. The corresponding maps of the $0.05$ and $0.95$ quantiles are shown in Figure~12 of the Supplementary material. The model estimates that the global warming trends have increased the number of records expected in the past decade almost two-fold, $1.93$ $(1.89,1.98)$, which suggests that only about half as many records would have been observed in those years over peninsular Spain if the process were stationary. By season the values are $1.88$ $(1.80,1.97)$ in winter, $1.81$ $(1.72 ,1.89)$ in spring, $2.13$ $(2.04,2.22)$ in summer, and $1.92$ $(1.83,2.01)$ in autumn. The number of records in the past decade is higher than in the stationary case everywhere, and this difference is significant for any point in the region. The percentage of area that has a significantly higher number of records is $100\%$ for winter, $90.1\%$ for spring, $98.4\%$ for summer, and $99.9\%$ for autumn. During this period, summer presents greater warming on average but also greater spatial variability compared to, e.g., winter.  
Analogous are the results for the number of records in the 21st century (2000--2021) in Figure~13 of the Supplementary material.

\begin{figure}[tb]
    \centering
    \includegraphics[width=.5\textwidth]{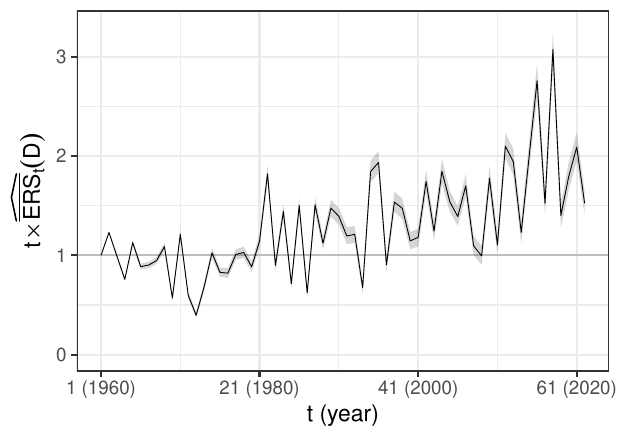}
    \caption{Posterior mean (black line) and $90\%$ CI (gray ribbon) of $t \times \widehat{\overline{\text{ERS}}}_{t}(D)$ against $t$. Expected value of $1$ under stationarity remarked for reference.}
    \label{fig:ERS}
\end{figure}

Moving to annual evolution, Figure~\ref{fig:ERS} shows the posterior mean and $90\%$ CI of $t \times \widehat{\overline{\text{ERS}}}_{t}(D)$ averaging in \eqref{eq:ERS} against years. There is substantial variability across years, but with a clear approximately increasing linear trend from around $t=34$ (year 1993), which is not compatible with a stationary climate. The ERS has been consistently and clearly higher than 1 since $t=49$ (year 2008). This plot provides evidence of the strong effect that climate change has had on the occurrence of records in the 21st century, and particularly in the past decade. Specifically, the times $t=56$ and $t=58$ (years 2015 and 2017) exceeded the expected number of records under stationarity by more than two and a half times. 
Figure~14 of the Supplementary material repeats the same results by season showing that ERS is not homogeneous within years. The most relevant results show that in winter, the increase in the number of records is evident only during the last seven years. In contrast, in summer, deviations from stationarity start earlier and reach  higher values; e.g., in summer 2003, the expected ERS under stationarity is multiplied by around $5$.

\begin{figure}[t]
    \centering
    \includegraphics[width=.24\textwidth]{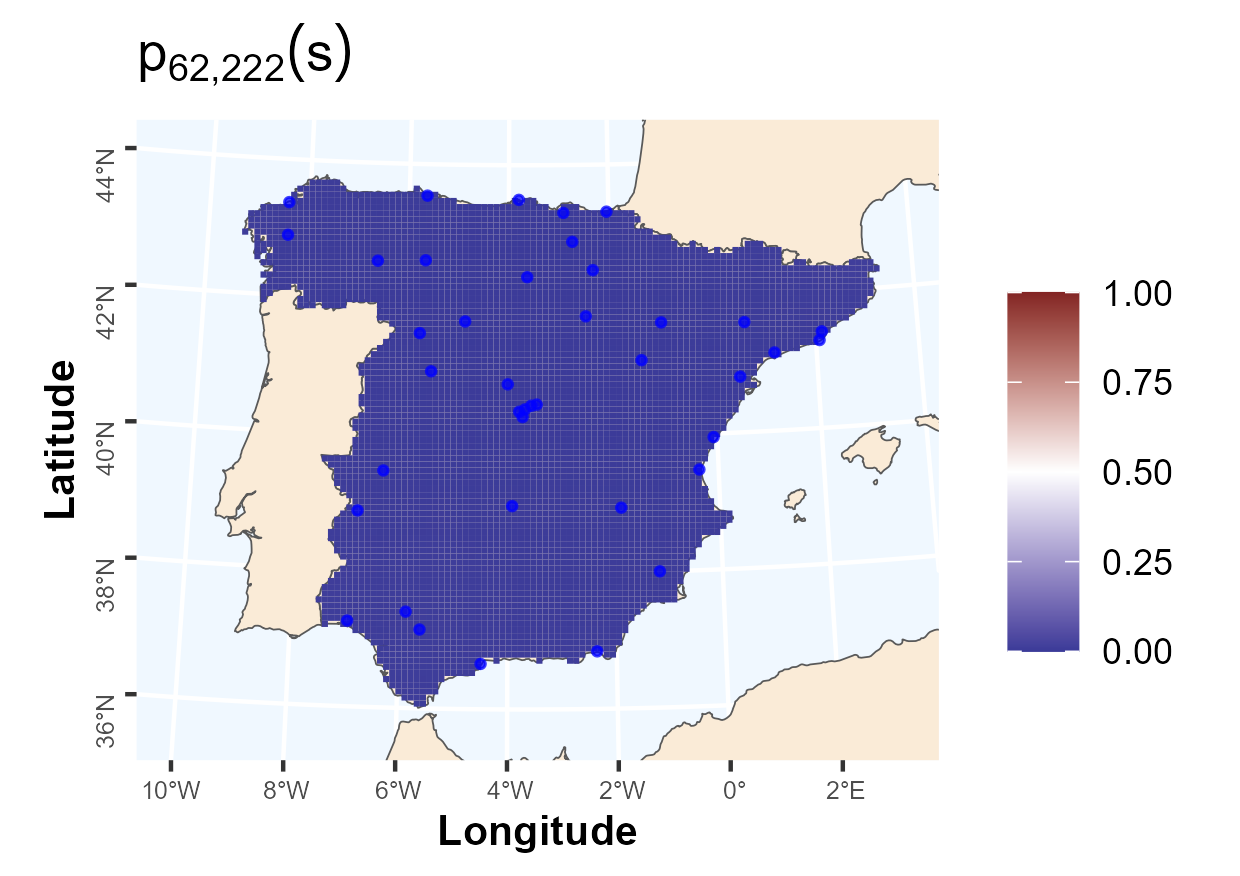}
    \includegraphics[width=.24\textwidth]{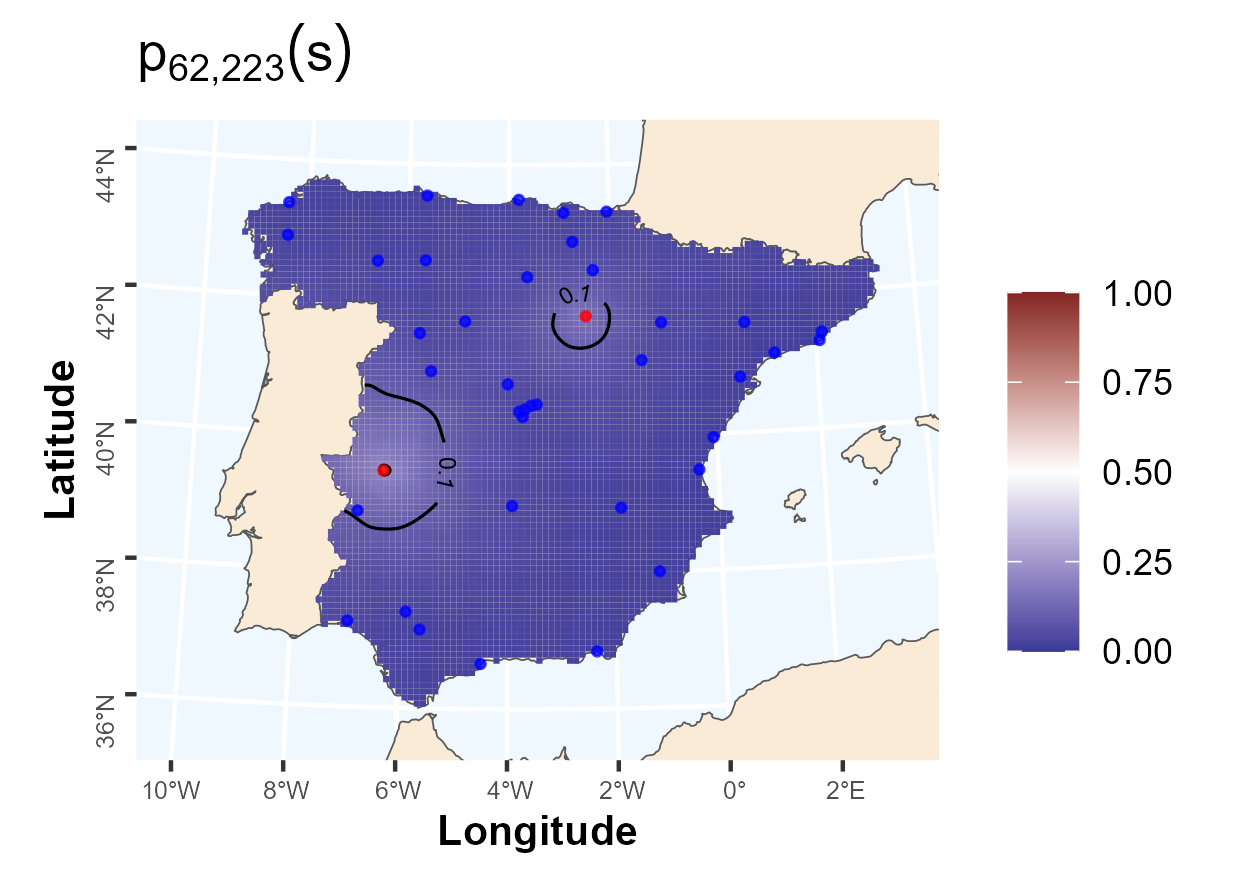}
    \includegraphics[width=.24\textwidth]{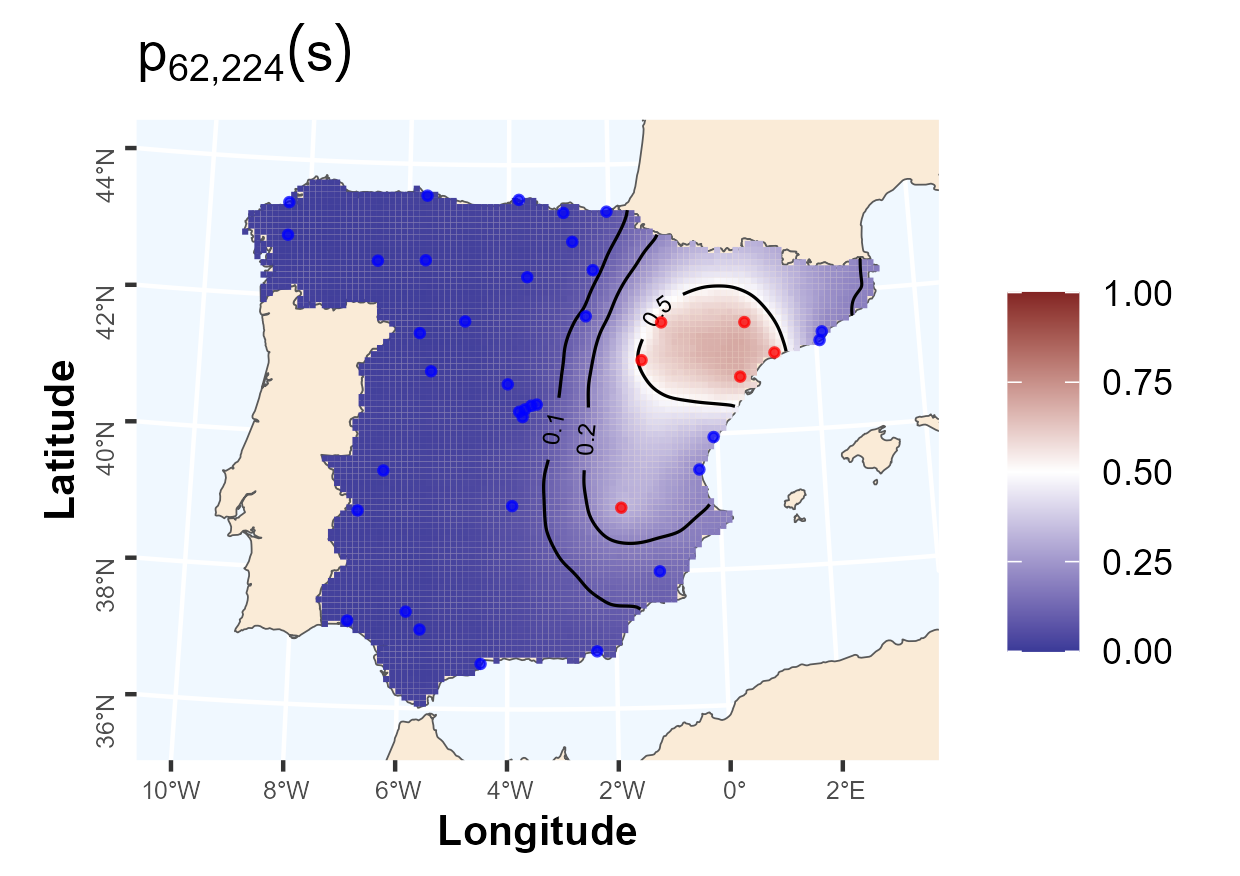}
    \includegraphics[width=.24\textwidth]{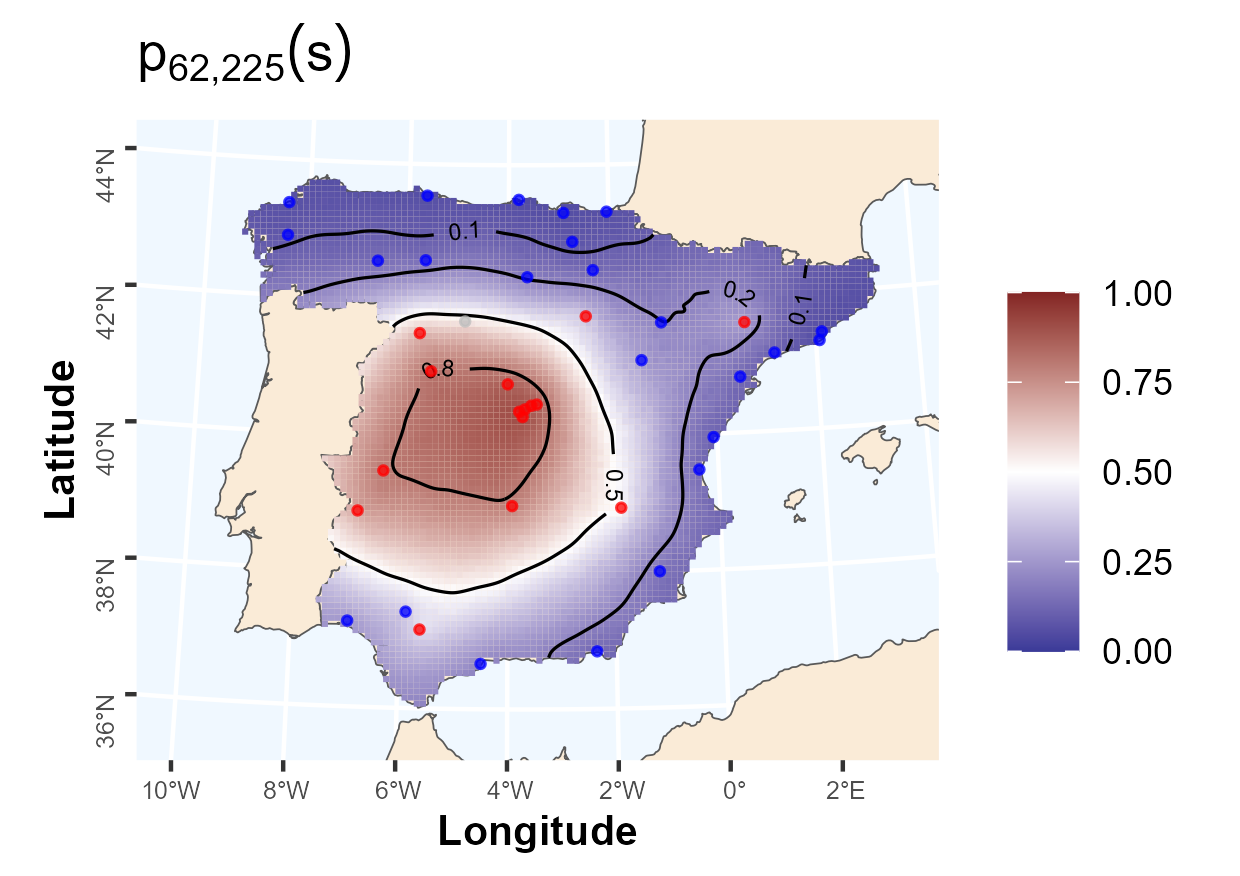}
    \includegraphics[width=.24\textwidth]{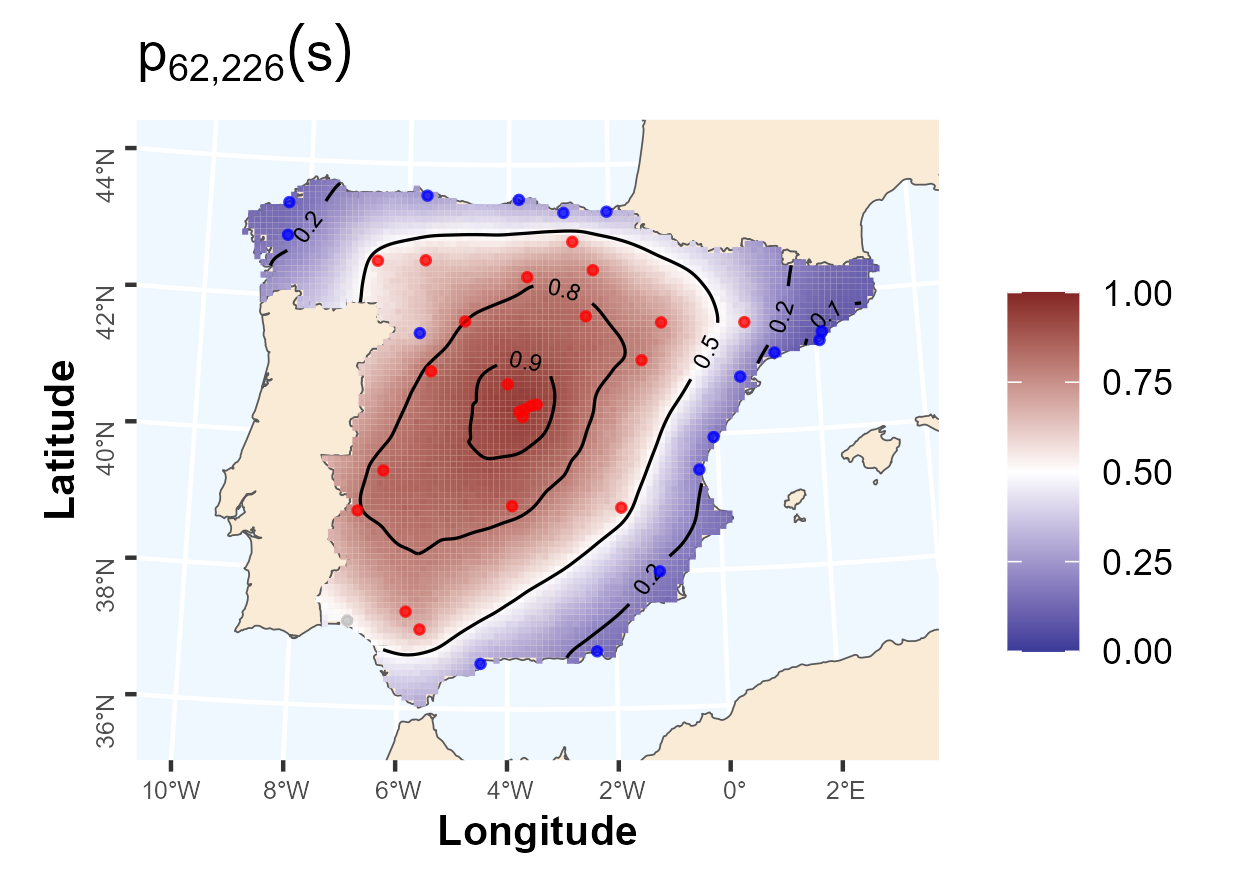}
    \includegraphics[width=.24\textwidth]{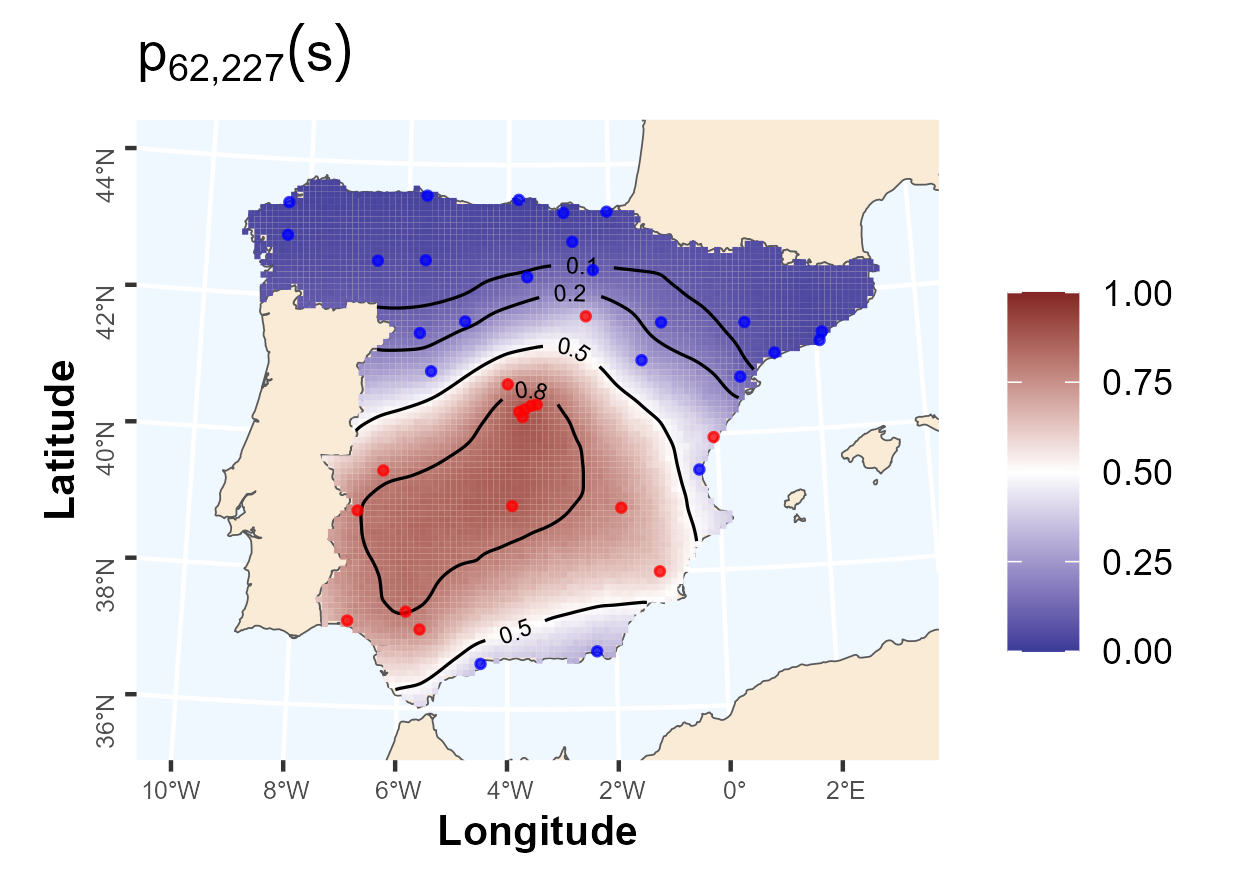}
    \includegraphics[width=.24\textwidth]{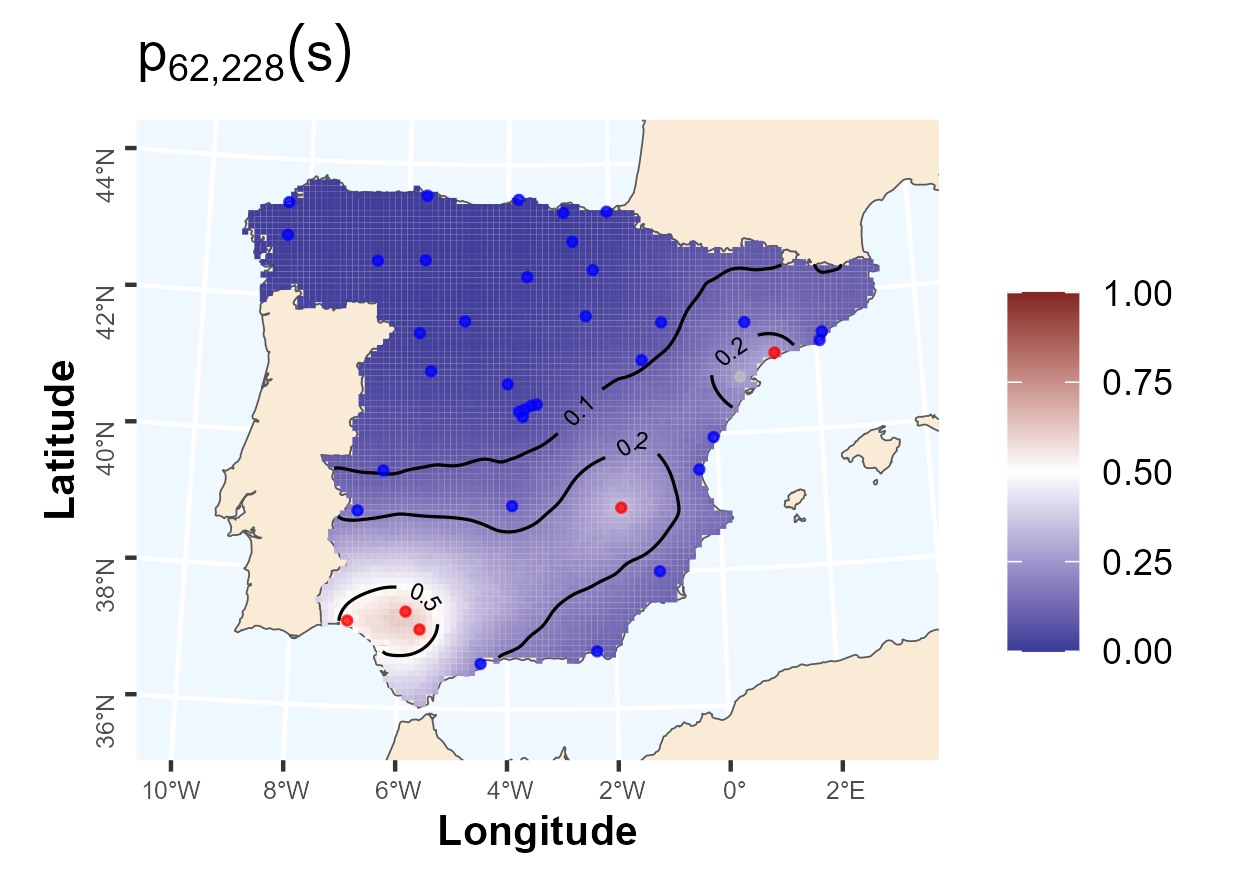}
    \includegraphics[width=.24\textwidth]{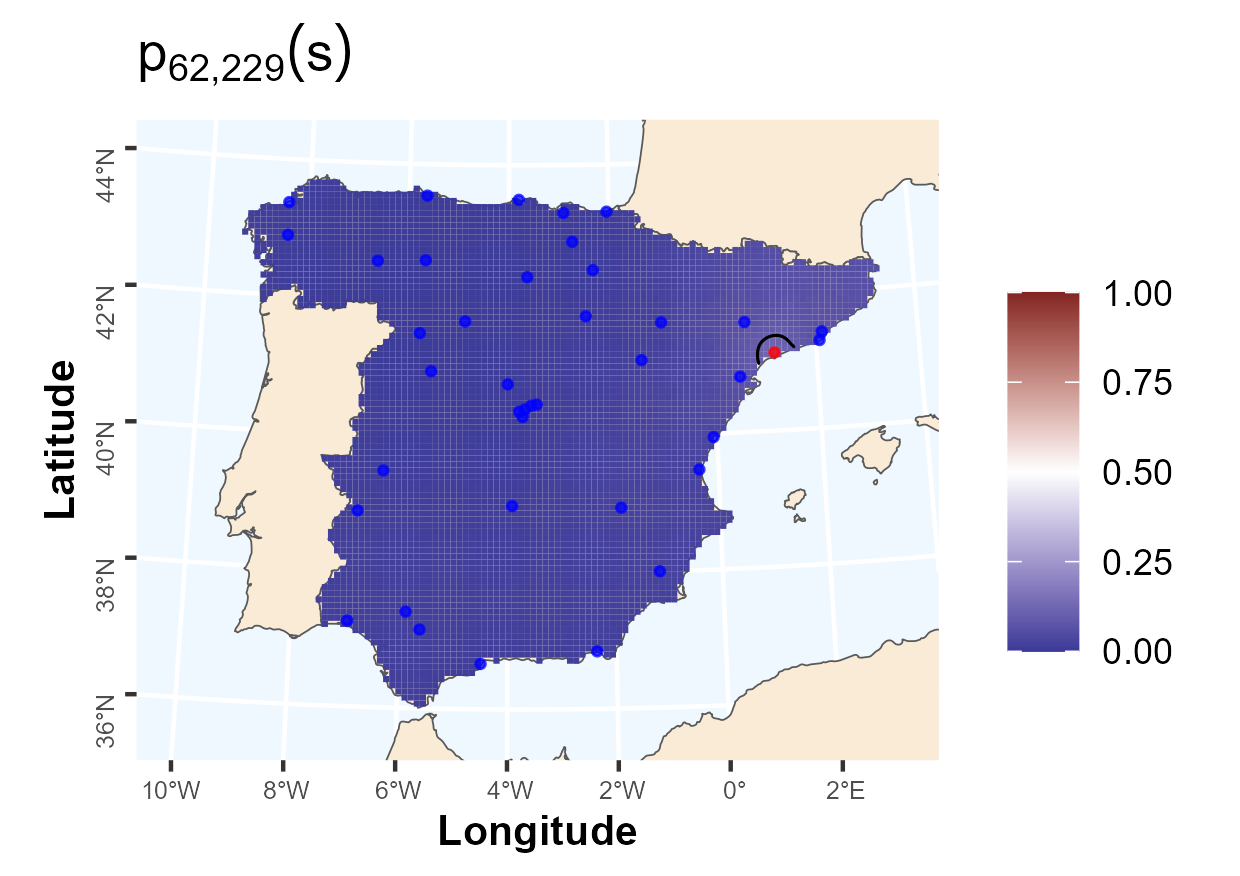}
    \caption{Maps of the posterior mean of $p_{62\ell}(\bs)$ on days $\ell=222,\ldots,229$ within year 2021. Observed indicators are points in blue, gray, and red to indicate non-records, tied records, and records, respectively. Contour lines indicate probabilities $0.1$, $0.2$, $0.5$, $0.8$, $0.9$.}
    \label{fig:heatwave}
\end{figure}

The model is also useful to obtain maps for the probability of record across days during a particular heatwave. To demonstrate the spatial and temporal extent of the August 14 ($\ell = 225$), 2021 absolute record and August 2021 heatwave over peninsular Spain, Figure~\ref{fig:heatwave} shows the posterior mean of the probabilities of record $p_{62\ell}(\bs)$ on days $\ell=222,\ldots,229$ within year 2021. This specific episode demonstrates the dynamics of persistence as well as the spatial structure of dependence. The beginning of the effects of the heatwave on the occurrence of records was observed on day 224, with posterior mean probabilities surpassing $0.5$ in the northeast. Over the following three days, the heatwave continued to evolve, with high probabilities observed across most regions, except for the coast, and with the area of high probabilities gradually diminished towards the southwest.

\section{Summary and future work} \label{sec:summary}

\paragraph{Summary.} We have taken up the first fully developed modeling attempt to analyze the incidence of record-breaking temperatures.  We have analyzed more than $60$ years of daily temperature data over peninsular Spain.  The modeling needed to effectively explain the incidence of record-breaking over this region during this period required careful specification of the probabilities of the indicator functions which define record-breaking sequences.  Key features include: explicit trend behavior, autoregression, significance of distance to the coast, useful interactions, spatial random effects, and very strong daily random effects. 

Our main findings are that the pattern of occurrence of records over peninsular Spain during 1960--2021 is not compatible with  stationary behavior of climate and that it accords with a global increase of temperature. More precisely, the total number of records in decade 2012--2021 has almost doubled compared to a stationary situation. In all the period from a spatial point of view, in $90.0\%$ of the region there is a significantly higher number of records compared to the stationary case, and there is no area where the number of records is significantly lower. Further, the increase in the number of records is not spatially homogeneous, being higher inland than coastal, and higher in the Mediterranean coast than in the Atlantic coast. The behavior is not homogeneous within year; during the past decade the highest increase has occurred in summer, and the lowest in spring.

\paragraph{Future work.} While we considered calendar day temperature records over peninsular Spain, the modeling proposed here can be useful for analyzing data from other parts of the world, other temporal scales, or for other climate signals such as ozone levels or rainfall.

There is interest in future incidence of record-breaking as a manifestation of future climate change. Such investigation will lead to analysis of future climate scenarios. Our approach here may prove useful though these scenarios are developed on spatial and temporal scales different from our current data, requiring substantial additional analysis. In particular, addressing this would require the inclusion of more information in the model, such as atmospheric covariates with future projections available or climate model simulations. Models that rely solely on time covariates for obtaining future projections may result in unsatisfactory extrapolation.

Another challenging problem with modeling calendar day records is that, for each day and location, we assume that the annual temperature series begins at the same year.  With temperature data having initial year which varies with say space and day, it would be attractive not to discard many early years of indicator variables in order to achieve alignment in the dataset.  One path for addressing varying record lengths could implement suitable posterior predictive ``imputation" of the missing indicators at the sites having later initiation.  Then, with replicates of the full series at each day at each site, the record breaking would be aligned and the predictive behavior could be explored.  Such analysis does not arise with the dataset used here but presents a possible future investigation.

\bigskip
\begin{center}
{\large\bf Supplementary material}
\end{center}

\begin{description}

\item[Title:] \textbf{Supplementary material} Document with five sections: 1. Data and exploratory analysis, 2. Distributions for Gibbs sampling, 3. Model comparison and checking, 4. MCMC convergence diagnostics, 5. Additional results. (.pdf file)

\item[\texttt{R}~package:] \texttt{R}~package \texttt{sprom} containing code to perform the spatial record occurrence modeling described in the manuscript. The package also contains a folder with data and a folder with \texttt{R} scripts to replicate the results in Section~\ref{sec:results}, including temperature data or covariates acquisition. (GNU zipped tar file)

\end{description}

\bigskip
\begin{center}
{\large\bf Acknowledgments}
\end{center}

This work was supported in part by MCIN/AEI/10.13039/501100011033 and Uni\'on Europea NextGenerationEU under Grants PID2020-116873GB-I00 and TED2021-130702B-I00; and Gobierno de Arag\'on under Research Group E46\_20R: Modelos Estoc\'asticos. Jorge Castillo-Mateo was supported by Gobierno de Arag\'on under Doctoral Scholarship ORDEN CUS/581/2020 and Mobility Scholarship ORDEN CUS/1668/2022 number MVE\_06\_23. Zeus Gracia-Tabuenca was supported by Ministerio de Universidades and Uni\'on Europea NextGenerationEU under Margarita Salas Scholarship RD 289/2021 UNI/551/2021. The authors thank Erin M. Schliep for useful discussions on Bayesian spatial binary regression and Jes\'us Abaurrea for valuable comments and insights.

\bigskip
\begin{center}
{\large\bf Disclosure Statement}
\end{center}

The authors report there are no competing interests to declare.

\bibliographystyle{agsm}

\bibliography{bibliography}

\end{document}


\def\spacingset#1{\renewcommand{\baselinestretch}%
{#1}\small\normalsize} \spacingset{1}


\if1\blind
{
  \title{\bf Supplementary material of the Manuscript: ``Spatio-temporal modeling for record-breaking temperature events in Spain''}
  \author{Jorge Castillo-Mateo \\
  Department of Statistical Methods, University of Zaragoza \\
  Alan E. Gelfand \\
  Department of Statistical Science, Duke University \\
  Zeus Gracia-Tabuenca \\ 
  Department of Statistical Methods, University of Zaragoza \\
  Jes\'us As\'in \\
  Department of Statistical Methods, University of Zaragoza \\
  Ana C. Cebri\'an \\
  Department of Statistical Methods, University of Zaragoza \\}
  \maketitle
} \fi

\if0\blind
{
  \bigskip
  \bigskip
  \bigskip
  \begin{center}
    {\LARGE\bf Supplementary material of the Manuscript: ``Spatio-temporal modeling for record-breaking temperature events in Spain''}
\end{center}
  \medskip
} \fi

This supplementary material contains additional exploratory analyses, a detailed MCMC algorithm, explicit details for model comparison and checking metrics, convergence diagnostics, and further results from the Manuscript.

\section{Data and exploratory analysis}

\subsection{Provenance of the dataset}

We considered as the initial dataset the stations from the ECA\&D with a minimum of 99.5\% reliable data over 1960--2021 in the Iberian Peninsula. There were 53 stations with this requirement over peninsular Spain. There were not Portuguese series included because data from 2021 was not available at the time of collection. In order to guarantee the quality of the final dataset an additional relative homogeneity control of the series was carried out, and Alicante was removed from the dataset. Finally, 12 series that included inputted data based on information from nearby series also included in the dataset were identified. These series were also removed to avoid the use of redundant information. Thus, the final dataset is made up of 40 daily temperature series of proven quality and homogeneity. This dataset has already been used in previous studies, in particular, \cite{castillo2023b} used it to analyze deviations from a stationary behavior in the occurrence of records; they excluded four of the five series located in Madrid area to avoid an over-representation of the region since they analyzed the series individually. The dataset includes the four centennial observing stations located in Spain and recognized by World Meteorological Office \citep{WMO2022b}: Barcelona (Fabra), Daroca, Madrid (Retiro), and Tortosa.

\subsection{Impact of missing data}

The 40 stations analyzed have, on average, a $0.07\%$ of missing values and eight stations have no missing values, 27 have 10 ($0.04\%$) or fewer missing values, and the station with the most missing values has 70 ($0.31\%$). The latter station is Sevilla (Airport), and 29 of the missing values correspond to 1965 and 18 to 1966; despite this, it was included in the dataset due to the city's significance as the most populated in the South of the Iberian Peninsula. To address this, we assigned a value of $-\infty$ to missing data, so that they are not considered records except if they appear at $t=1$. 

The following simulation study shows that the impact of this missing data on the results is negligible. This study simulates $40 \times 365$ independent time series of length $62$ from the Gaussian LDM, 
\begin{equation*}
   Y_t = \hat{c} t + \epsilon_t, \quad \epsilon_t \sim N(0, \hat{\sigma}^2), \qquad t = 1,\ldots, 62.
\end{equation*}
We used the estimated parameters $\hat c = 0.035$ and $\hat \sigma = 3.56$, which were obtained averaging across days and sites the simple linear regression estimates from the daily temperature data across years. We then compute the number of record indicators that are different between the simulated series and the same series after introducing $649$ missing values in the same position as observed in the ECA\&D stations. By repeating the simulation $10,000$ times, we found that the number of different indicators when missing values were introduced is $165$ on average and between $(136,197)$ in $90\%$ of the simulations. In addition, the total number of records in the $40 \times 365$ series with missing data increased by $40$ $(18,62)$. These values are reduced, respectively, to $17$ $(9,26)$ and $2$ $(-5,9)$ between 1991--2021. The largest of these quantities represents a negligible $0.33\%$ of the number of non-trivial records, and the effect of missing data is even smaller in the second half of the period, which is the period of greatest interest.

\subsection{Exploratory analysis}

The exploratory analysis described in Section~2.3 of the Manuscript aims to implement a data-driven variable selection to include covariates to capture deviations from stationarity (annual trends), persistence, seasonality, and spatial variability. This section includes additional details to those included in the Manuscript. For convenience, \texttt{trend1} and \texttt{trend2} are respectively the first and second degree orthogonal polynomials of $\log(t - 1)$, and \texttt{lag1} and \texttt{lag2} are the first and second autoregressive terms $I_{t,\ell-1}(\bs)$ and $I_{t,\ell-2}(\bs)$ with \texttt{lag1:lag2} being their interaction.

\subsubsection{Persistence}

\paragraph{First-order autoregression.} The joint distribution $[I_{t\ell}(\bs), I_{t,\ell-1}(\bs)]$ can be expressed in terms of $2 \times 2$ tables. To analyze the evolution of persistence across years, summing across space and days within year, we obtain tables that are used to compute the empirical log OR's, $LOR_t$, defined in the Manuscript. Table~\ref{Tabla1} is a generic example of these tables.

\begin{table}[tb]
\begin{center}
\caption{A generic $2 \times 2$ table for the joint distribution $[I_{t\ell}(\bs), I_{t,\ell-1}(\bs)]$ across years.}
\label{Tabla1}
\begin{threeparttable}
\begin{tabular}{|c|cc|c|}
 \hline
 \backslashbox{CD}{PD} & $0$ & $1$ & $0$ or $1$ \\
  \hline
 $0$ & $n_{t,00}$ & $n_{t,01}$ & $n_{t,0.}$ \\
 $1$ & $n_{t,10}$ & $n_{t,11}$ & $n_{t,1.}$ \\
  \hline
 $0$ or $1$ & $n_{t,.0}$ & $n_{t,.1}$ & $n_{t}$    \\
 \hline
\end{tabular}
 \begin{tablenotes}
   \footnotesize
   \item Current day (CD), previous day (PD).
 \end{tablenotes}
\end{threeparttable}
\end{center}
\end{table}

\paragraph{Second-order autoregression.} The joint distribution $[I_{t\ell}(\bs), I_{t,\ell-1}(\bs), I_{t,\ell-2}(\bs)]$ can be expressed in terms of $2 \times 2 \times 2$ tables. As above, to analyze the evolution across years, we compute a table for each $t$, obtained by summing over space and days within year. The two layers of a generic table corresponding to two days ago with no record ($0$) and record ($1$) are shown in Table~\ref{Tabla2}. The notation $n_{t,jkv}$ generalizes that used in $2\times 2$ tables. In particular, for $j,k,v \in \{0,1\}$ and $I_{t,-1}(\bs) \equiv I_{t-1,364}(\bs)$,
\begin{equation*}
    n_{t,jkv} = \sum_{i=1}^{36} \sum_{\ell=1}^{365} \bone(I_{t\ell}(\bs_i) = j, I_{t,\ell-1}(\bs_i) = k, I_{t,\ell-2}(\bs_i) = v).
\end{equation*}

\begin{table}[tb]
\begin{center}
\caption{A generic $2 \times 2 \times 2$ table for the joint distribution $[I_{t\ell}(\bs), I_{t,\ell-1}(\bs), I_{t,\ell-2}(\bs)]$ across years.}
\label{Tabla2}
\begin{threeparttable}
\begin{tabular}{cc*{3}{S[table-format=3]}} 
  \toprule 
  & & \multicolumn{3}{c}{PD} \\ 
  \cmidrule{3-5} 
  \multirow{-2.5}{*}{\makecell{2-PD}} & \multirow{-2.5}{*}{\makecell{CD}} & {$0$} & {$1$} & {$0$ or $1$}\\ 
  \midrule 
$0$ & $0$ & $n_{t,000}$ & $n_{t,010}$ & $n_{t,0.0}$ \\
    & $1$ & $n_{t,100}$ & $n_{t,110}$ & $n_{t,1.0}$ \\
    & $0$ or $1$ & $n_{t,.00}$ & $n_{t,.10}$ & $n_{t,..0}$ \\ \addlinespace
$1$ & $0$ & $n_{t,001}$ & $n_{t,011}$ & $n_{t,0.1}$ \\
    & $1$ & $n_{t,101}$ & $n_{t,111}$ & $n_{t,1.1}$ \\
    & $0$ or $1$ & $n_{t,.01}$ & $n_{t,.11}$ & $n_{t,..1}$ \\
  \bottomrule 
\end{tabular} 
 \begin{tablenotes}
   \footnotesize
   \item Current day (CD), previous day (PD), two days ago (2-PD).
 \end{tablenotes}
\end{threeparttable}
\end{center}
\end{table}

These tables are used to estimate the four conditional probabilities of record in a day given the occurrence or not of a record one and two days ago; i.e., $p_{t \mid 11}$, $p_{t \mid 01}$, $p_{t \mid 10}$, and $p_{t \mid 00}$. The following log OR's compare the probabilities $p_{t \mid 11}$ and $p_{t \mid 01}$, and $p_{t \mid 10}$ and $p_{t \mid 00}$, respectively,
\begin{equation*}
    LOR_{t,..1} = \log\frac{(n_{t,001}+0.5)(n_{t,111}+0.5)}{(n_{t,101}+0.5)(n_{t,011}+0.5)}, \quad \text{and} \quad 
    LOR_{t,..0} = \log\frac{(n_{t,000}+0.5)(n_{t,110}+0.5)}{(n_{t,100}+0.5)(n_{t,010}+0.5)}.
\end{equation*}

\subsubsection{Seasonality}

\begin{figure}[tb]
    \centering
    \includegraphics[width=0.55\textwidth]{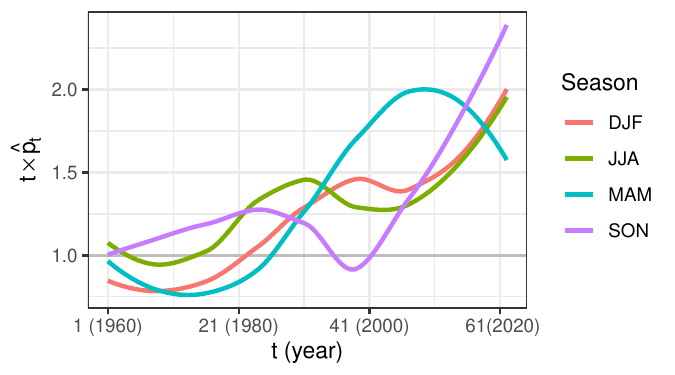}
    \caption{LOESS curves of $t \times \hat{p}_{t}$ against $t$ (year) by season.}
    \label{fig:EDA:2nd:LORt}
\end{figure}

The need to allow a different annual evolution within year is supported by Figure~\ref{fig:EDA:2nd:LORt}.  It shows the  evolution across years of the estimated probability at year $t$ multiplied by $t$ separated by season.  Different patterns are observed in each season, specially in spring, where a decreasing trend in the underlying trend is observed from about time $t = 46$ (year 2005). 

\subsubsection{Spatial variability}

\paragraph{Potential spatial covariates.}
The extension of peninsular Spain is relatively small. However, it includes a wide variety of climates, mostly due to its complex relief, the different characteristics of inland and coastal areas, and the different influence of the Atlantic Ocean in the West and the Mediterranean Sea in the East. These characteristics suggest the need for including spatial covariates in the model to capture at least some of the existing spatial variability. Several covariates were considered. Latitude and longitude are common spatial variables, however, their range in the peninsula, 36$^{\circ}$ to 43.8$^{\circ}$N for latitude and 9.3$^{\circ}$W to 3.3$^{\circ}$E for longitude, is not large enough to have a relevant influence. Then, we considered as potential regressors, elevation and distance to the coast, $\elev(\bs)$ and $\dist(\bs)$, respectively. 

To study the relationship between the occurrence of records and those covariates, we implement several exploratory analyses based on 36 logit models fitted separately for each observed site. The responses of the models are the non-trivial record indicator series in each site and the covariates are the quadratic time trend terms, one harmonic and the interaction with the trend terms, first and second order autoregressive terms as well as their product, and their interaction with the first trend term.

\begin{figure}[t]
    \centering
    \includegraphics[width=0.8\linewidth]{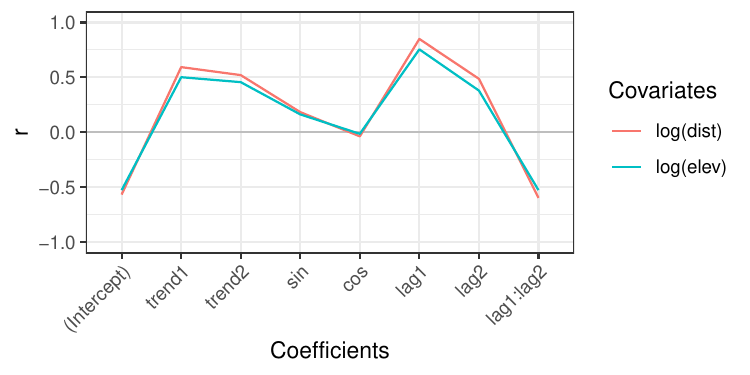}
    \caption{Pearson's $r$ for the relationship of the spatial covariates log elevation and log distance to the coast with estimated coefficients of the local logit models for each observed site.}
    \label{fig:EDA:supp:glmcoefcovs}
\end{figure}

The Pearson correlation coefficients between the estimated regression coefficients of the model fitted for each location and the spatial covariates in that location are computed. Figure~\ref{fig:EDA:supp:glmcoefcovs} shows these correlation coefficients with $\log(\elev(\bs))$ and $\log(\dist(\bs))$, for the main effects of the model. The correlation coefficients are quite similar in both cases, but slightly higher in absolute value for $\log(\dist(\bs))$. The similar results obtained for the spatial covariates suggest that they provide the same information, and in particular, their pairwise Pearson correlation coefficient equals $0.79$. Consequently, $\log(\dist(\bs))$ was the only spatial covariate considered in the following exploratory analyses.

\begin{figure}
    \centering
    \includegraphics[width=.45\textwidth]{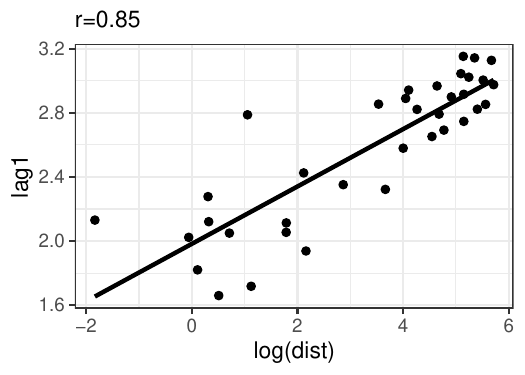}
    \includegraphics[width=.45\textwidth]{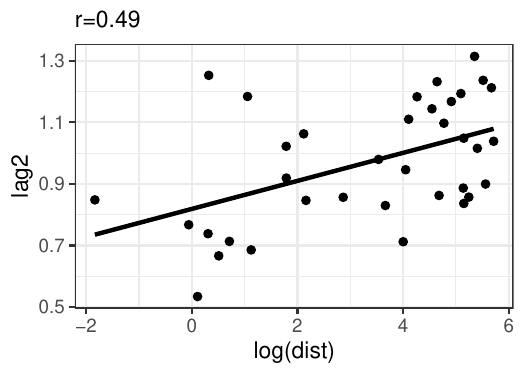}
    \includegraphics[width=.45\textwidth]{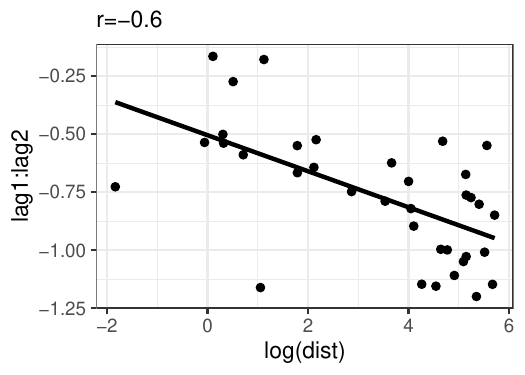}
    \caption{Estimated coefficients of the persistence terms in local logit models for each observed site $\bs$ against $\log(\dist(\bs))$.}
    \label{fig:EDA:supp:glmlogdist}
\end{figure}

\paragraph{Relationship between distance to the coast and persistence.} 
Here, the intent is to study whether persistence in the occurrence of records is spatially homogeneous. In particular, find differences in coastal and inland locations. To capture this spatial difference, interactions between persistence terms and $\dist(\bs)$ are considered. To study the need for those interactions, we explore graphically the relationship between the estimated coefficients of the persistence terms in the previous 36 local logit models and different functions of $\dist(\bs)$. Figure~\ref{fig:EDA:supp:glmlogdist} shows those coefficients against $\log(\dist(\bs))$, and the linear relation suggests to include the interaction terms $\log(\dist(\bs)) \times I_{t,\ell-1}(\bs)$, $\log(\dist(\bs)) \times I_{t,\ell-2}(\bs)$, and $\log(\dist(\bs)) \times I_{t,\ell-1}(\bs) \times I_{t,\ell-2}(\bs)$.

\subsubsection{Global logit models as a variable selection tool}

In order to complement the exploratory graphical tools, logit models with linear predictors including different temporal and spatial fixed effects are fitted. The aim is to fit global models for all the observed sites, using maximum likelihood estimation, that allows to compare models with different regressors using the AIC as a goodness of fit measure. This comparison is used to assess the need for the different temporal and spatial fixed effects suggested by the exploratory. Many models were compared, but Table~\ref{table:EDA:glm:aic} summarizes the fit of the main models considered. For example, the model with a cubic trend provides a worse AIC than the previous nested model with a quadratic trend.

\begin{sidewaystable}
\begin{center}
\caption{Nested fixed effects models, degrees of freedom (DOF), and AIC.}
\label{table:EDA:glm:aic}
\begin{threeparttable}
\begin{tabular}{llcc} 
 \toprule 
  Model & Linear predictor & DOF & AIC\\
 \midrule
  Stationary            & $\text{offset}(-\log(t-1)) - 1$ & 0 & 343369.2\\
  Linear trend          & $\log(t-1)$ & 2 & 342238.3\\
  Quadratic trend       & $\text{poly}(\log(t-1),2)$ & 3 & 340112.8\\
   $\ \ +$ 1st-order AR & $\ \ + I_{t,\ell-1}(\bs), \log(t-1) \times I_{t,\ell-1}(\bs)$ & 5 & 295425.6\\ 
   $\ \ +$ 2nd-order AR & $\ \ + I_{t,\ell-2}(\bs), I_{t,\ell-1}(\bs) \times I_{t,\ell-2}(\bs), \log(t-1) \times I_{t,\ell-2}(\bs), \log(t-1) \times I_{t,\ell-1}(\bs) \times I_{t,\ell-2}(\bs)$ & 9 & 293101.5\\ 
   $\ \ +$ seasonal terms & $\ \ + \sin_{\ell}, \cos_{\ell}, \sin_{\ell} \times \trendONE_{t}, \cos_{\ell} \times \trendONE_{t}, \sin_{\ell} \times \trendTWO_{t}, \cos_{\ell} \times \trendTWO_{t}$ & 15 & 292821.8\\ 
   $\ \ +$ $\log(\dist(\bs))$-trend int. & $\ \ +\log(\dist(\bs)), \log(\dist(\bs)) \times \trendONE_{t}, \log(\dist(\bs)) \times \trendTWO_{t}$ & 18 & 292781.9\\ 
 $\ \ +$ $\log(\dist(\bs))$-AR int. & $\ \ + \log(\dist(\bs)) \times I_{t,\ell-1}(\bs), \log(\dist(\bs)) \times I_{t,\ell-2}(\bs), \log(\dist(\bs)) \times I_{t,\ell-1}(\bs) \times I_{t,\ell-2}(\bs)$ & 21 & 292030.7\\ 
 Cubic trend & $\text{poly}(\log(t-1),3)$ & 4 & 340113.8\\ 
 \bottomrule
\end{tabular}
 \begin{tablenotes}
   \footnotesize
   \item Autoregression (AR), interaction (int.).
 \end{tablenotes}
\end{threeparttable}
\end{center}
\end{sidewaystable}

\clearpage

\section{Distributions for Gibbs sampling} \label{sec:MCMC}

It is convenient to denote parameters in vector notation as follows. First, let $\bI^{*}$ denote the vector of indicators associated with any $r$-tied record. Let the vector of data including tied records be $\bI = (\bI_{2,1}^\top,\ldots,\bI_{T,365}^\top)^\top$ with $\bI_{t\ell} = (I_{t\ell}(\bs_1), \ldots I_{t\ell}(\bs_n))^\top$, and equivalently define the vectors $\bW$, $\bW_{t\ell}$, $\bY$, $\bY_{t\ell}$, $\blam$, and $\blam_{t\ell}$ from $w_{t\ell}(\bs)$ and the latent variables $Y_{t\ell}(\bs)$ and $\lambda_{t\ell}(\bs)$, respectively.  Define the matrices $\bX_{t\ell} = (\bx_{t\ell}^{\top}(\bs_1),\ldots,\bx_{t\ell}^{\top}(\bs_n))^{\top}$ and $\bLambda_{t\ell} = \diag(\lambda_{t\ell}(\bs_1),\ldots,\lambda_{t\ell}(\bs_n))$. The steps in every iteration of the MCMC algorithm are as follows:
\begin{enumerate}
    \item Sample the values of $\bI^{*}$ independently from $\text{Bernoulli}(1 / r)$.
    \item Sample the values of the latent variables $Y_{t\ell}(\bs)$ from their  truncated normal full conditional distribution,
    \begin{equation*}
        Y_{t\ell}(\bs) \mid \bbeta, w_{t\ell}(\bs),\lambda_{t\ell}(\bs),I_{t\ell}(\bs),I_{t,\ell-1}(\bs),I_{t,\ell-2}(\bs) \sim TN\left(\bx_{t\ell}(\bs) \bbeta + w_{t\ell}(\bs), \lambda_{t\ell}(\bs), (a, b)\right),
    \end{equation*}
    where $(a,b) = (0,\infty)$ if $I_{t\ell}(\bs) = 1$ and $(a,b) = (-\infty,0)$ otherwise. Use $\bbeta_{\ell}$ instead of $\bbeta$ when $\ell=1,2$. The truncated normal distribution is easy to sample with an algorithm that mixes the usual inverse transform method and the algorithm proposed by \cite{robert1995}.
    \item Sample the values of the variance parameters $\lambda_{t\ell}(\bs)$ from their full conditional distribution. It does not have a standard form, but an efficient procedure that does not require tuning to update these parameters is to include Metropolis-Hastings steps using KS proposals \citep[see][for an efficient and exact KS sampling algorithm]{devroye1981}. Note that sampling from the KS distribution avoids the evaluation of the KS density which is an infinite series. In any case, this is the main bottleneck of the algorithm.
    \item With an abuse of notation and exclusively to fit the model improving the mixing of the algorithm, we implement hierarchical centering \citep{gelfand1995} of the daily effects $w_{t\ell}$ on the global intercept $\beta_{0}$, and leave $\bbeta$ and $\bX_{t\ell}$ with the $k$ covariates. Analogously, we center $w_{t\ell}$ on their respective intercepts $\beta_{0,\ell}$ for days $\ell = 1,2$. Given the normal prior for the regression coefficients, $\bbeta \sim N_k(\bmu_{\bm{\beta}}, \bSigma_{\bm{\beta}})$, sample $\bbeta$ from its normal full conditional distribution,
    \begin{equation*}
    \begin{aligned}
        \bbeta &\mid \bW, \blam, \bY, \bI \sim N_k(\hat{\bbeta}, \bOmega_{\bm{\beta}}); \\
        \hat{\bbeta} &= \bOmega_{\bm{\beta}} \left(\sum_{t=2}^{T} \sum_{\ell=3}^{365} \bX_{t\ell}^{\top} \bLambda_{t\ell}^{-1} \left(\bY_{t\ell} - \bW_{t\ell}\right) + \bSigma_{\bm{\beta}}^{-1} \bmu_{\bm{\beta}}\right), \\
        \bOmega_{\bm{\beta}}^{-1} &= \sum_{t=2}^{T} \sum_{\ell=3}^{365} \bX_{t\ell}^{\top} \bLambda_{t\ell}^{-1} \bX_{t\ell} + \bSigma_{\bm{\beta}}^{-1}. \\
    \end{aligned}
    \end{equation*}
    Similarly sample $\bbeta_{\ell}$ for days $\ell=1,2$.
    \item The Gaussian process prior on $w_{t\ell}(\bs)$ defines it as a stochastic process for which the values at the $n$ observed locations come from the normal distribution, $\bW_{t\ell} \sim N_n(\bmu_{\bm{\bW}_{t\ell}}, \bSigma_{\bm{\bW}})$, where $\bmu_{\bm{\bW}_{t\ell}} = w_{t\ell} \bone_n$ and $\bSigma_{\bm{\bW}} = \sigma_{0}^{2} \bR_{\phi_0}$ with $\bR_{\phi_0}$ the correlation matrix where the $j$th element in the $i$th row is $\exp\{-\phi_{0} \lvert\lvert \bs_i - \bs_j \rvert\rvert\}$ and $\lvert\lvert \bs_i - \bs_j \rvert\rvert$ is the distance between $\bs_i$ and $\bs_j$. Then, sample $\bW_{t\ell}$ from its normal full conditional distribution,
    \begin{equation*}
    \begin{aligned}
        \bW_{t\ell} &\mid \bbeta, \blam, \bY, \bI, w_{t\ell}, \sigma_{0}^{2}, \phi_{0} \sim N_n(\widehat{\bW}_{t\ell}, \bOmega_{\bm{W}_{t\ell}}); \\
        \widehat{\bW}_{t\ell} &= \bOmega_{\bm{W}_{t\ell}} \left(\bLambda_{t\ell}^{-1} \left(\bY_{t\ell} - \bX_{t\ell} \bbeta\right) + \bSigma_{\bm{W}}^{-1} \bmu_{\bm{W}_{t\ell}}\right), \\
        \bOmega_{\bm{W}_{t\ell}}^{-1} &= \bLambda_{t\ell}^{-1} + \bSigma_{\bm{W}}^{-1}. \\
    \end{aligned}
    \end{equation*}
    Similarly sample from the distributions obtained for $\ell=1,2$, using $\bbeta_{\ell}$ instead of $\bbeta$ and $\sigma_{0,\ell}^{2}$ instead of $\sigma_{0}^{2}$.
    \item Given the prior specification of the daily and global intercepts, $w_{t\ell} \sim N(\beta_{0}, \sigma_{1}^{2})$ and $\beta_{0} \sim N(\mu_{\beta_{0}}, \sigma_{\beta_{0}}^{2})$, sample them from their normal full conditional distribution, where ``$\cdots$'' denotes ``all the other parameters'',
    \begin{equation*}
    \begin{aligned}
        w_{t\ell} &\mid \cdots \sim N(\hat{w}_{t\ell}, \omega_{1w}^{2});
        &\beta_{0} &\mid \cdots \sim N(\hat{\beta}_{0}, \omega_{\beta_{0}}^{2}); \\
        \hat{w}_{t\ell} &= \omega_{1w}^{2} \left(\frac{\bone_n^{\top} \bR_{\phi_{0}}^{-1} \bW_{t\ell}}{\sigma_{0}^{2}} + \frac{\beta_{0}}{\sigma_{1}^{2}}\right),
        &\hat{\beta}_{0} &= \omega_{\beta_{0}}^{2} \left(\frac{1}{\sigma_{1}^{2}} \sum_{t=2}^{T} \sum_{\ell=3}^{365}  w_{t\ell} + \frac{\mu_{\beta_0}}{\sigma_{\beta_{0}}^{2}}\right), \\
        \omega_{1w}^{-2} &= \frac{\bone_n^{\top} \bR_{\phi_{0}}^{-1} \bone_n}{\sigma_{0}^{2}} + \frac{1}{\sigma_{1}^{2}},
        &\omega_{\beta_{0}}^{-2} &= \frac{(T - 1) 363}{\sigma_{1}^{2}} + \frac{1}{\sigma_{\beta_0}^{2}}.\\
    \end{aligned}
    \end{equation*}
    Similarly sample from $w_{t\ell}$ and $\beta_{0,\ell}$ when $\ell=1,2$.
    \item Sample the value of the decay parameter $\phi_{0}$ from its non-standard full conditional distribution,
    \begin{equation*}
    \begin{aligned}
        \relax [\phi_{0} \mid \cdots] \propto 
        [\phi_{0}] &\times
        \lvert\bR_{\phi_{0}}\rvert^{-\frac{(T - 1) 365}{2}} 
        \exp\left\{\frac{-1}{2\sigma_{0}^{2}} \sum_{t=2}^{T} \sum_{\ell=3}^{365} (\bW_{t\ell} - \bmu_{\bm{W}_{t\ell}})^{\top} \bR_{\phi_{0}}^{-1} (\bW_{t\ell} - \bmu_{\bm{W}_{t\ell}}) \right\} \\
        &\times \exp\left\{\frac{-1}{2} \sum_{t=2}^{T} \sum_{\ell=1}^{2} \frac{1}{\sigma_{0,\ell}^{2}} (\bW_{t\ell} - \bmu_{\bm{W}_{t\ell}})^{\top} \bR_{\phi_{0}}^{-1} (\bW_{t\ell} - \bmu_{\bm{W}_{t\ell}}) \right\},
    \end{aligned}
    \end{equation*}
    where $[\phi_{0}]$ is its prior distribution. A simple procedure to update this parameter is to include a random walk Metropolis-Hastings step in the log scale with a normal proposal and a variance parameter tuned such that the acceptance rate is close to $33\%$. We implement a version of the adaptive method described in Section~3 of \cite{roberts2009} during the burn-in iterations to reach the desired acceptance rate without the need of manual tuning.
    \item Given the gamma prior distribution for the precision parameters, $1/\sigma_{0}^{2},1/\sigma_{1}^{2} \sim G(a_{\sigma}, b_{\sigma})$, sample them from their gamma full conditional distribution,
    \begin{equation*}
    \begin{aligned}
        1/\sigma_{0}^{2} \mid \cdots
        &\sim G\left(  
        \frac{n (T - 1) 363}{2} + a_{\sigma}, \frac{1}{2} \sum_{t=2}^{T} \sum_{\ell=3}^{365} (\bW_{t\ell} - \bmu_{\bm{W}_{t\ell}})^{\top} \bR_{\phi_{0}}^{-1} (\bW_{t\ell} - \bmu_{\bm{W}_{t\ell}}) + b_{\sigma}
        \right),\\
        1/\sigma_{1}^{2} \mid \cdots 
        &\sim G\left(  
        \frac{(T - 1) 363}{2} + a_{\sigma}, \frac{1}{2} \sum_{t=2}^{T} \sum_{\ell=3}^{365} (w_{t\ell} - \beta_{0})^{2} + b_{\sigma}
        \right).\\
    \end{aligned}
    \end{equation*}
    Again, similarly sample from $\sigma_{0,\ell}^{2}$ and $\sigma_{1,\ell}^{2}$ when $\ell=1,2$.
\end{enumerate}

\section{Model comparison and checking}

\subsection{Cross-validation} 

To be more explicit, the $10$ random groups for the cross-validation are: 1) Valencia, Almer\'ia, Gij\'on, Lleida; 2) San Sebasti\'an, Bilbao, Logro\~no, Daroca; 3) Tortosa, Murcia, Madrid (Torrej\'on), Mor\'on; 4) Ciudad Real, Sevilla, Madrid (Barajas), Valladolid; 5) Madrid (Retiro), Zaragoza, Barcelona (Fabra), Zamora; 6) M\'alaga, Albacete, Santiago, Vitoria; 7) Salamanca, Burgos, C\'aceres, Madrid (Getafe); 8) Badajoz, Navacerrada, Soria, Reus; 9) Barcelona (Aeropuerto), Santander, Huelva, Madrid (Cuatrovientos); 10) Coru\~na, Ponferrada, Le\'on, Castell\'on.

The third metric is on the data space rather than the probability space to account for uncertainty in the number of records. In the cross-validation, for each hold-out group we can obtain $K$ posterior predictive sequences, $\{N_{t\ell}^{(k)}(\bs) : k=1,\ldots,K \}$, of the number of records up to year $t$ for day $\ell$ and location $\bs$. We calculate an absolute deviation (AD) statistic, $\lvert N_{t\ell}(\bs) - N_{t\ell}^{(k)}(\bs) \rvert$, comparing any model-generated sequence realization with the observed sequence realization. We average the AD's over days $\ell$, locations $\bs$, and the $K$ posterior samples and plot this average versus $t$.  We can see annual variation with smaller values indicating better performance.

\subsection{DIC} 

Given the data $\bI$ with associated probabilities $\bp$ and the Bernoulli likelihood,
\begin{equation*}
    [\bI \mid \bp] = \prod_{i=1}^{n} \prod_{t=2}^{T} \prod_{\ell=1}^{365} p_{t\ell}(\bs_i)^{I_{t\ell}(\bs_i)} (1 - p_{t\ell}(\bs_i))^{1 - I_{t\ell}(\bs_i)},
\end{equation*}
the deviance $\D(\bp) = -2 \log([\bI \mid \bp])$ of the model is given by
\begin{equation*}
    \D(\bp) = -2  \sum_{i=1}^{n} \sum_{t=2}^{T} \sum_{\ell=1}^{365} \big( I_{t\ell}(\bs_i) \log p_{t\ell}(\bs_i) + (1 - I_{t\ell}(\bs_i)) \log (1 - p_{t\ell}(\bs_i)) \big).
\end{equation*}
The deviance of the model measures the variability linked to the likelihood, and is a random variable usually summarized by the posterior mean deviance $\hat{\D} = E[\D(\bp) \mid \bI]$.

The DIC is the sum of two components, one for quantifying the model fit and the other for penalizing its complexity. The first component is measured through $\hat{\D}$ while the model complexity is measured through the effective number of parameters,
\begin{equation*}
    p_{\D} = E[\D(\bp) \mid \bI] - \D(E[\bp \mid \bI]) = \hat{\D} - \D(\hat{\bp}),
\end{equation*}
and the DIC is defined as
\begin{equation*}
    \DIC = \hat{\D} + p_{\D}.
\end{equation*}
A smaller DIC indicates a better model fit.

\subsection{Probability integral transform} 

The PIT histogram proposed by \cite{czado2009} can be used to check for deviations from uniformity; U-shaped histograms suggest under-dispersed predictive distributions, inverse-U shaped histograms point at their over-dispersion, and skewed histograms occur when central tendencies are biased. Values near $0$ or $1$ (observed values that are outside of the CI) are denoted as outliers, and a high proportion of outliers suggest poor calibration of the model.

This PIT histogram adapted to our setting is as follows. We want to apply this to our features of interest like ratios or ERS's. For simplicity denote a feature by $X_{i}$ for $i=1,\ldots,n$, where $i$ can index time, space or both. These features can attain a countable number of real numbers, say $x_k$ for $k = 1,2,\ldots$ such that $x_{k} < x_{k+1}$. We define the function
\begin{equation*}
    F^{(i)}(u \mid x_{k}) = 
    \begin{cases} 
      0, & \text{if } u \le P_{k-1}^{(i)}, \\
      \frac{u - P_{k-1}^{(i)}}{P_{k}^{(i)} - P_{k-1}^{(i)}}, & \text{if } P_{k-1}^{(i)} \le u \le P_{k}^{(i)}, \\
      1, & \text{if } P_{k}^{(i)} \le u,
   \end{cases}
\end{equation*}
where $P_{0}^{(i)} = 0$, and $P_{k}^{(i)}$ is the value that the posterior cdf of $X_{i}$ attains at $x_{k}$. Aggregating in time and/or space, 
\begin{equation*}
    \bar{F}(u) = \frac{1}{n} \sum_{i=1}^{n} F^{(i)}(u \mid X_{i}), \quad 0 \le u \le 1.
\end{equation*}
Finally, for $J = 10$, we compute
\begin{equation*}
    f_{j} = \bar{F}\left(\frac{j}{J}\right) - \bar{F}\left(\frac{j - 1}{J}\right)
\end{equation*}
for equally spaced bins $j=1,\ldots,J$, plot a histogram with height $f_{j}$ for bin $j$, and check for uniformity.

\section{MCMC convergence diagnostics}

\begin{figure}[tbh]
    \centering
    \includegraphics[width=.24\textwidth]{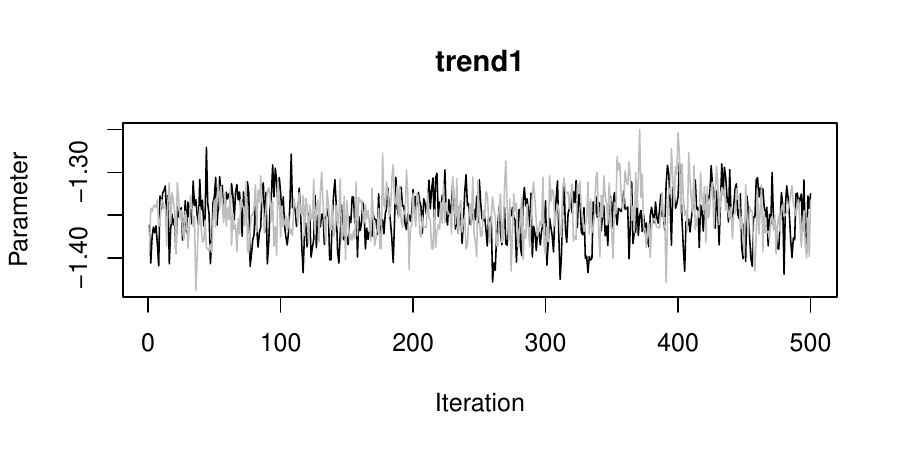}
    \includegraphics[width=.24\textwidth]{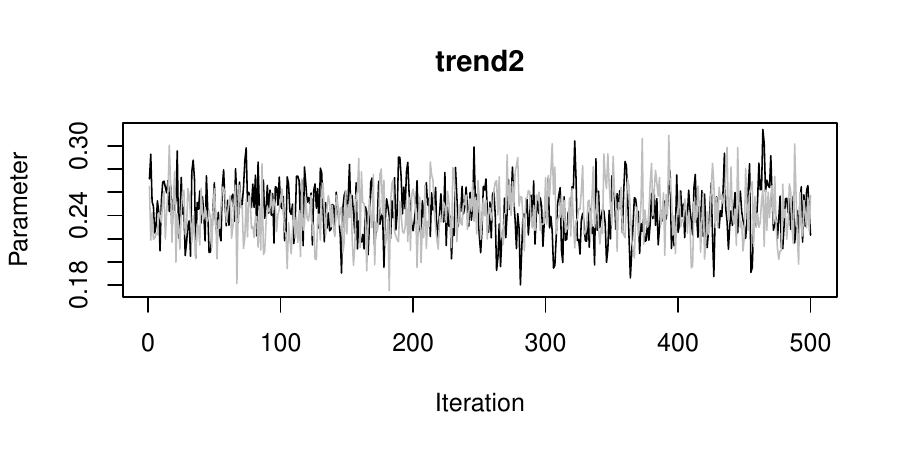}
    \includegraphics[width=.24\textwidth]{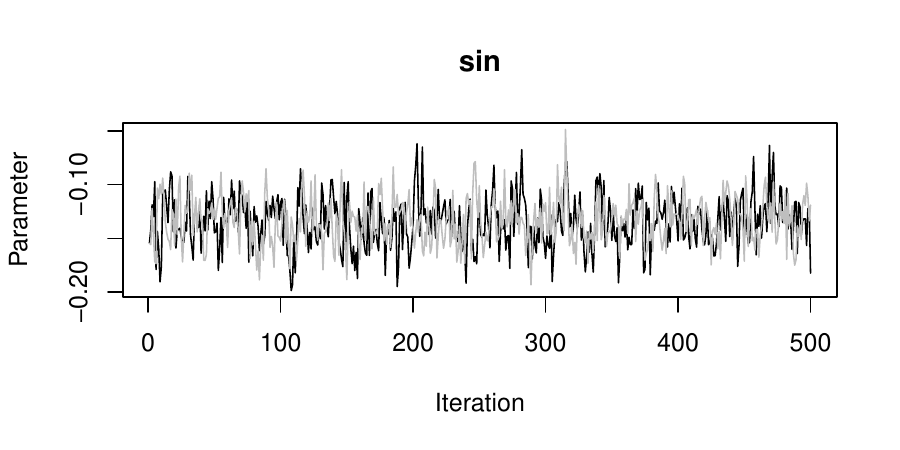}
    \includegraphics[width=.24\textwidth]{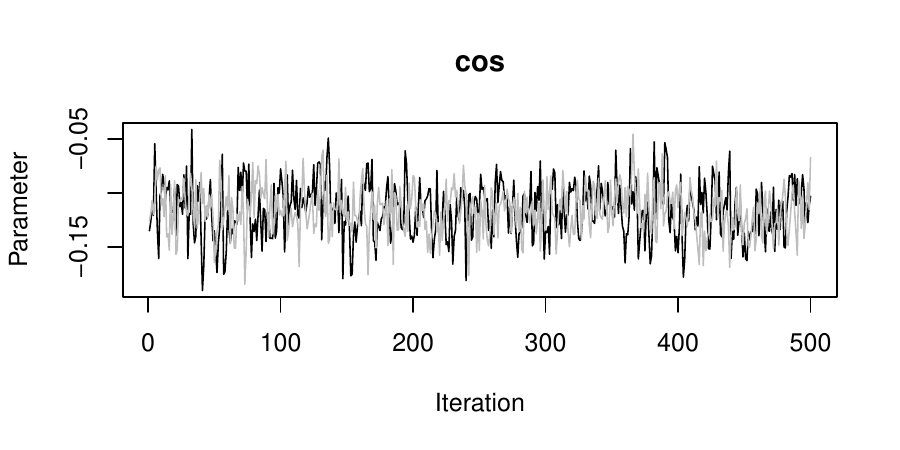}
    \includegraphics[width=.24\textwidth]{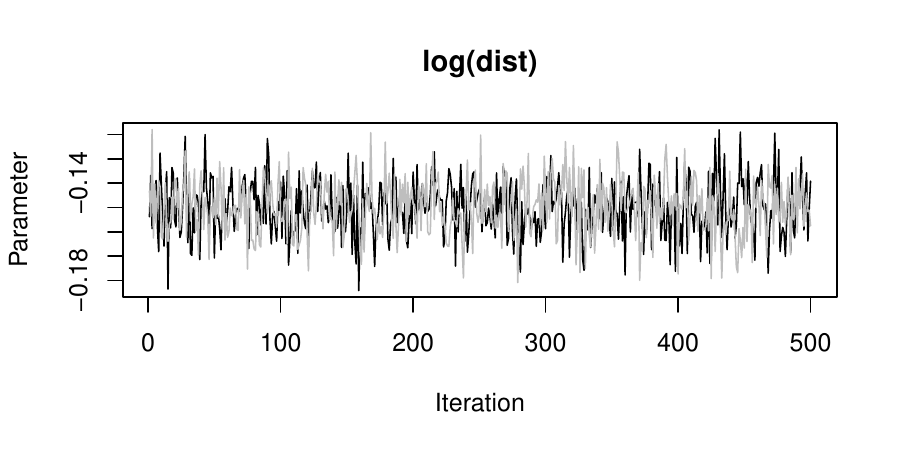}
    \includegraphics[width=.24\textwidth]{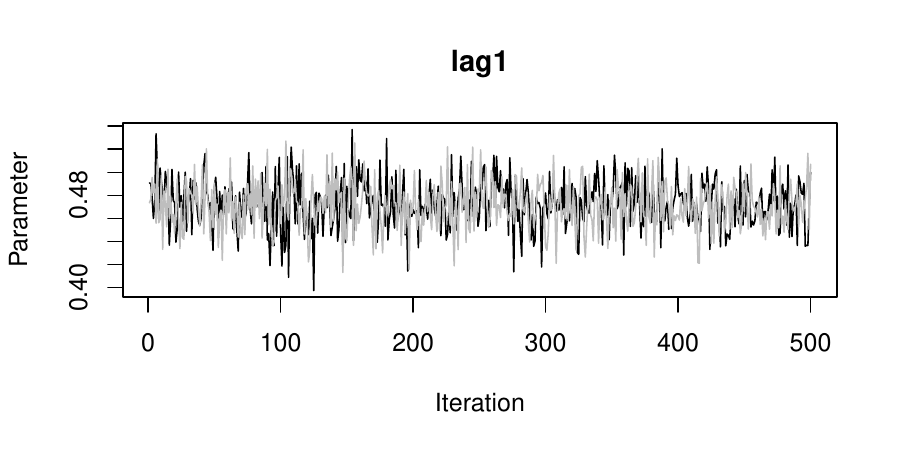}
    \includegraphics[width=.24\textwidth]{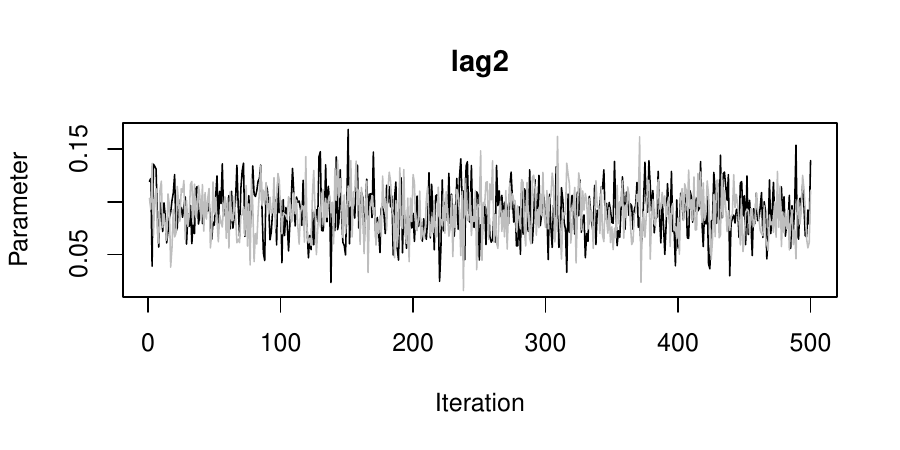}
    \includegraphics[width=.24\textwidth]{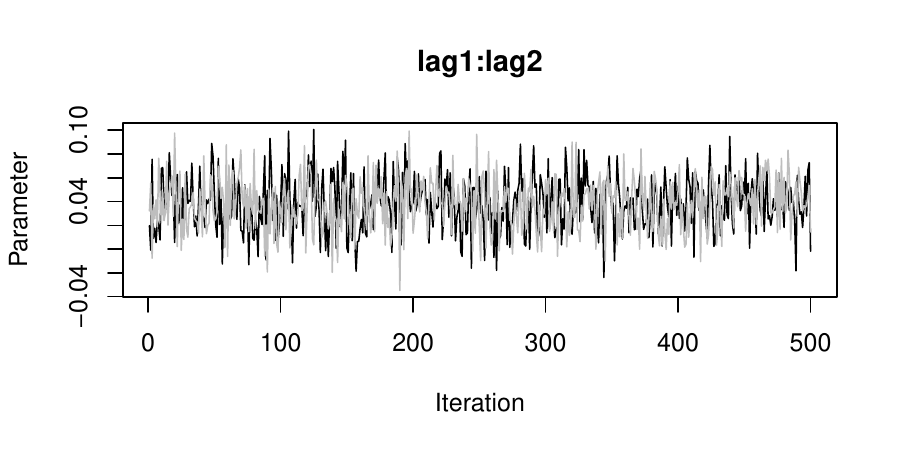}
    \includegraphics[width=.24\textwidth]{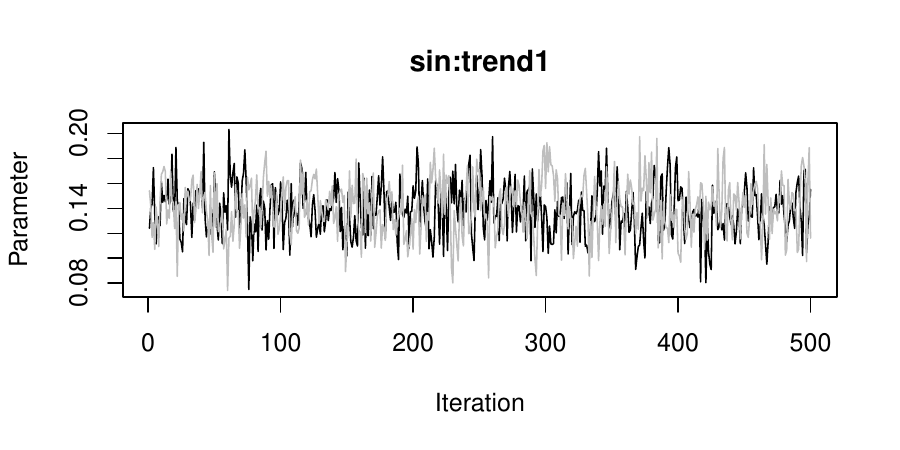}
    \includegraphics[width=.24\textwidth]{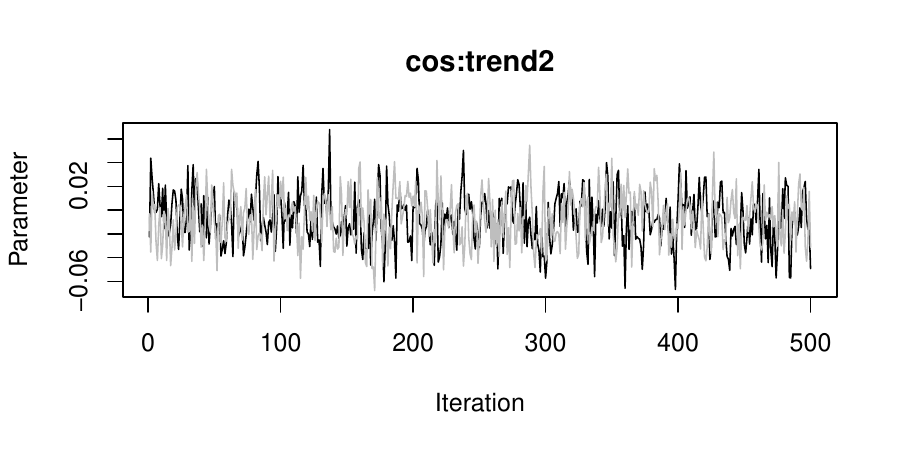}
    \includegraphics[width=.24\textwidth]{SUPP_traceplotsinepolytrend21.pdf}
    \includegraphics[width=.24\textwidth]{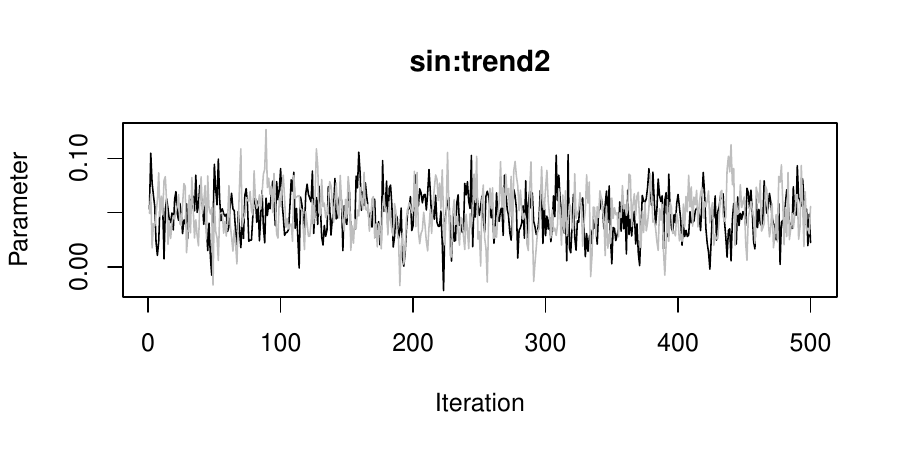}
    \includegraphics[width=.24\textwidth]{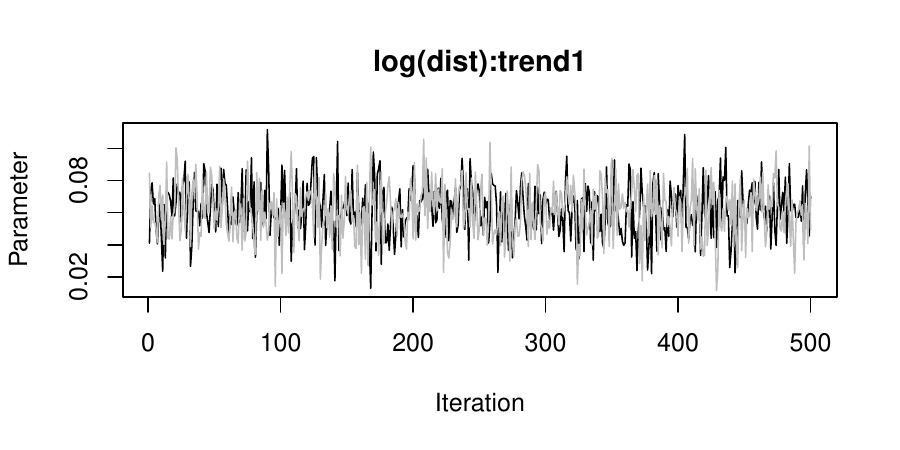}
    \includegraphics[width=.24\textwidth]{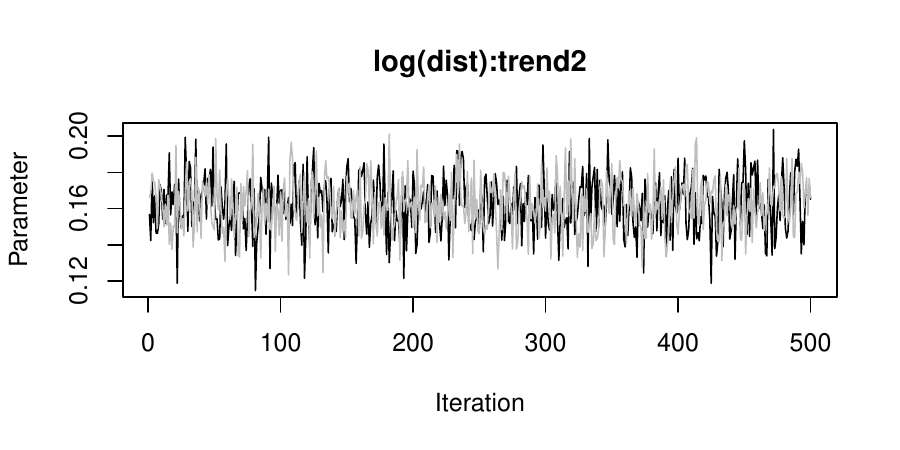}
    \includegraphics[width=.24\textwidth]{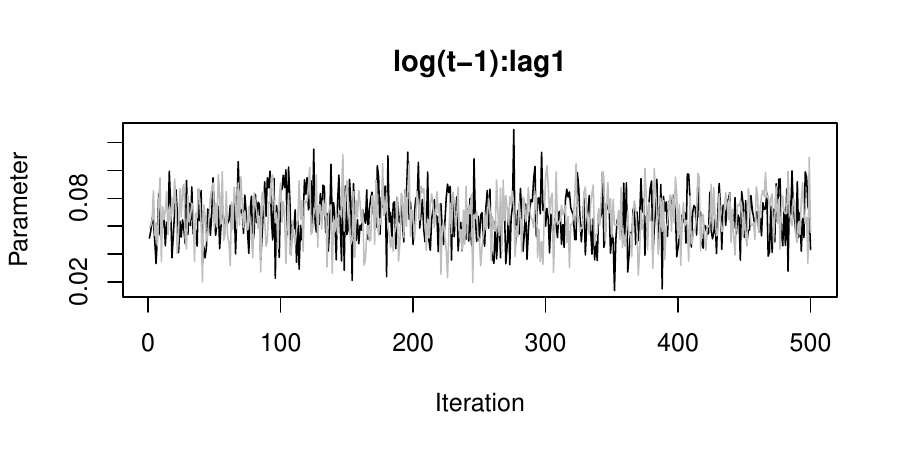}
    \includegraphics[width=.24\textwidth]{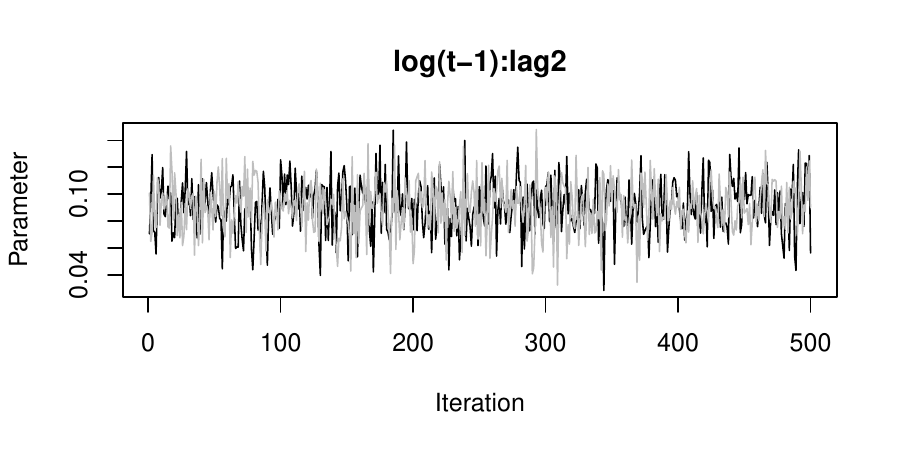}
    \includegraphics[width=.24\textwidth]{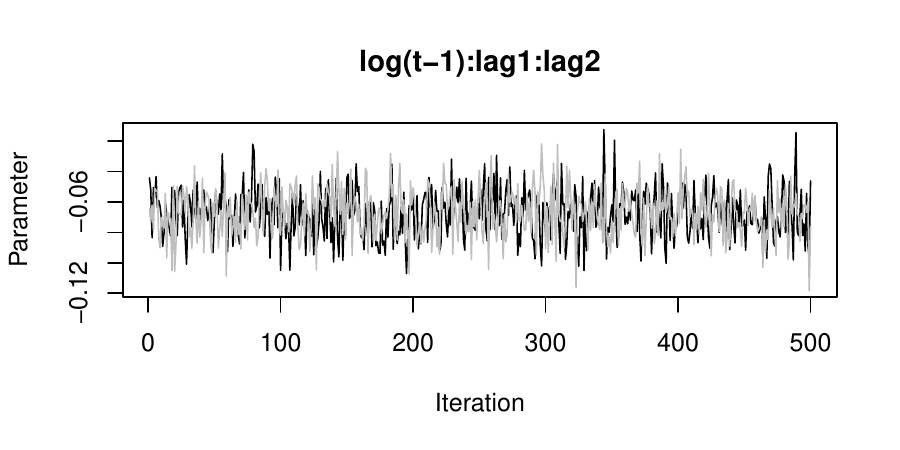}
    \includegraphics[width=.24\textwidth]{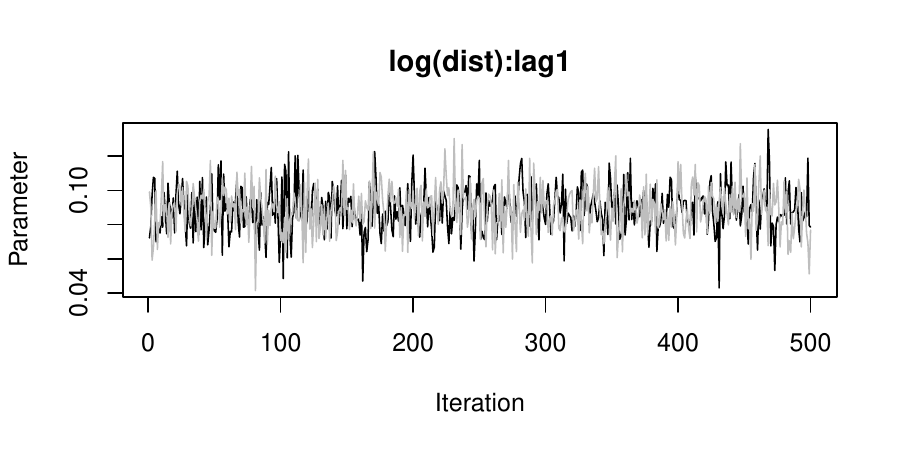}
    \includegraphics[width=.24\textwidth]{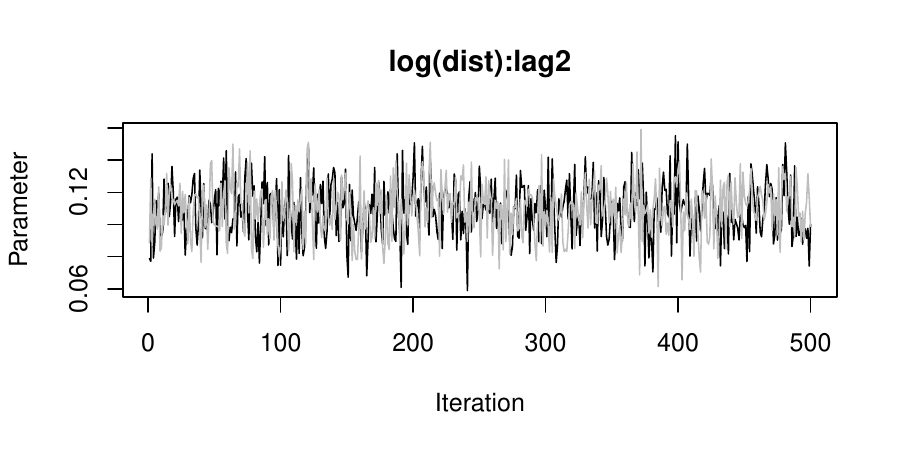}
    \includegraphics[width=.24\textwidth]{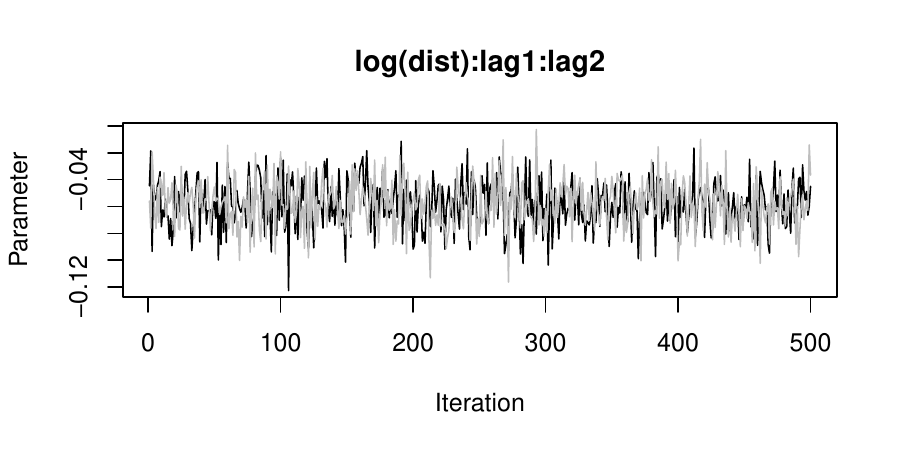}
    \caption{Trace plots for the parameters in $\bm{\beta}$ in the full model.}
    \label{fig:traceplot1}
\end{figure}

\begin{figure}[tbh]
    \centering
    \includegraphics[width=.24\textwidth]{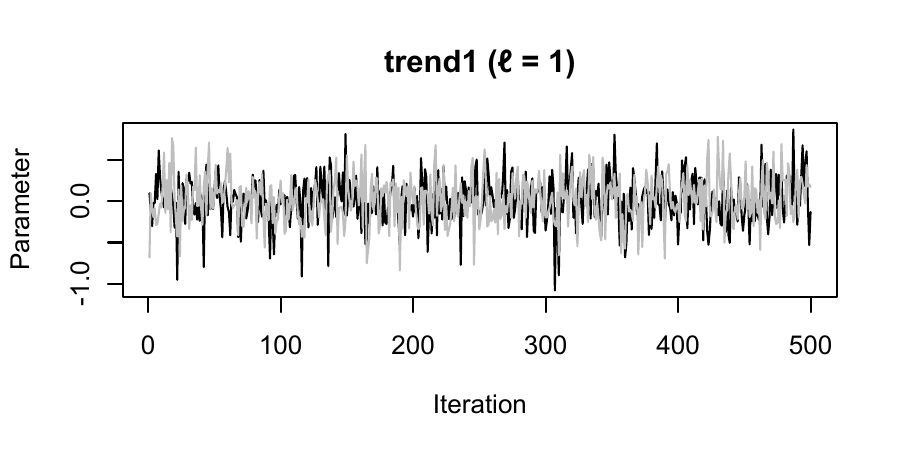}
    \includegraphics[width=.24\textwidth]{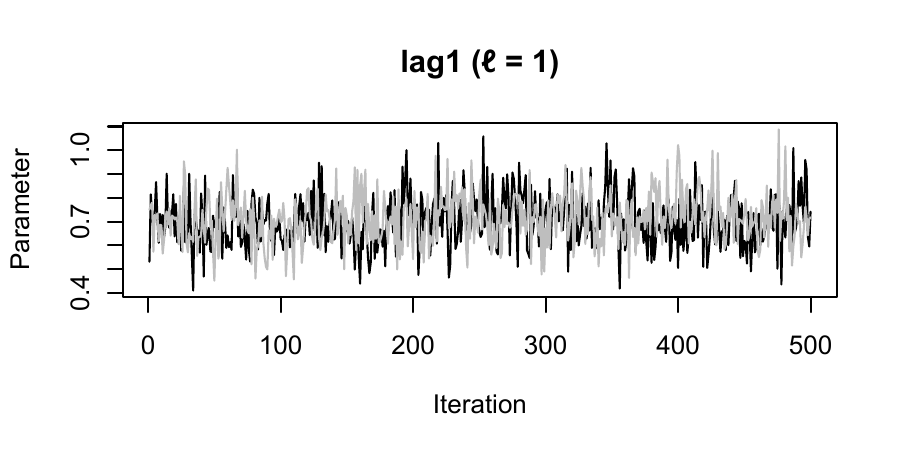}
    \includegraphics[width=.24\textwidth]{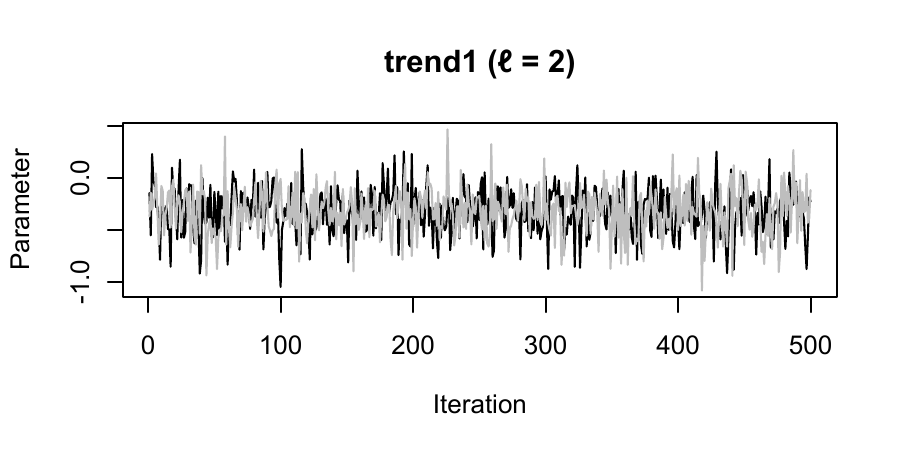}
    \includegraphics[width=.24\textwidth]{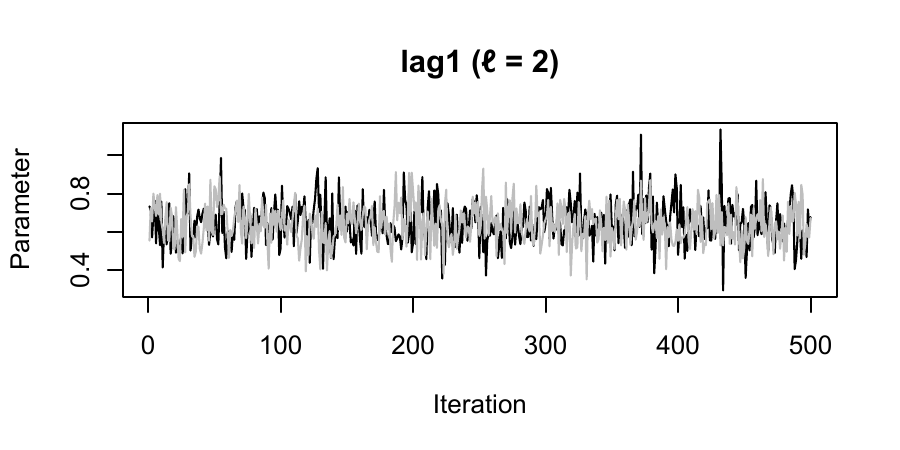}
    \caption{Trace plots for the parameters in $\bm{\beta}_{1}$ and $\bm{\beta}_{2}$ in the full model.}
    \label{fig:traceplot2}
\end{figure}

\begin{figure}[tbh]
    \centering
    \includegraphics[width=.24\textwidth]{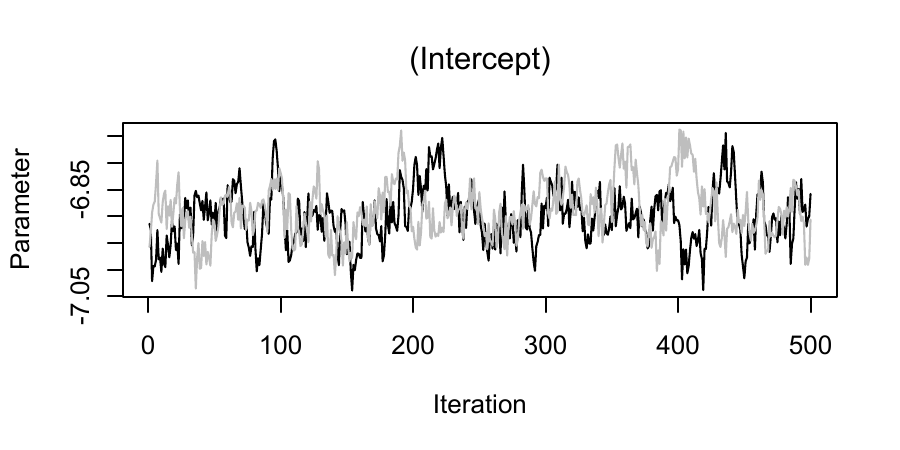}
    \includegraphics[width=.24\textwidth]{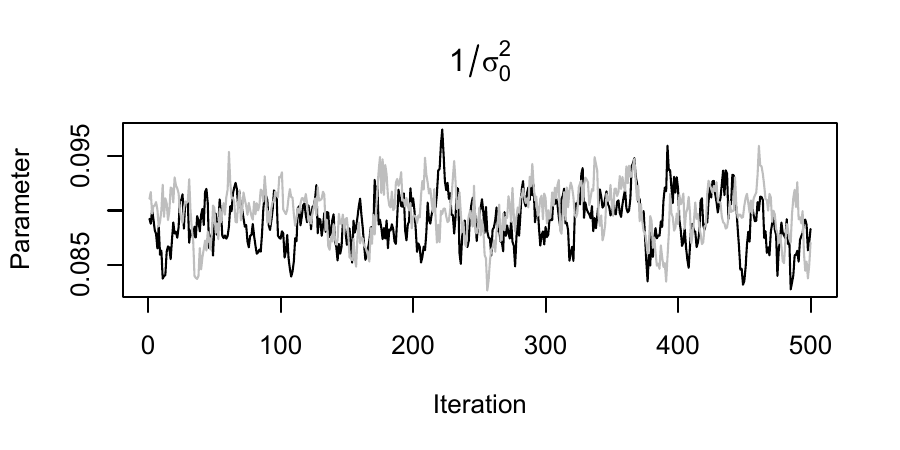}
    \includegraphics[width=.24\textwidth]{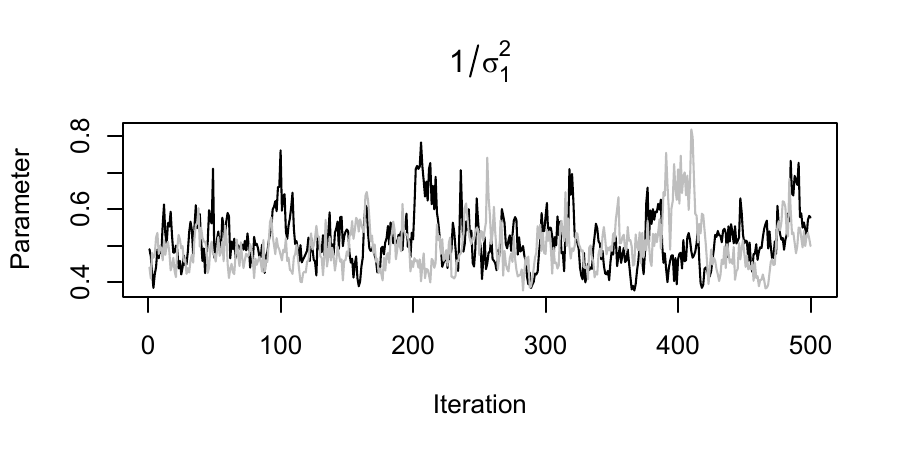} 
    \includegraphics[width=.24\textwidth]{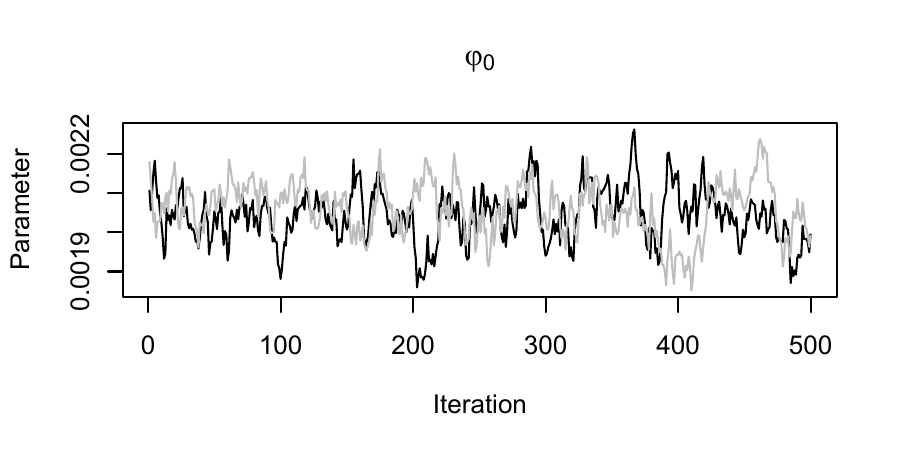} \\
    \includegraphics[width=.24\textwidth]{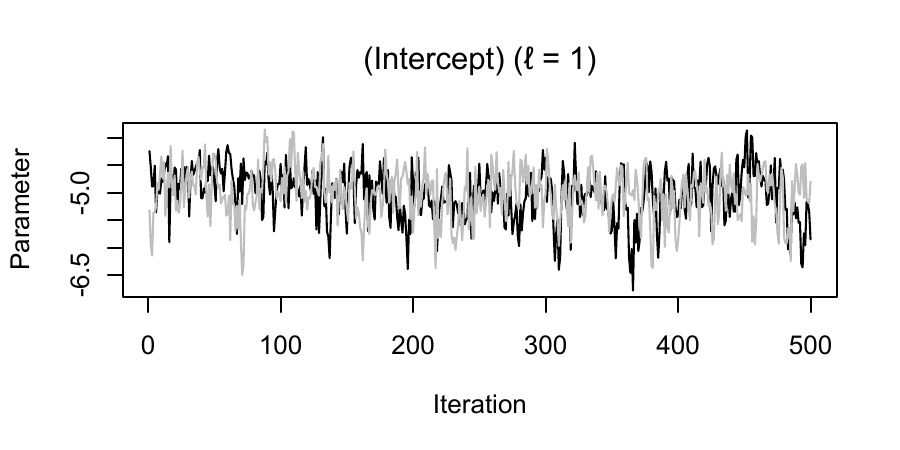}
    \includegraphics[width=.24\textwidth]{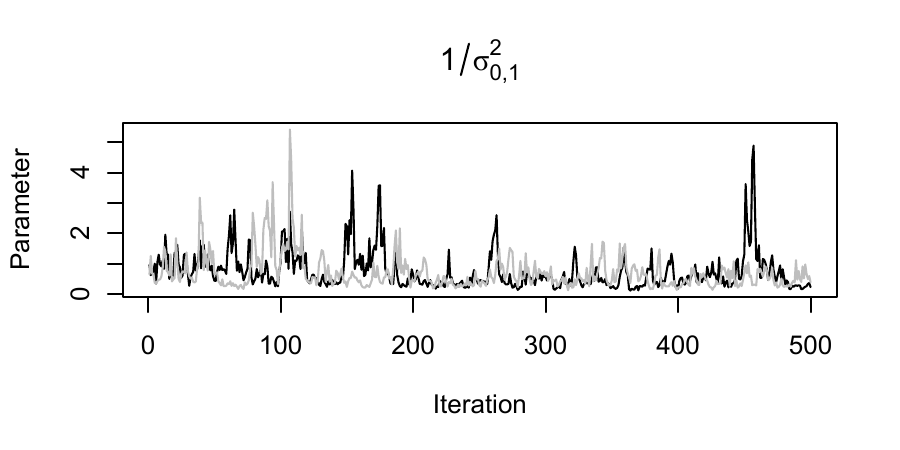}
    \includegraphics[width=.24\textwidth]{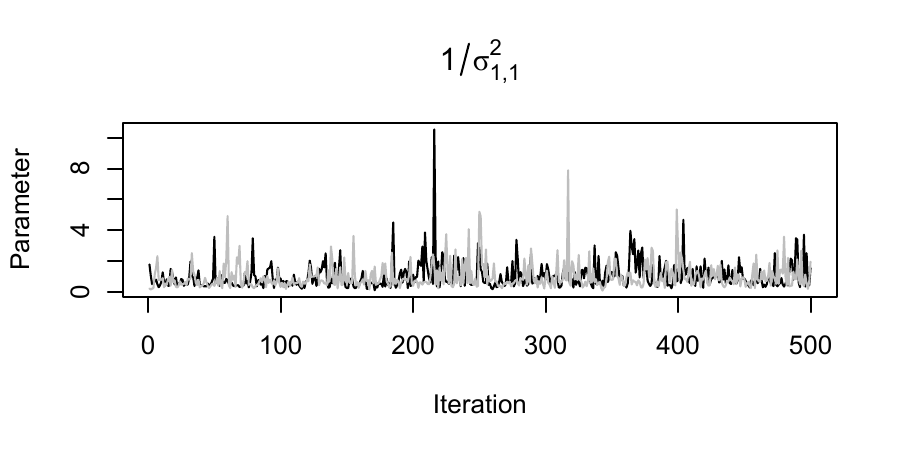} \\
    \includegraphics[width=.24\textwidth]{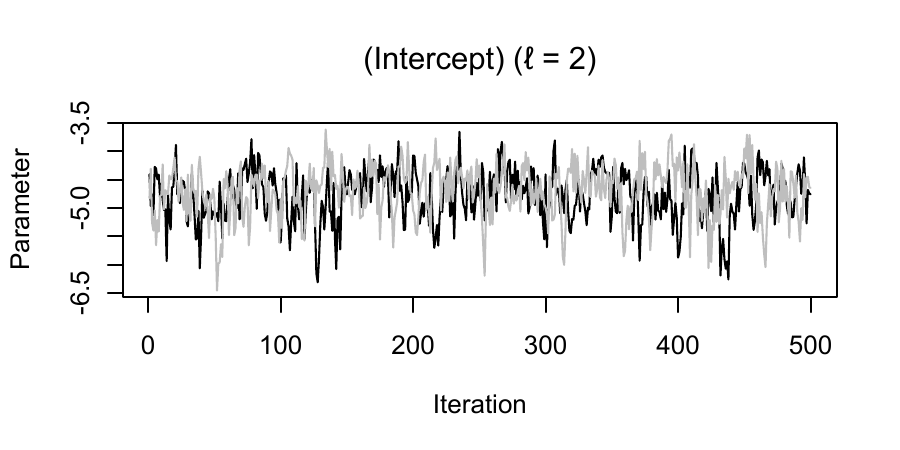}
    \includegraphics[width=.24\textwidth]{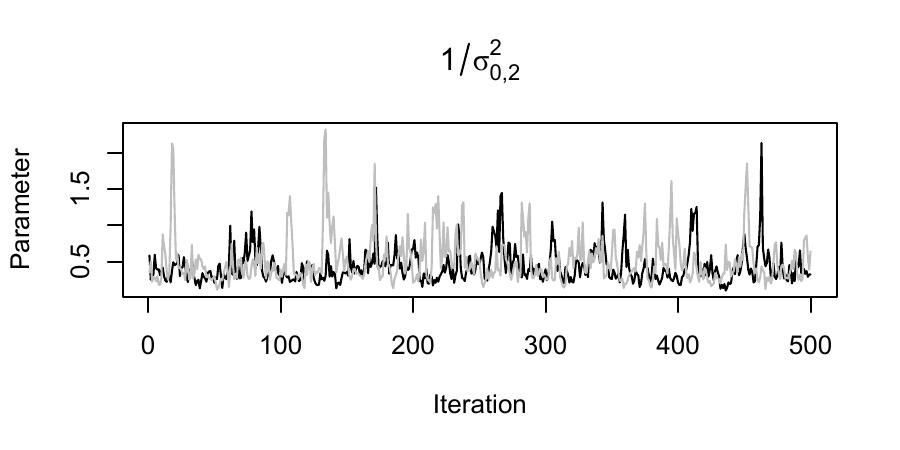}
    \includegraphics[width=.24\textwidth]{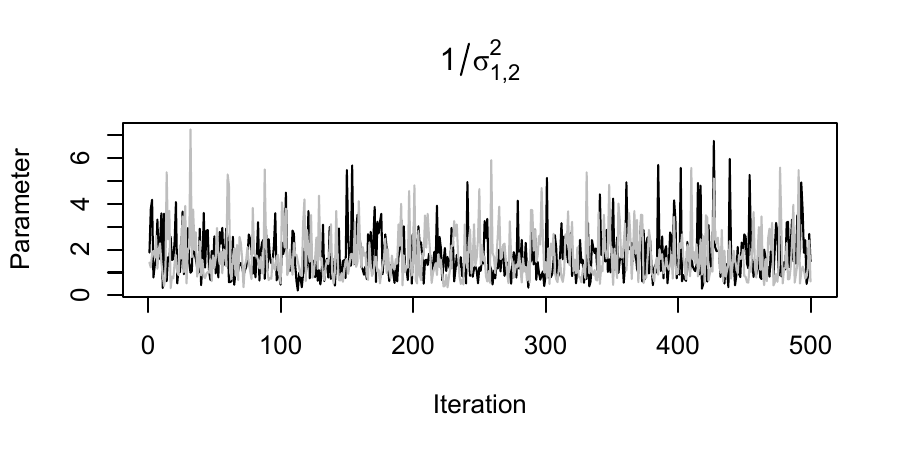}
    \caption{Trace plots for the hyperparameters $\beta_{0}$, $1 / \sigma_{0}^{2}$, $1 / \sigma_{1}^{2}$, $\phi_{0}$, $\beta_{0,1}$, $1 / \sigma_{0,1}^{2}$, $1 / \sigma_{1,1}^{2}$, $\beta_{0,2}$, $1 / \sigma_{0,2}^{2}$, and $1 / \sigma_{1,2}^{2}$, in the full model.}
    \label{fig:traceplot3}
\end{figure}

Convergence was monitored by usual trace plots (see Figures~\ref{fig:traceplot1}, \ref{fig:traceplot2}, and \ref{fig:traceplot3}), and marginal and multivariate potential scale reduction factors. The trace plots show good agreement between the two chains in black and gray. The potential scale reduction factors were obtained via the \texttt{R}~package \texttt{coda} \citep{coda} and values below $1.1$ are in agreement with the convergence of the algorithm.

\clearpage

\section{Additional results} \label{sec:results}

\subsection{Model comparison and checking}

\begin{figure}[tb]
    \centering
    \includegraphics[width=0.8\textwidth]{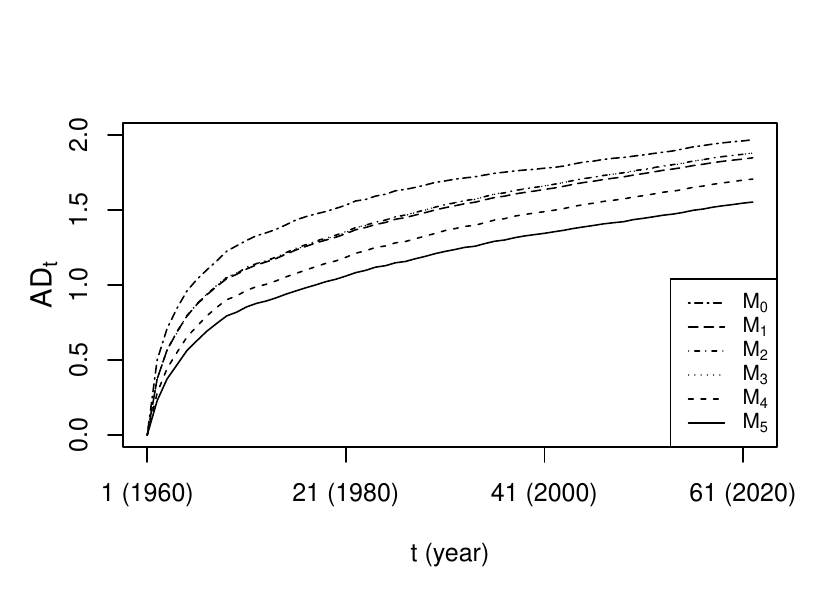}
    \caption{Annual AD metric in 10-fold cross-validation for the different nested models.}
    \label{fig:AD}
\end{figure}

\paragraph{Model comparison.} Figure~\ref{fig:AD} shows the AD metric across years for the six models. The lowest values reached by the full model indicate that this model has the best model performance compared with the other five models. Table~\ref{tab:DIC} shows DIC values, including deviance and complexity penalty measures for the six models compared. The increase in model complexity is clear as richer components are included in the models, but the improvement in model fit is more remarkable, meaning that lower DIC's are associated with more complex models.

\begin{table}[tb]
\begin{center}
\caption{The different nested models compared by DIC.} \label{tab:DIC}
\begin{tabular}{clccc} 
 \toprule 
 Model & Linear predictor & DIC & $\hat{\D}$ & $p_{\D}$ \\
 \midrule
 $M_{0}$ & $\eta_{t\ell}(\bs) = - \log(t - 1)$ &  $393805.4$ & $393805.4$ & $0$ \\
 $M_{1}$ & $ \eta_{t\ell}(\bs) = \bx_{t\ell}(\bs) \bbeta$ & $333193.6$ & $331798.4$ & $1395.2$ \\
 $M_{2}$ & $\eta_{t\ell}(\bs) = \bx_{t\ell}(\bs) \bbeta + w(\bs)$ & $333130.3$ & $331686.3$ & $1444.0$ \\
 $M_{3}$ & $\eta_{t\ell}(\bs) = \bx_{t\ell}(\bs) \bbeta + w(\bs) + w_{t}$ & $331142.9$ & $329646.3$ & $1496.6$ \\
 $M_{4}$ & $\eta_{t\ell}(\bs) = \bx_{t\ell}(\bs) \bbeta + w(\bs) + w_{t\ell}$ & $251653.5$ & $241949.7$ & $9703.8$ \\
 $M_{5}$ & $\eta_{t\ell}(\bs) = \bx_{t\ell}(\bs) \bbeta + w_{t\ell}(\bs)$ & $182680.2$ & $148885.7$ & $33794.5$ \\ 
 \bottomrule
\end{tabular}
\end{center}
\end{table}

\begin{figure}[tb]
    \centering
    \includegraphics[width=\textwidth]{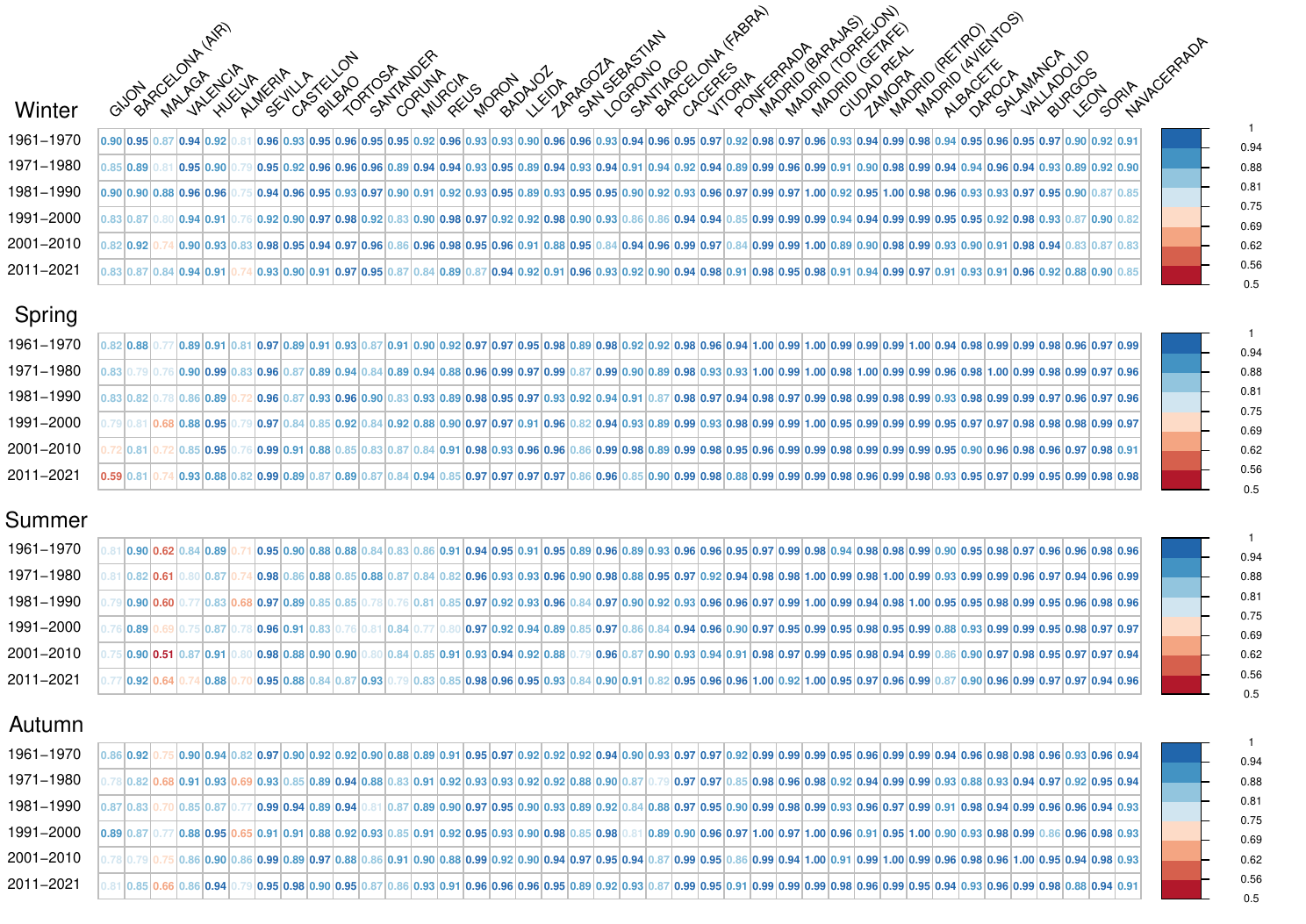}
    \caption{AUC metric in 10-fold cross-validation for the full model $M_{5}$ disaggregated by decade, season, and location (sorted from lower to higher elevation). Values higher (lower) than $0.75$ are in blue (red) scale.}
    \label{fig:AUC}
\end{figure}

Figure~\ref{fig:AUC} shows AUC values for the full model disaggregated by decade (the last period are 11 years), season, and location. This is useful to know periods of time or spatial regions that the model reproduces better or worse. The model has a very similar performance throughout the six decades, perhaps slightly better for the first one for obvious reasons. The seasons also show a similar behavior except for summer, which seems more difficult to reproduce, possibly because, as seen in the results below, it presents greater spatial variability. The greatest differences in performance appear between regions. While $71\%$ of AUC's are higher than $0.9$ and $94\%$ are higher than $0.8$, there are two locations that have relatively low values for most time periods. The worst captured is M\'alaga, especially the summer period with most AUC's between $0.6$ and $0.7$ for this season, followed by Almer\'ia also with some AUC's below $0.75$. Both weather stations are adjacent in space and situated along the coast, but they are surrounded by the highest mountain range on the peninsula, potentially blocking the passage of winds and creating a localized climate. Gij\'on also presents some specific inconsistencies in recent decades in spring; it may be because it presents an unusually low number of records for this time period. On the other hand, there are locations that, due to being located in a valley or sharing information with other very nearby stations, have practically all AUC's very close to $1$, as is the case of the Madrid region.

\begin{figure}[t]
    \centering
    \includegraphics[width=0.4\textwidth]{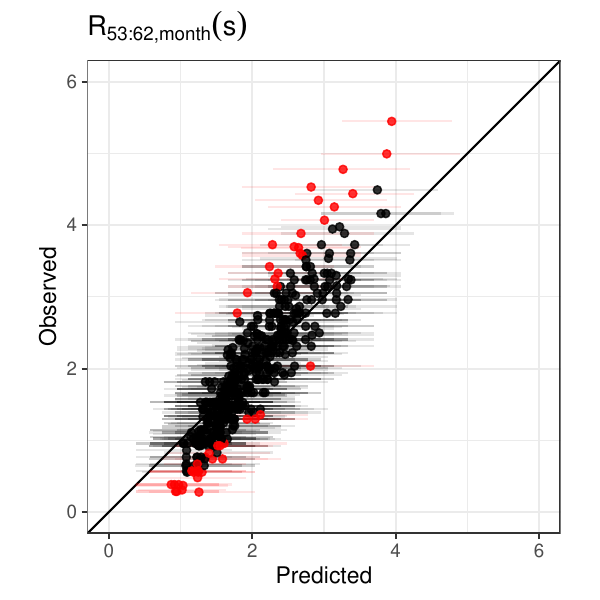}
    \includegraphics[width=0.4\textwidth]{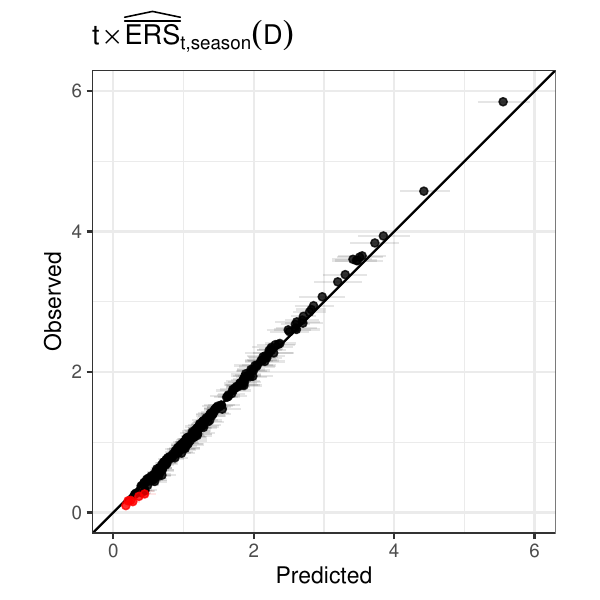}
    \includegraphics[width=0.4\textwidth]{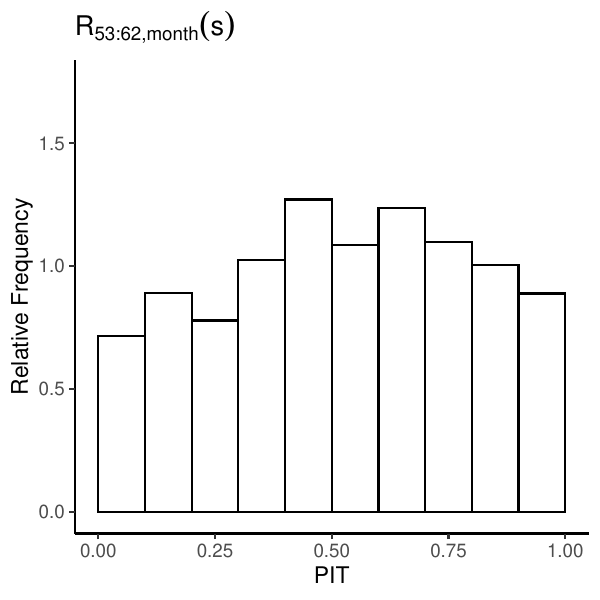}
    \includegraphics[width=0.4\textwidth]{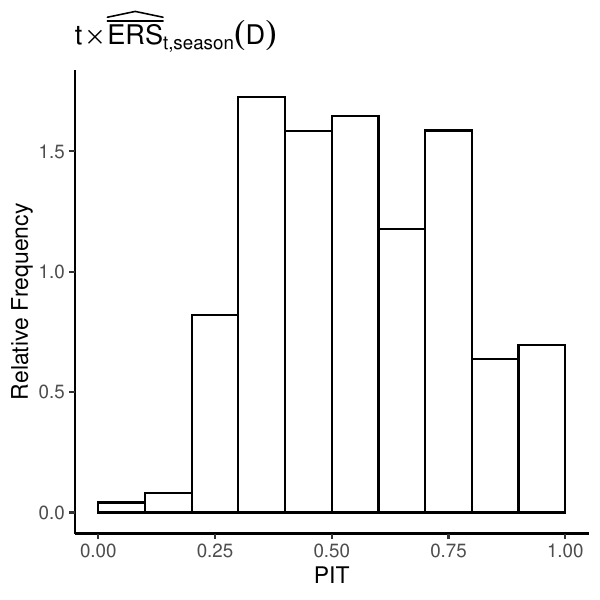}
    \caption{Scatter plots of the observed values versus the posterior mean and $90\%$ CI for the corresponding predictive distribution (top), and PIT histograms (bottom) for $R_{53:62,month}(\bs)$ (left) and $t \times \widehat{\overline{\text{ERS}}}_{t,season}(D)$ (right). Red indicates CI's not containing the observed value.}
    \label{fig:check}
\end{figure}

\paragraph{Model checking.} The top left plot in Figure~\ref{fig:check} shows the scatter plot of the observed values of $R_{53:62,month}(\bs)$ in (3) of the Manuscript for each observed location and $month$ going through the $12$ months versus a summary for the corresponding predictive distribution.  The bottom plot shows the associated PIT histogram. The same is shown on the right for $t \times \widehat{\overline{\text{ERS}}}_{t,season}(D)$ averaging in (4) of the Manuscript (defined over the grid of $40$ observed locations) for $t=2,\ldots,T$ going through the $4$ seasons. It is clear for the scatter plots that the predictions are very close to the observed values, with exceptionally sharp CI's for the ERS's. With regard to the histograms, it seems that the ratios are well-calibrated, with a $90\%$ coverage of $90.2\%$; while the ERS's suggest a slight over-dispersion with most PIT values around the middle of the range, and a $90\%$ coverage of $97.1\%$.

\subsection{Posterior distribution of the model parameters}

\paragraph{Coefficients with non-scaled covariates and raw polynomials.}
Table~\ref{tab:coefs} summarizes the posterior mean and $90\%$ CI of the $\bm{\beta}$ coefficients in the full model expressed in terms of the original non-scaled covariates and raw polynomials  which are easier for interpretation. Just to complete the model specification, Table~\ref{tab:coefs12} shows the posterior mean and $90\%$ CI of the coefficients $\bm{\beta}_{1}$ and $\bm{\beta}_{2}$ in the autoregressive model that specifies the initial condition for $I_{t1}(\bs)$ and $I_{t2}(\bs)$.

\begin{table}[p]
\begin{center}
\caption{Posterior mean and $90\%$ CI of $\bm\beta$ coefficients in the full model with non-scaled covariates and raw polynomials.} \label{tab:coefs}
\begin{tabular}{lrc}
  \toprule
  Covariate of  & Mean & $90\%$ CI \\ 
  \midrule
  $\log(t-1)$               & $-2.6531$ & $(-2.8301,-2.4876)$ \\
  $[\log(t-1)]^2$             & $ 0.2268$ & $( 0.1946, 0.2614)$ \\
  $\sin_{\ell}$                    & $-0.1899$ & $(-0.2424,-0.1366)$ \\
  $\cos_{\ell}$                    & $-0.1640$ & $(-0.2183,-0.1099)$ \\
  $\log(\dist(\bs))$               & $-0.0711$ & $(-0.0801,-0.0627)$ \\
  $I_{t,\ell-1}(\bs)$              & $ 1.8835$ & $( 1.7429, 2.0212)$ \\
  $I_{t,\ell-2}(\bs)$              & $ 0.3606$ & $( 0.2169, 0.5143)$ \\
  $I_{t,\ell-1}(\bs) \times I_{t,\ell-2}(\bs)$  & $ 0.2160$ & $( 0.0047, 0.4303)$ \\
  $\sin_{\ell}\times \log(t-1)$       & $-0.1078$ & $(-0.3433, 0.1095)$ \\
  $\cos_{\ell}\times \log(t-1)$       & $-0.0585$ & $(-0.2857, 0.1656)$ \\
  $\sin_{\ell}\times [\log(t-1)]^2$   & $ 0.0651$ & $( 0.0206, 0.1128)$ \\
  $\cos_{\ell}\times [\log(t-1)]^2$   & $-0.0095$ & $(-0.0562, 0.0366)$ \\
  $\log(\dist(\bs))\times \log(t-1)$  & $-0.1682$ & $(-0.1945,-0.1419)$ \\
  $\log(\dist(\bs))\times [\log(t-1)]^2$  & $ 0.0366$ & $( 0.0310, 0.0422)$ \\
  $\log(t-1)\times I_{t,\ell-1}(\bs)$  & $ 0.1006$ & $( 0.0565, 0.1444)$ \\
  $\log(t-1)\times I_{t,\ell-2}(\bs)$  & $ 0.1381$ & $( 0.0884, 0.1860)$ \\
  $\log(t-1)\times I_{t,\ell-1}(\bs) \times I_{t,\ell-2}(\bs)$    & $-0.1692$ & $(-0.2354,-0.1049)$ \\
  $\log(\dist(\bs)) \times I_{t,\ell-1}(\bs)$          & $ 0.0850$ & $( 0.0645, 0.1055)$ \\
  $\log(\dist(\bs)) \times I_{t,\ell-2}(\bs)$         & $ 0.1040$ & $( 0.0780, 0.1311)$ \\
  $\log(\dist(\bs)) \times I_{t,\ell-1}(\bs) \times I_{t,\ell-2}(\bs)$     & $-0.0767$ & $(-0.1159,-0.0380)$ \\
  \bottomrule
\end{tabular}
  
\vspace*{1cm}

\caption{Posterior mean and $90\%$ CI of the  coefficients $\bm{\beta}_{1}$ and $\bm{\beta}_{2}$  in the autoregressive model  that specifies the  initial condition for $I_{t1}(\bs)$ and $I_{t2}(\bs)$.} \label{tab:coefs12}
\begin{tabular}{lrc}
  \toprule
  Covariate of & Mean & $90\%$ CI \\ 
  \midrule
  $\log(t-1)$ $(\ell=1)$        & $ 0.0205$ & $(-0.4845,0.5115)$ \\
  $I_{t-1,365}(\bs)$ $(\ell=1)$ & $ 2.8220$ & $( 2.1701,3.5888)$ \\
  $\log(t-1)$ $(\ell=2)$        & $-0.3611$ & $(-0.7753,0.0275)$ \\
  $I_{t,1}(\bs)$ \ \ \ \ \ $(\ell=2)$ & $ 2.5666$ & $( 1.8855,3.2728)$ \\
  \bottomrule
\end{tabular}
\end{center}
\end{table}

\begin{figure}[t]
    \centering
    \includegraphics[width=0.45\textwidth]{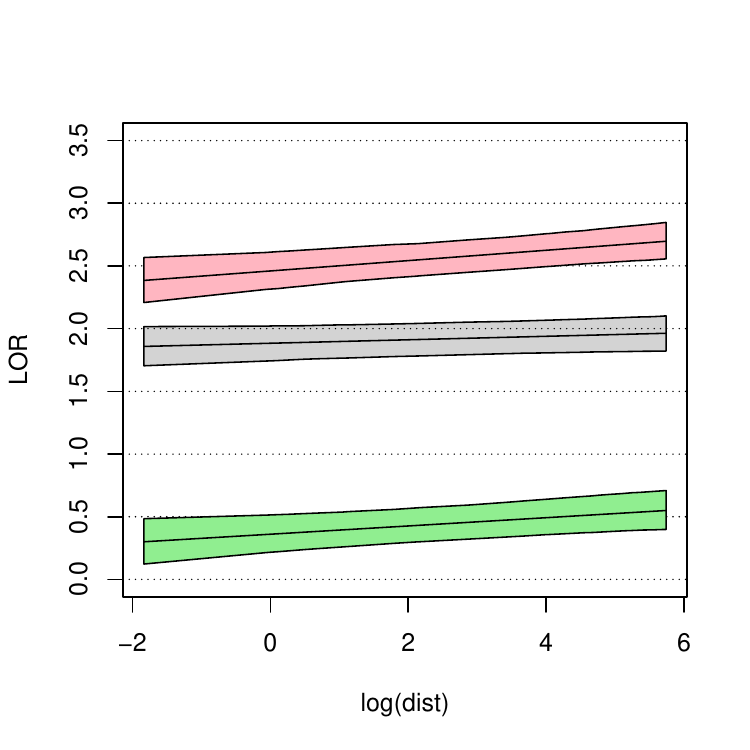}
    \includegraphics[width=0.45\textwidth]{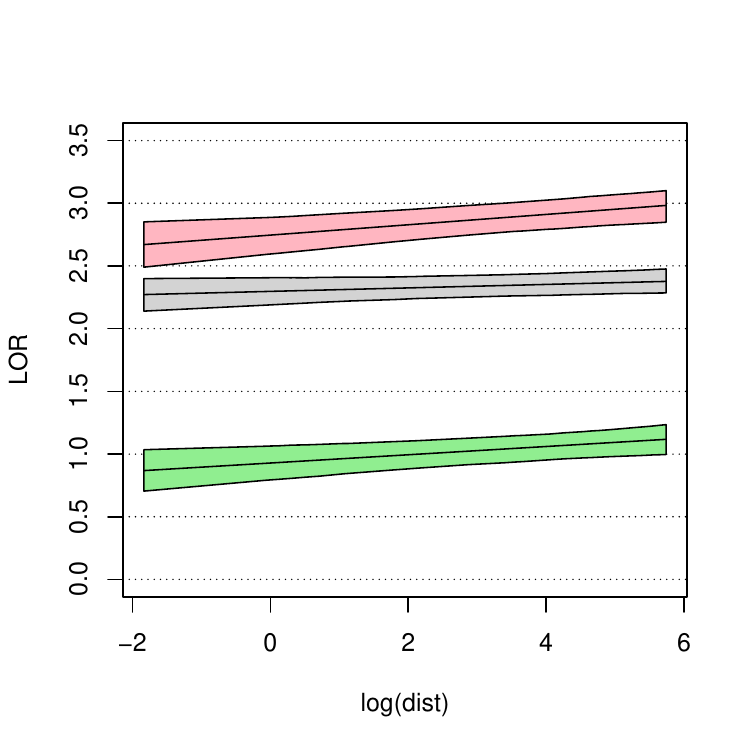}
    \caption{Posterior mean and $90\%$ CI of persistence coefficients against $\log(\dist(\bs))$ when $t=2$ (left) and $t=62$ (right). Gray for $I_{t,\ell-1}(\bs)=1, I_{t,\ell-2}(\bs)=0$, green for $I_{t,\ell-1}(\bs)=0, I_{t,\ell-2}(\bs)=1$, and red for $I_{t,\ell-1}(\bs)=1, I_{t,\ell-2}(\bs)=1$.}
    \label{fig:LOR:lag1:lag2}
\end{figure}

\paragraph{Persistence effect.}
To visualize the effect of persistence,  and its relationship with location and evolution across years, Figure~\ref{fig:LOR:lag1:lag2} displays the posterior mean and $90\%$ CI of the  persistence effect against $\log(\dist(\bs))$, for two years  $t=2$ and $t=62$, in three scenarios defined by the occurrence  of records or not in the two previous days.  For year $t=2$,  the occurrence of  records in the two  previous days increases the linear predictor, with respect to the occurrence of no records,  by approximately $2.5$, by around $2$ when there is only a record in the previous day, and by $0.5$ when there is a record only two days ago. These persistence effects are increasing over time, and for the last year studied ($t=62$), the corresponding increases are approximately $2.75$, $2.30$, and $1.0$, respectively. In the three scenarios, inland areas have higher values than coastal areas. However, the gradient associated with distance to the coast is flatter when there is only a record in the previous day.

\begin{table} [tb]
\begin{center}
\caption{Posterior mean and $90\%$ CI of persistence coefficients, for Barcelona (Fabra) and Madrid (Retiro) when $t = 2$ and $62$.}
\label{tab:LOR:lag1:lag2}
\begin{tabular}{cccccccc}
  \cline{3-8}
  & & \multicolumn{2}{c}{$I_{t,\ell-1}(\bs)=1$} & \multicolumn{2}{c}{$I_{t,\ell-1}(\bs)=0$} & \multicolumn{2}{c}{ $I_{t,\ell-1}(\bs)=1$} \\ 
  & & \multicolumn{2}{c}{$I_{t,\ell-2}(\bs)=0$} &  \multicolumn{2}{c}{$I_{t,\ell-2}(\bs)=1$} & \multicolumn{2}{c}{$I_{t,\ell-2}(\bs)=1$} \\ \cmidrule{3-8}
  & & Mean & $90\%$ CI & Mean & $90\%$ CI & Mean & $90\%$ CI\\ \midrule
  \multirow{2}{*}{$t=2$} & Barcelona (Fabra) & $1.91$ & $(1.78,2.04)$ & $0.43$ & $(0.30,0.57)$ & $2.55$ & $(2.42,2.68)$ \\
  & Madrid (Retiro) & $1.96$ & $(1.82,2.10)$ & $0.55$ & $(0.40,0.71)$ & $2.70$ & $(2.56,2.85)$ \\ \midrule
  \multirow{2}{*}{$t=62$}& Barcelona (Fabra) & $2.33$ & $(2.24,2.42)$ & $1.00$ & $(0.89,1.11)$ & $2.83$ & $(2.71,2.95)$ \\
  & Madrid (Retiro) & $2.38$ & $(2.28,2.48)$ & $1.12$ & $(1.00,1.23)$ & $2.98$ & $(2.85,3.10)$ \\ \bottomrule
\end{tabular}
\end{center}
\end{table}

To see more clearly the spatial effect, Table~\ref{tab:LOR:lag1:lag2} compares the posterior mean and $90\%$ CI of the persistence effect in a coastal location,  Barcelona (Fabra) and an inland location Madrid (Retiro), again in $t=2,62$. In all the cases, the posterior mean is lower in Barcelona than in Madrid although the CI's  slightly overlap.

\paragraph{Model hyperparameters.}
Table~\ref{tab:my_label} summarizes the posterior mean and $90\%$ CI of all the  hyperparameters in the full model (intercept for the models with scaled covariates).

\begin{table}[tb]
\begin{center}
\caption{Posterior mean and $90\%$ CI of the hyperparameters in the full model (intercepts for the model with scaled covariates).} \label{tab:my_label}
\begin{tabular}{crc}
  \toprule
  Parameter & Mean & $90\%$ CI \\ 
  \midrule
  $\beta_{0}$     & $-6.8954$ & $(-6.9779,-6.7988)$ \\
  $\phi_{0}$      & $ 0.0021$ & $( 0.0019, 0.0022)$ \\
  $\sigma_{0}$    & $3.3449$ & $(3.2777,3.4212)$ \\
  $\sigma_{1}$    & $1.4153$ & $(1.2529,1.5579)$ \\
  $\beta_{0,1}$   & $-5.0094$ & $(-5.8408,-4.3431)$ \\
  $\sigma_{0,1}$  & $1.3899$ & $(0.7396,2.0971)$ \\
  $\sigma_{1,1}$  & $1.1947$ & $(0.6362,1.8688)$ \\
  $\beta_{0,2}$   & $-4.7404$ & $(-5.5458,-4.0730)$ \\
  $\sigma_{0,2}$  & $1.6073$ & $(0.9815,2.2262)$ \\
  $\sigma_{1,2}$  & $0.8663$ & $(0.5162,1.3591)$ \\
  \bottomrule
\end{tabular}
\end{center}
\end{table}

\subsection{Model-based analysis over peninsular Spain}

\begin{figure}[t]
    \centering
    \includegraphics[width=.32\textwidth]{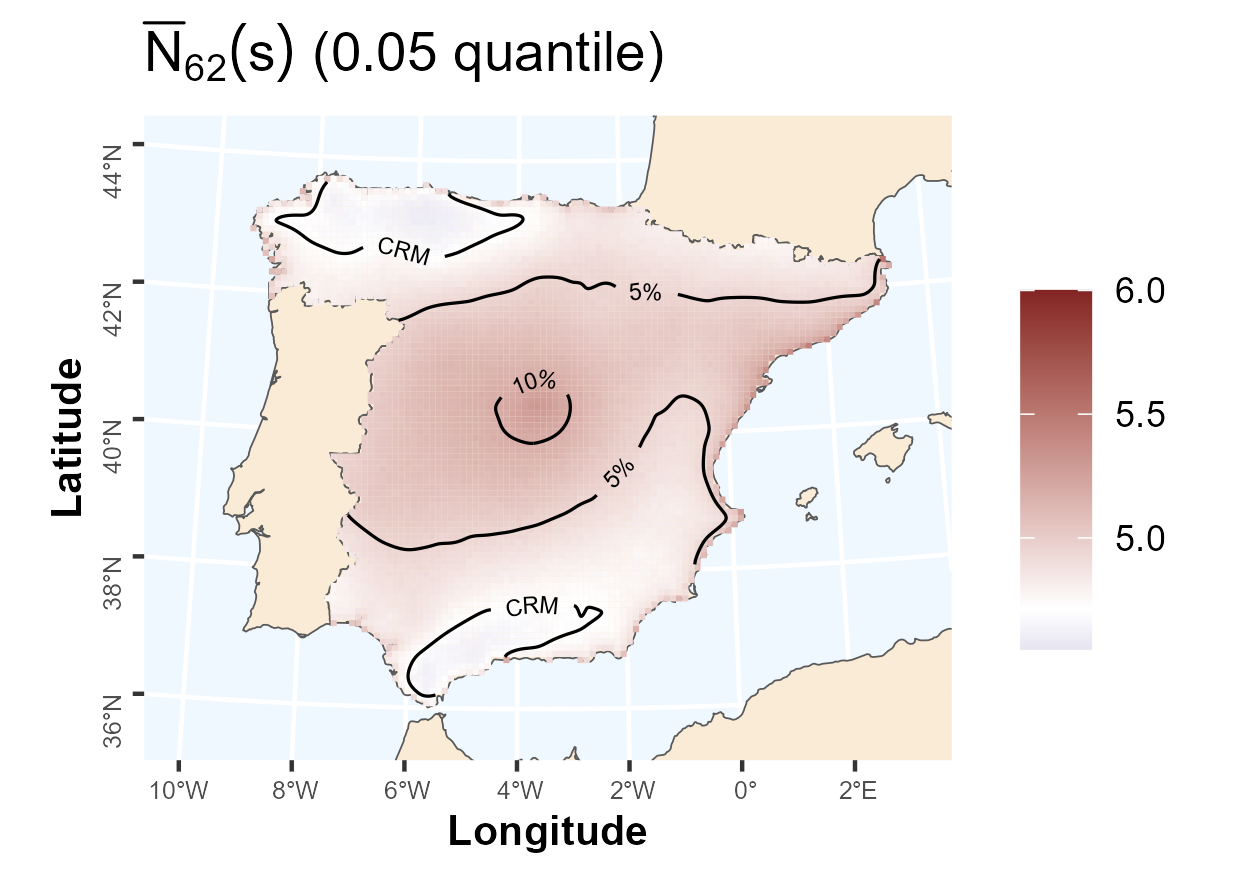}
    \includegraphics[width=.32\textwidth]{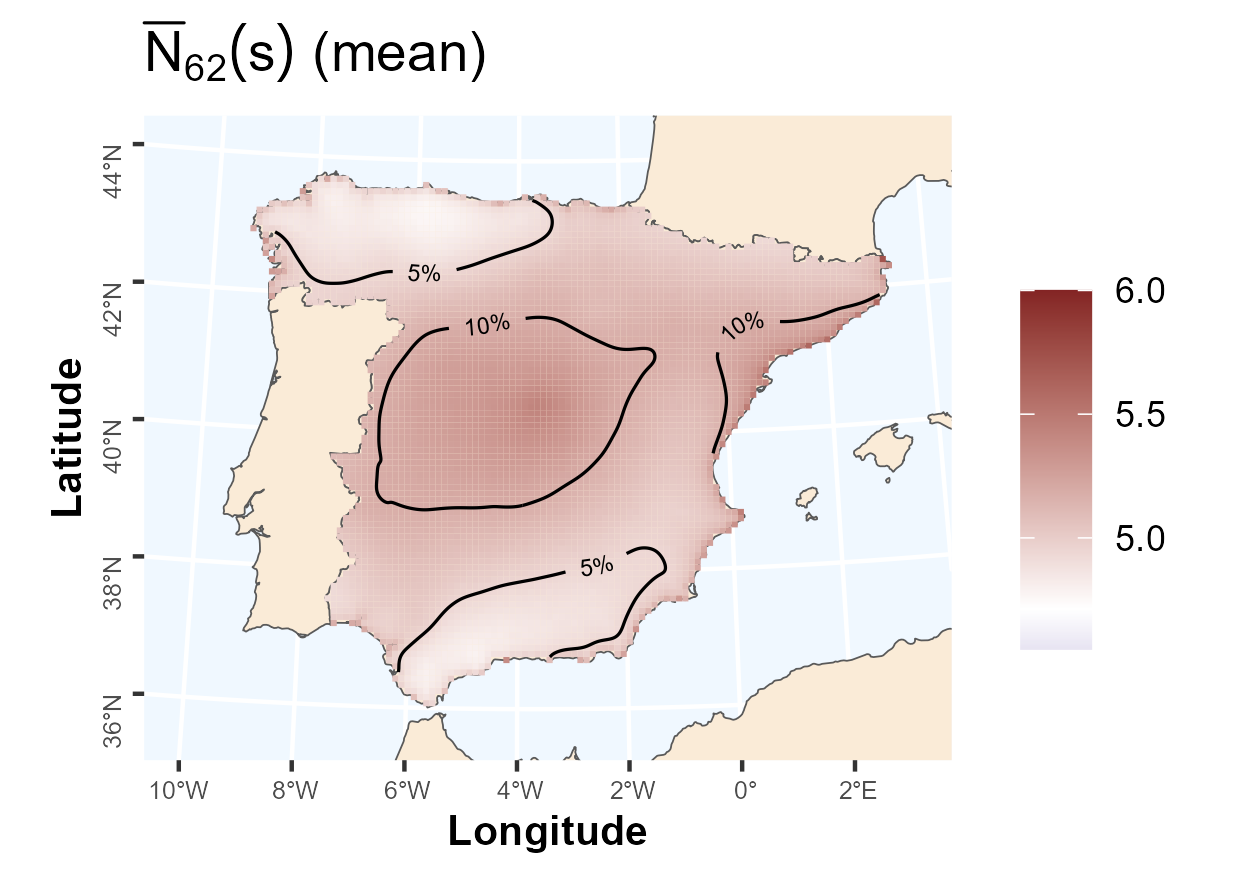}
    \includegraphics[width=.32\textwidth]{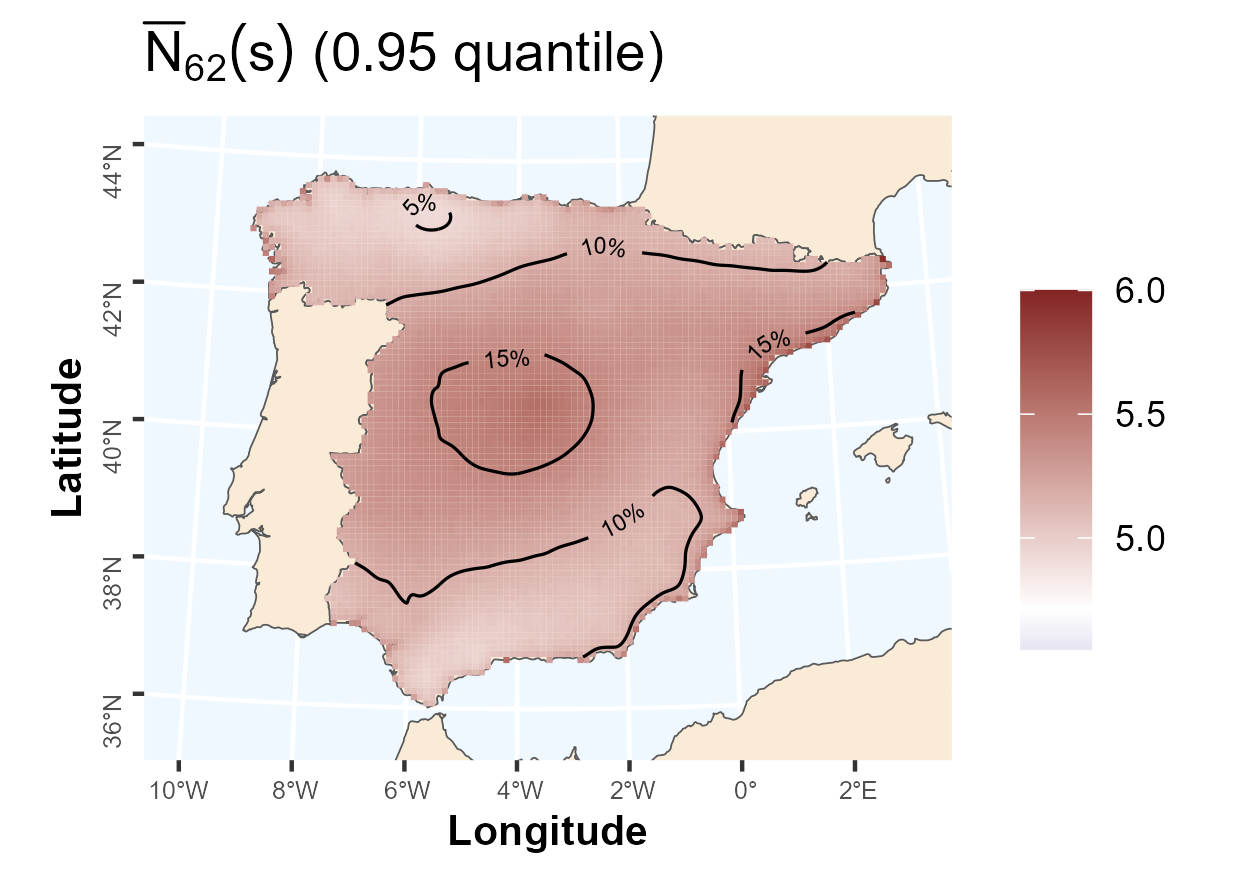}
    \caption{Map of the posterior $0.05$ quantile (left), mean (center), and $0.95$ quantile of $\bar{N}_{62}(\bs)$. The stationary reference value $4.71$ is in white with blue indicating lower values and red indicating higher values. Contour lines indicate deviations from stationarity by $5\%$ intervals.}
    \label{fig:N}
\end{figure}

\paragraph{Number of records.}
Figure~\ref{fig:N} shows three maps of the posterior mean and the borders of the $90\%$ CI (i.e., $0.05$ and $0.95$ quantiles) of $\bar{N}_{62}(\bs)$, i.e., the average across the 365 days of the number of records across years up to 2021.  The average number of records expected under a stationary climate is $E_0[\bar{N}_{62}(\bs)] = \sum_{t=1}^{62} 1 / t \approx 4.71$. The posterior mean of the number of records goes from $4.76$ to $5.73$ in space with a block average of $5.08$. The total number of records is higher than the stationary case in a $98.5$ $(97.3, 99.5)$ $\%$ of the region, with the center and northeast being the most affected regions. The Atlantic and southern Mediterranean coasts seem to have a number of records compatible with the stationary case. Analyzing point by point, $90.0\%$ of the region shows a significantly higher number of records compared to the stationary case, while no points exhibit a significantly lower number of records.

\begin{figure}[p]
    \centering
    \includegraphics[width=.45\textwidth]{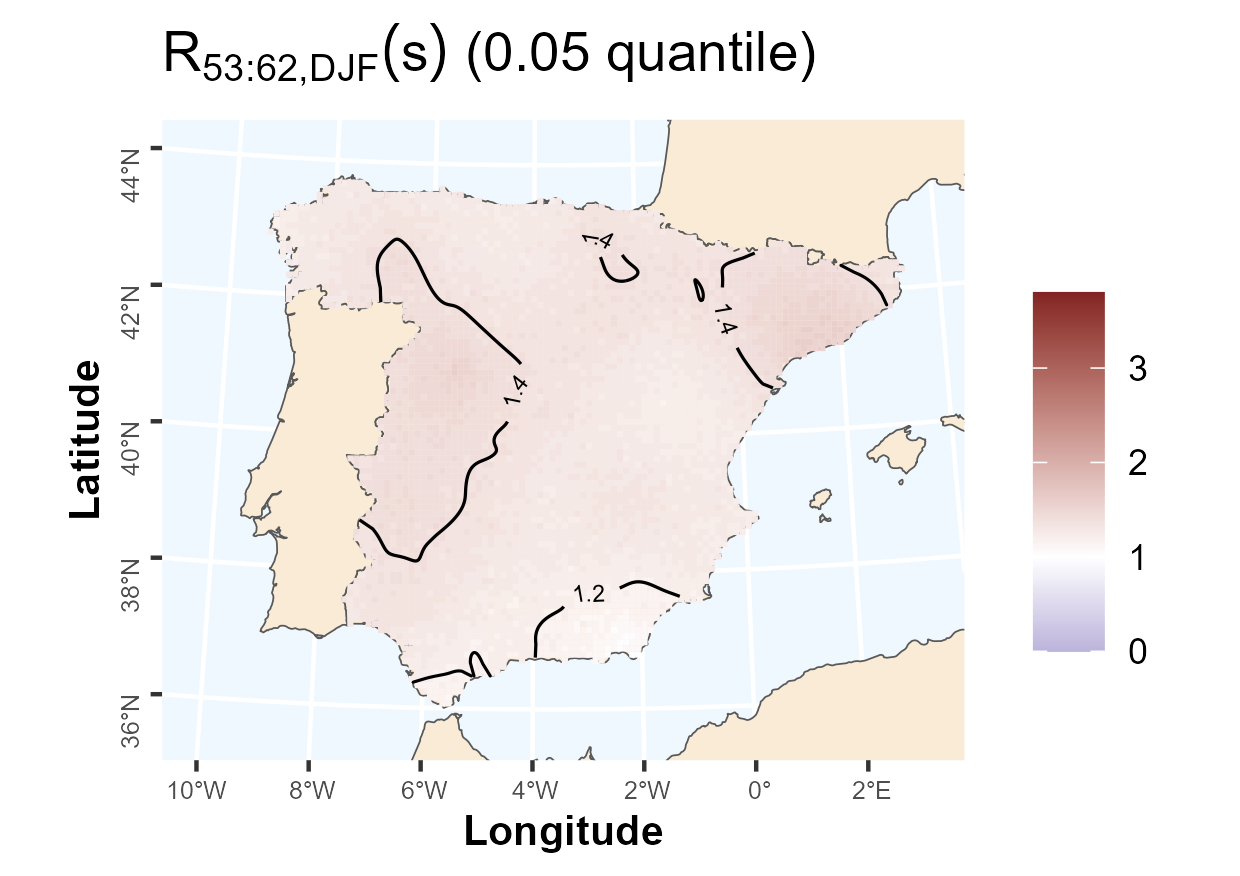}
    \includegraphics[width=.45\textwidth]{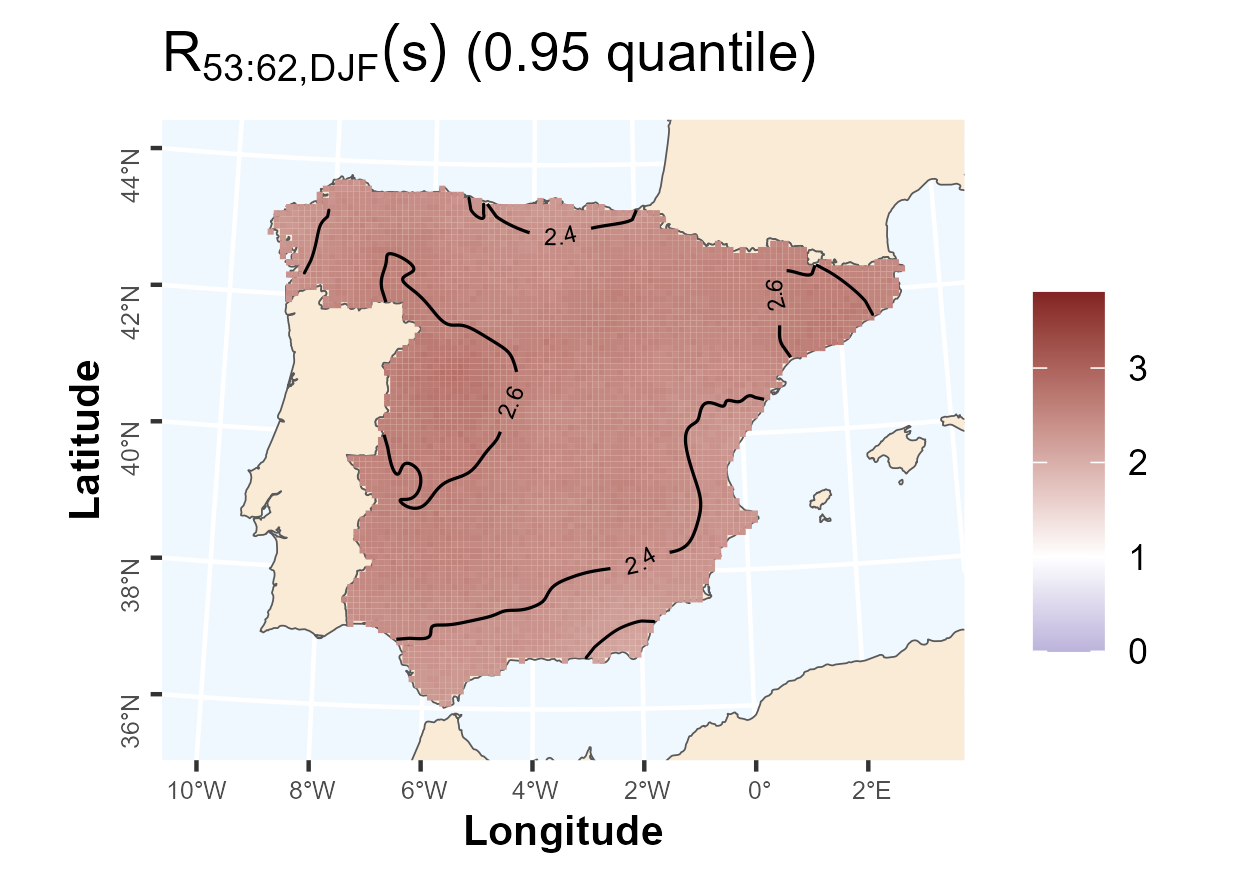} \\
    \includegraphics[width=.45\textwidth]{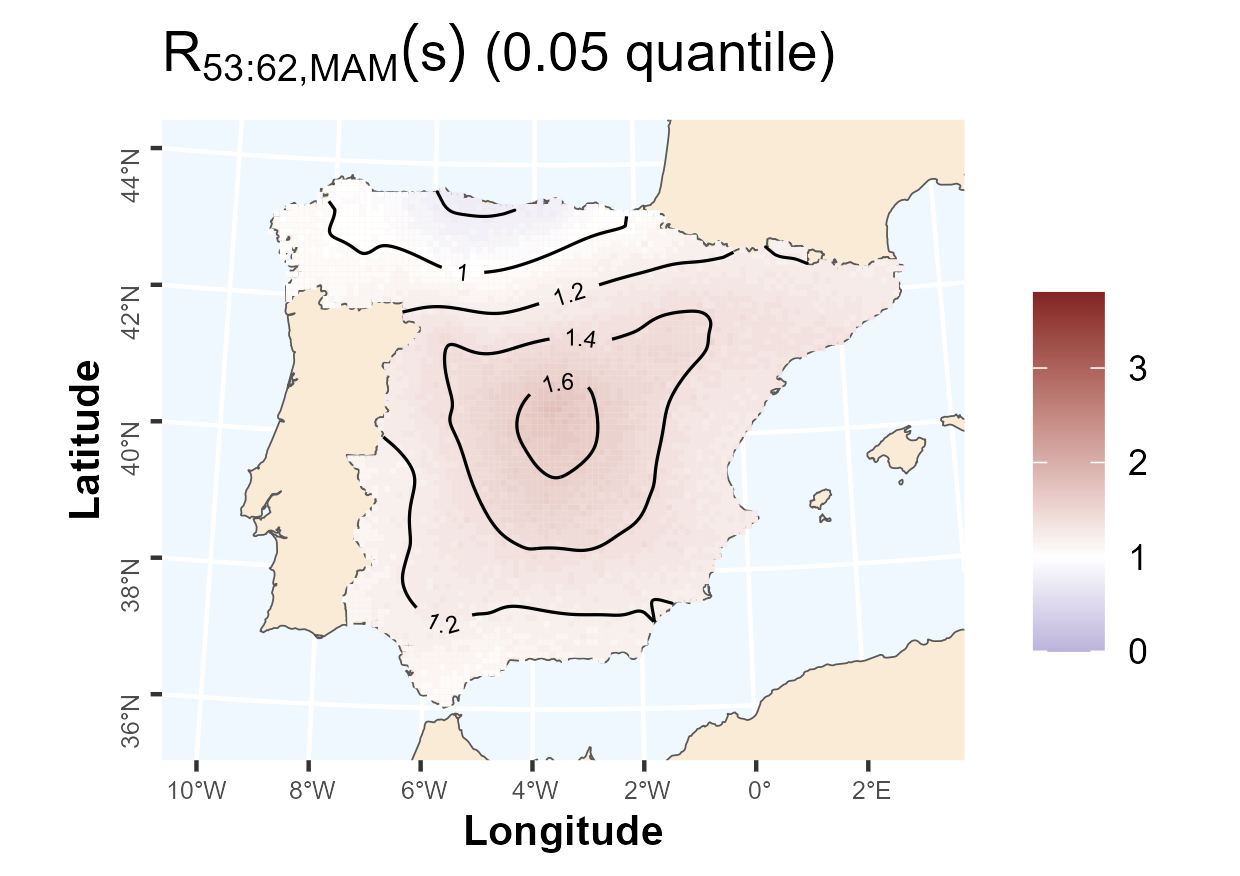}
    \includegraphics[width=.45\textwidth]{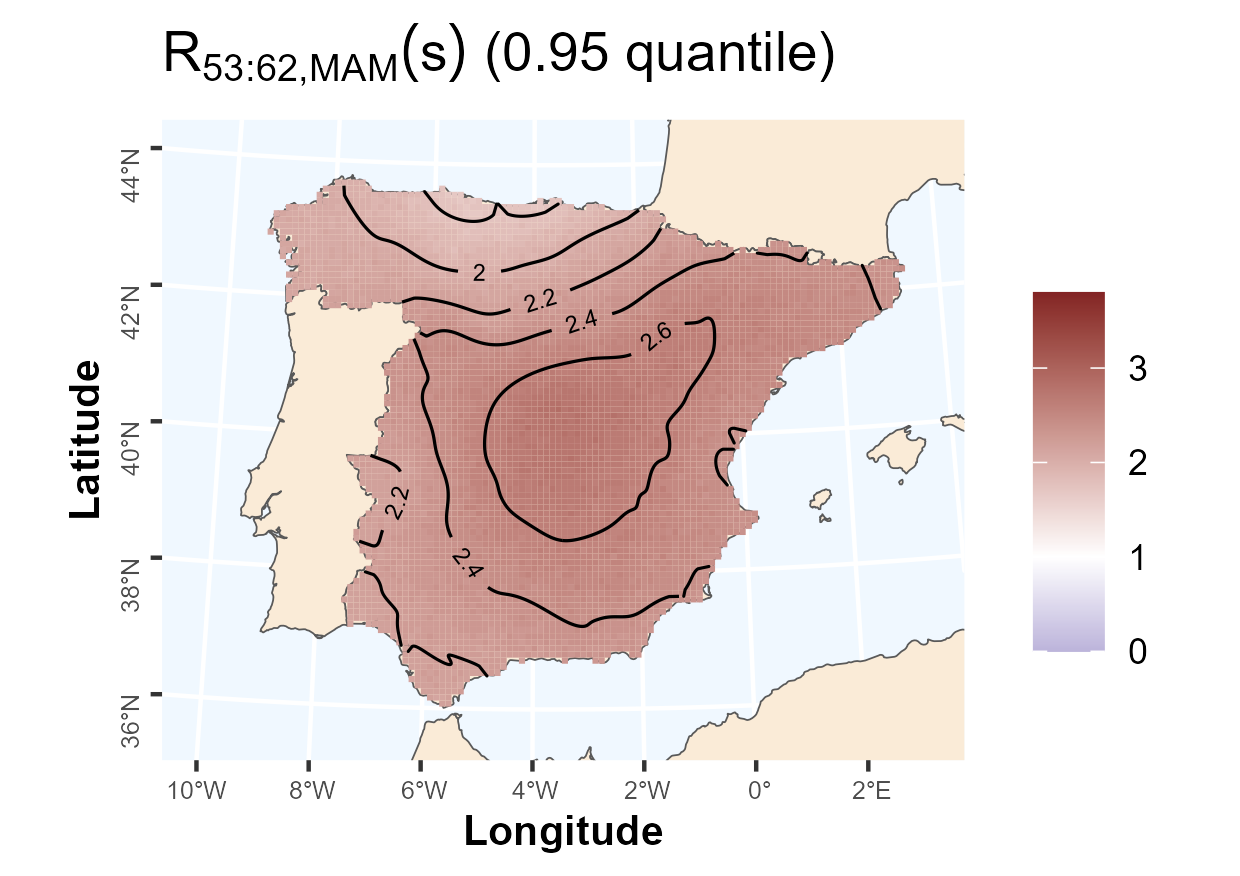} \\
    \includegraphics[width=.45\textwidth]{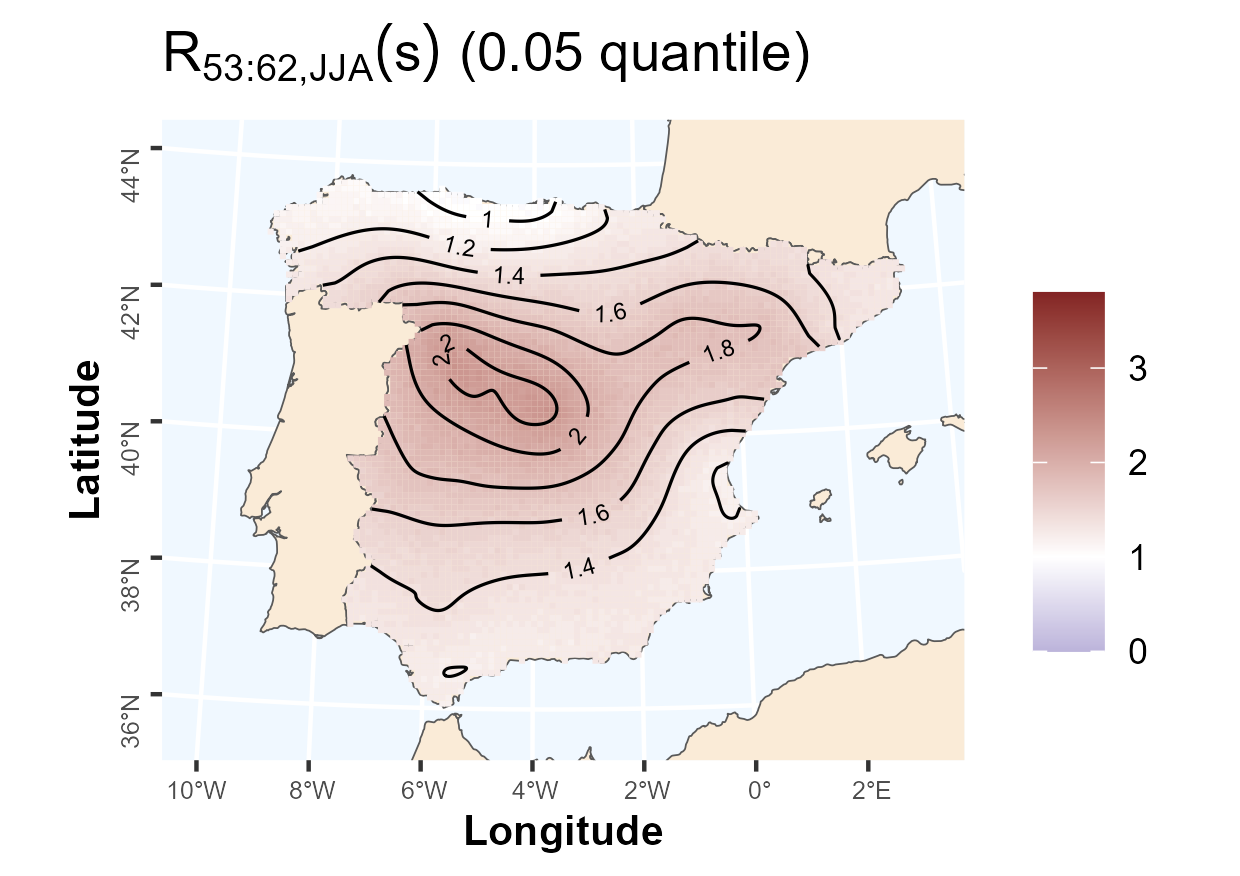}
    \includegraphics[width=.45\textwidth]{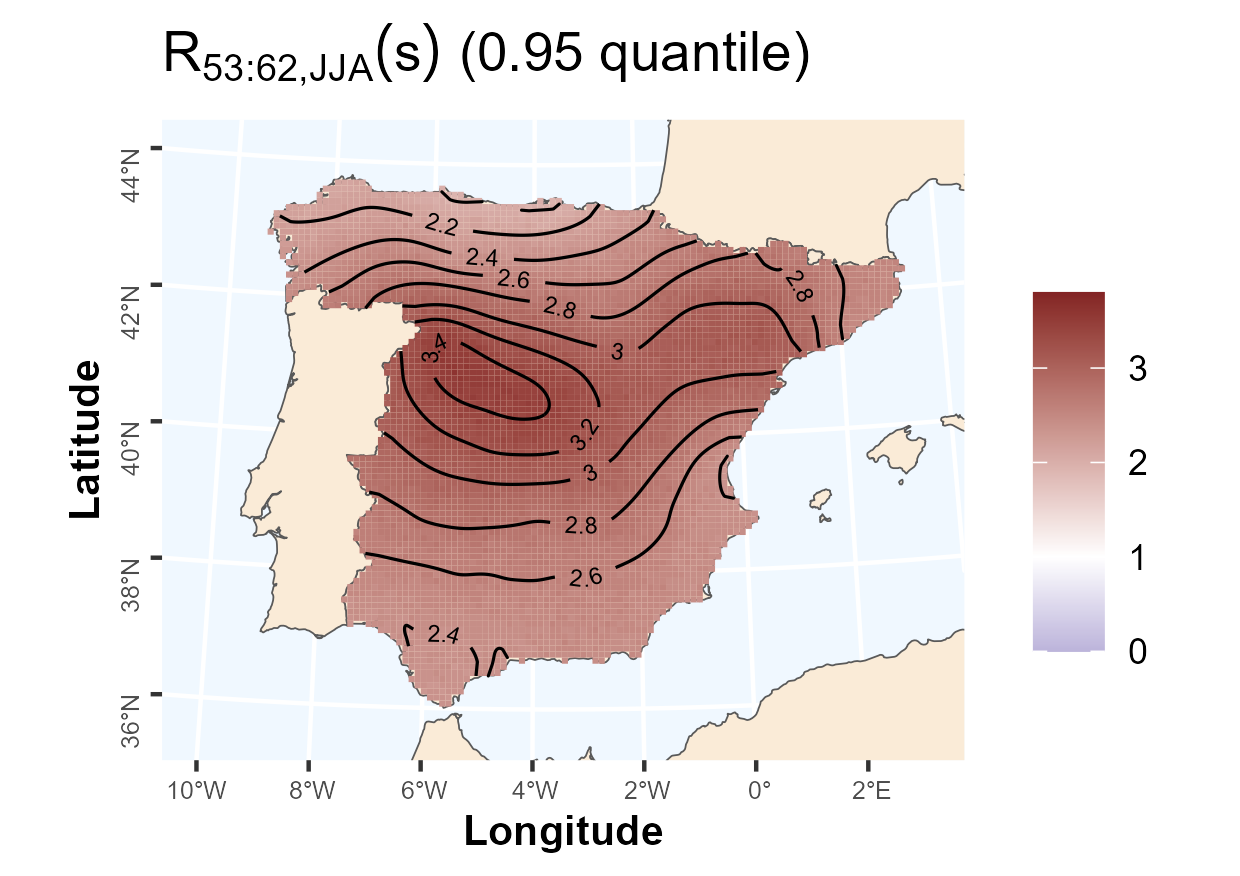} \\
    \includegraphics[width=.45\textwidth]{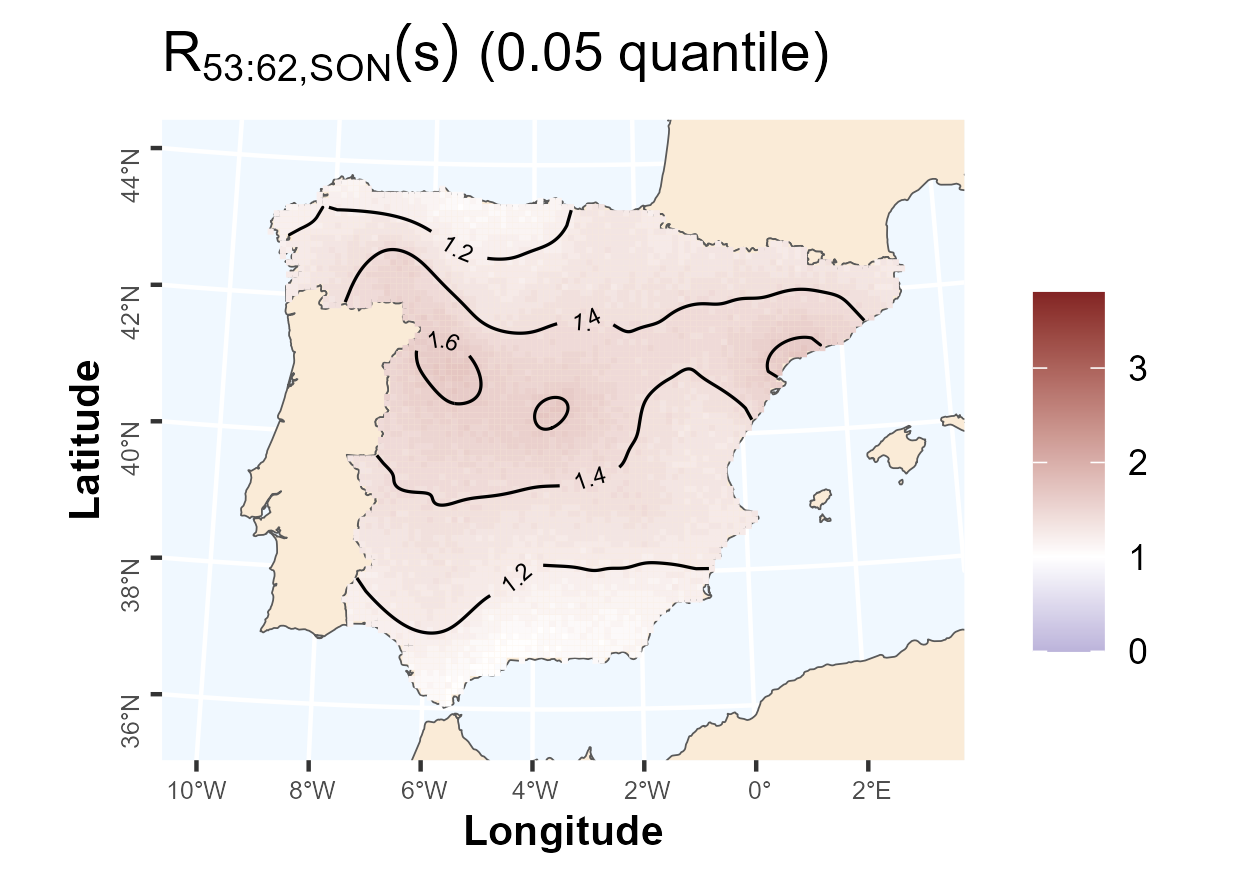}
    \includegraphics[width=.45\textwidth]{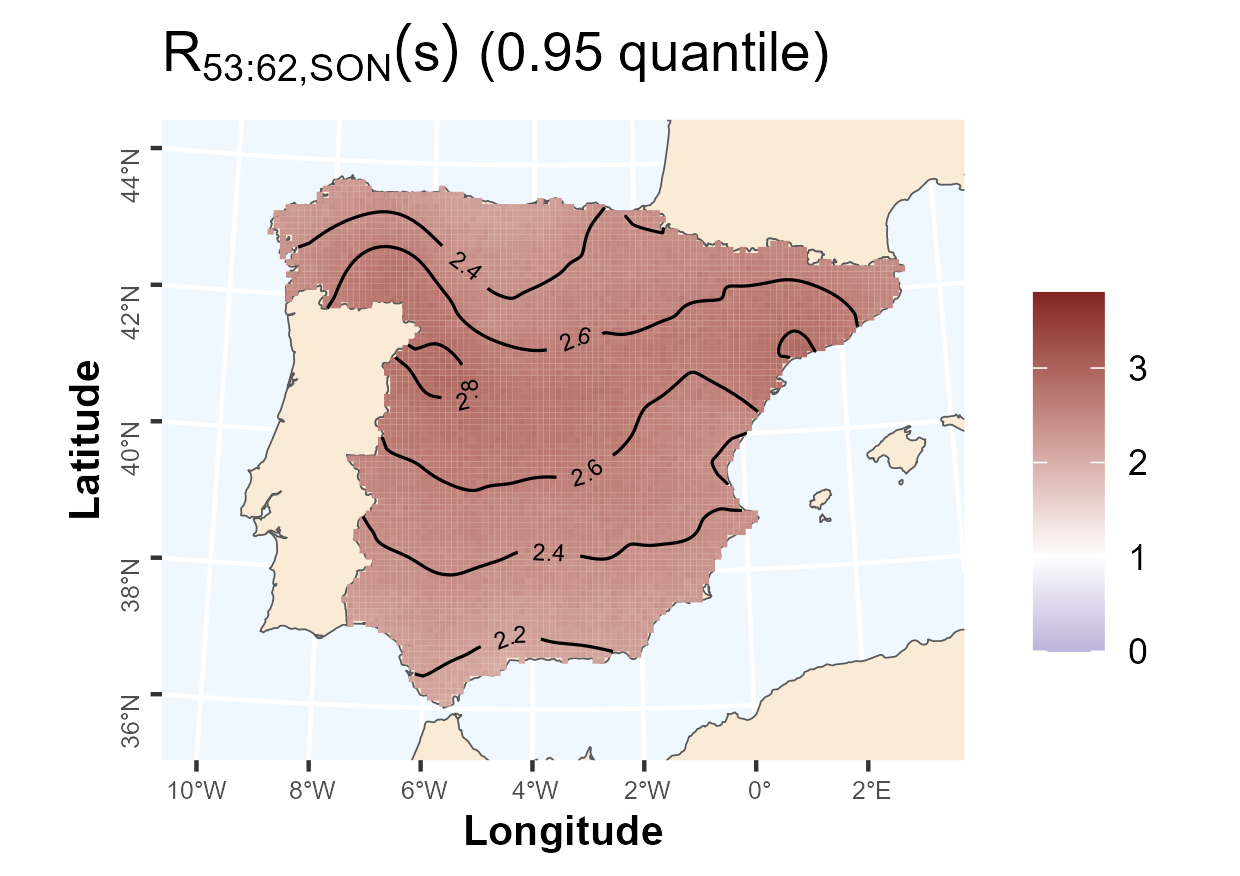}
    \caption{Map of the posterior $0.05$ (left) and $0.95$ (right) quantiles of $R_{53:62,DJF}(\bs)$ (winter, first row), $R_{53:62,MAM}(\bs)$ (spring, second row), $R_{53:62,JJA}(\bs)$ (summer, third row), and $R_{53:62,SON}(\bs)$ (autumn, fourth row). Contour lines indicate ratios between intervals of length $0.2$.}
    \label{fig:lastDecadeQuantiles}
\end{figure}

Figure~\ref{fig:lastDecadeQuantiles} shows the maps of the posterior $0.05$ and $0.95$ quantiles of the ratios $R_{53:62,season}(\bs)$ with years from 2012 to 2021 and the period of days in each $season$ (winter DJF, spring MAM, summer JJA, and autumn SON). These ratios compare the average number of records predicted by the model and the expected number under the stationary case in the considered periods of time. Note that the considered quantiles define the borders of the $90\%$ CI's of the ratios.

\begin{figure}[p]
    \centering
    \includegraphics[width=.32\textwidth]{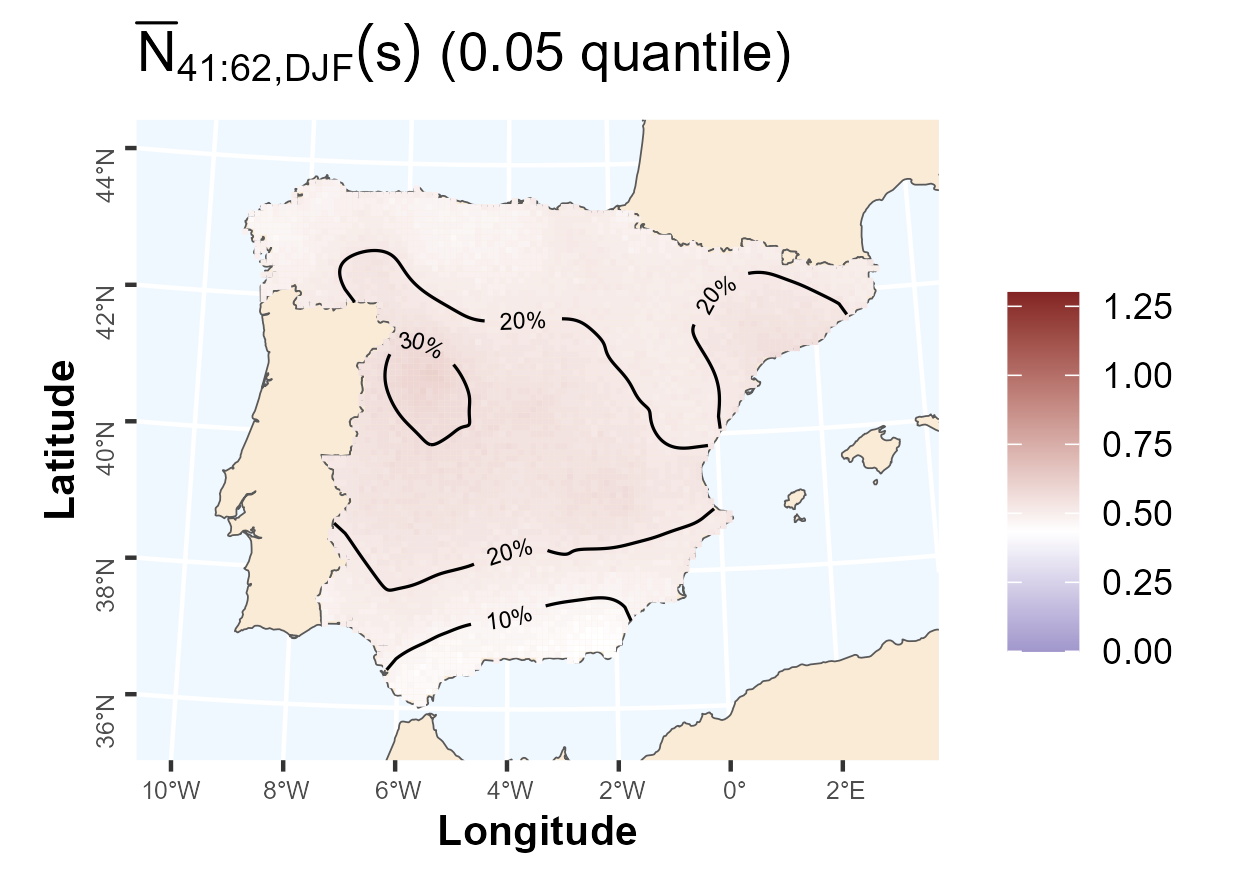}
    \includegraphics[width=.32\textwidth]{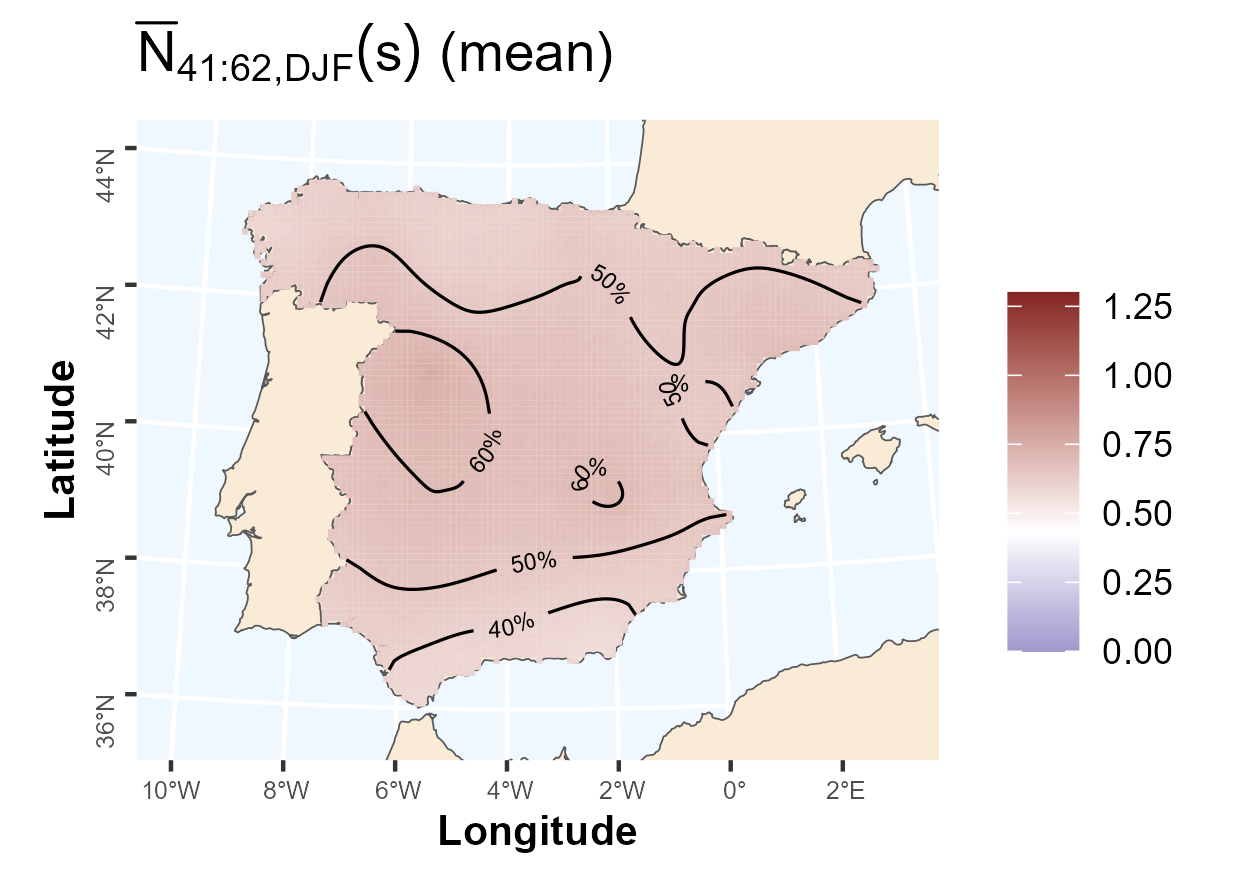}
    \includegraphics[width=.32\textwidth]{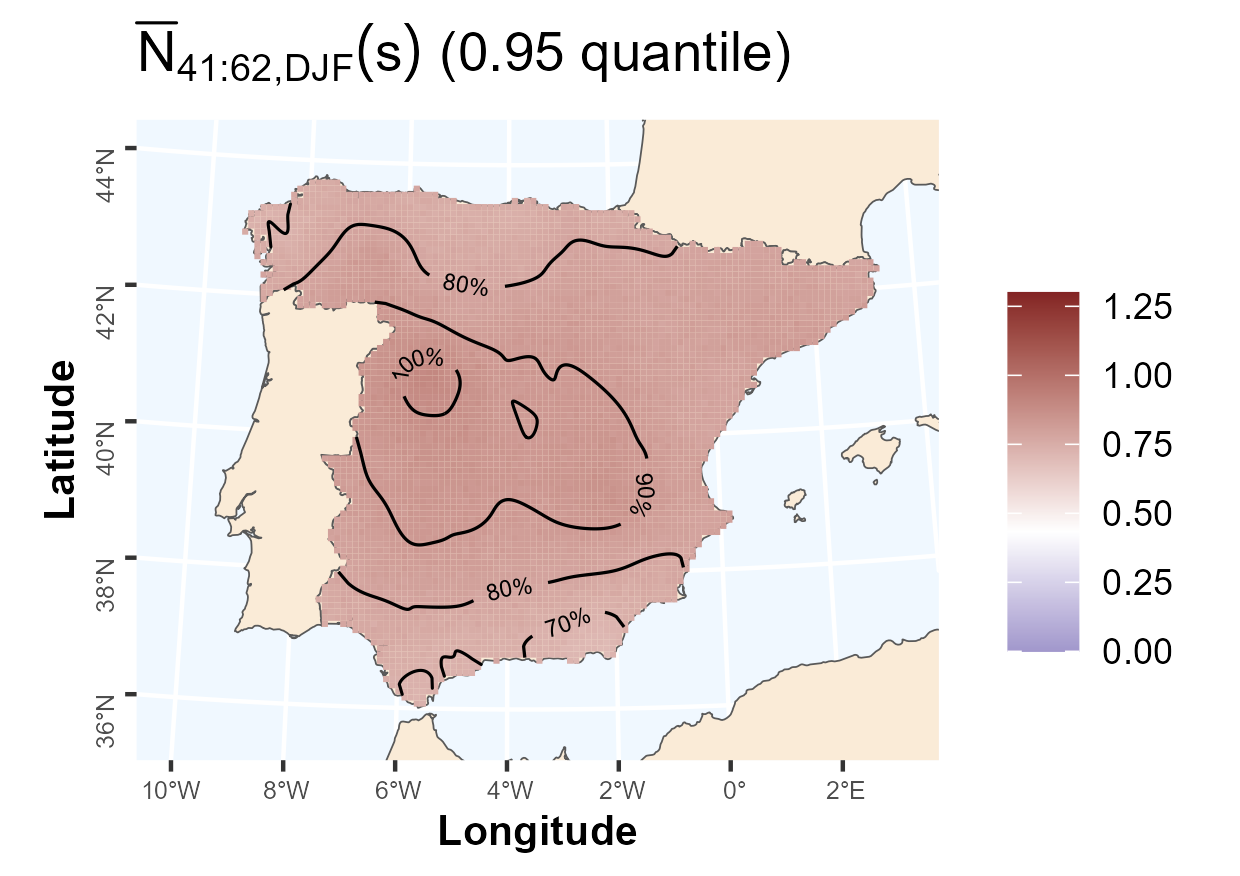}
    \includegraphics[width=.32\textwidth]{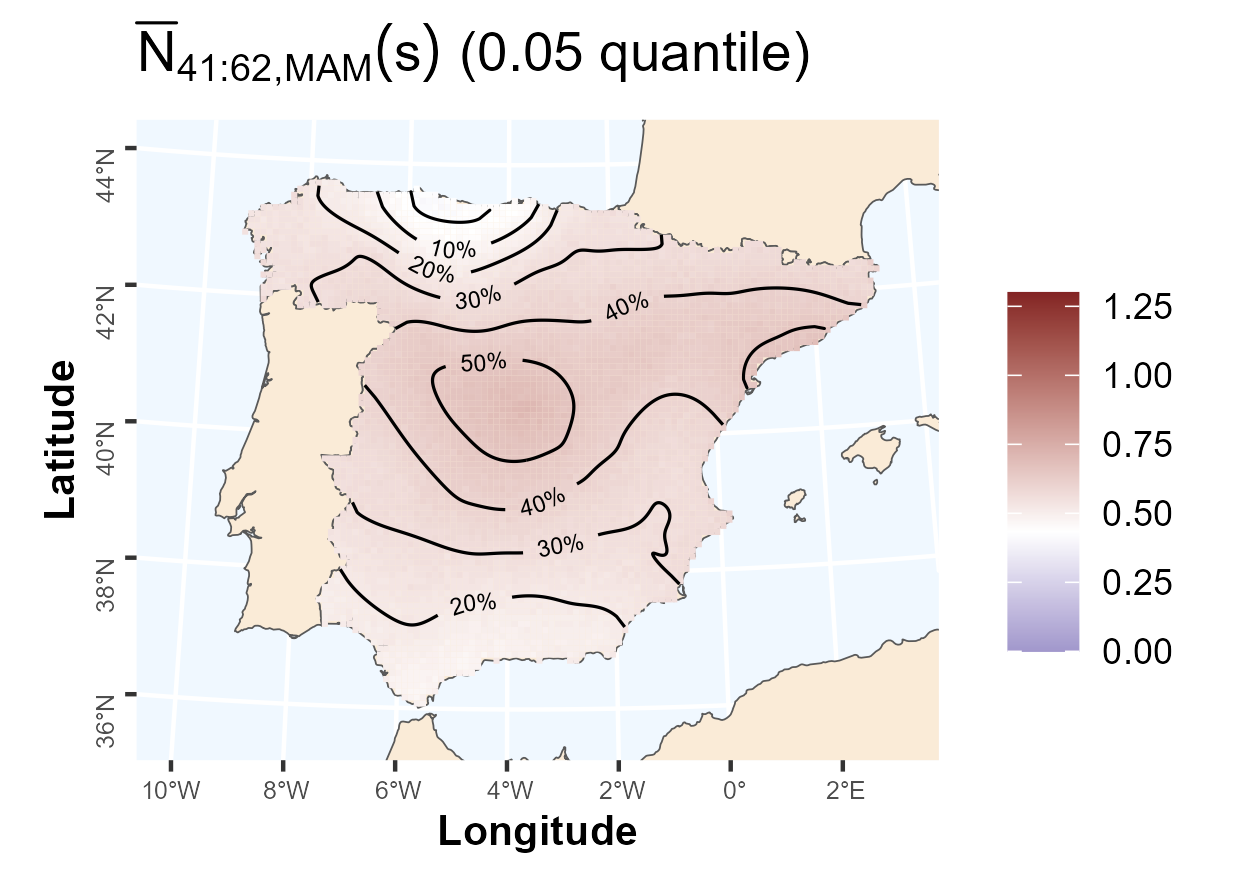}
    \includegraphics[width=.32\textwidth]{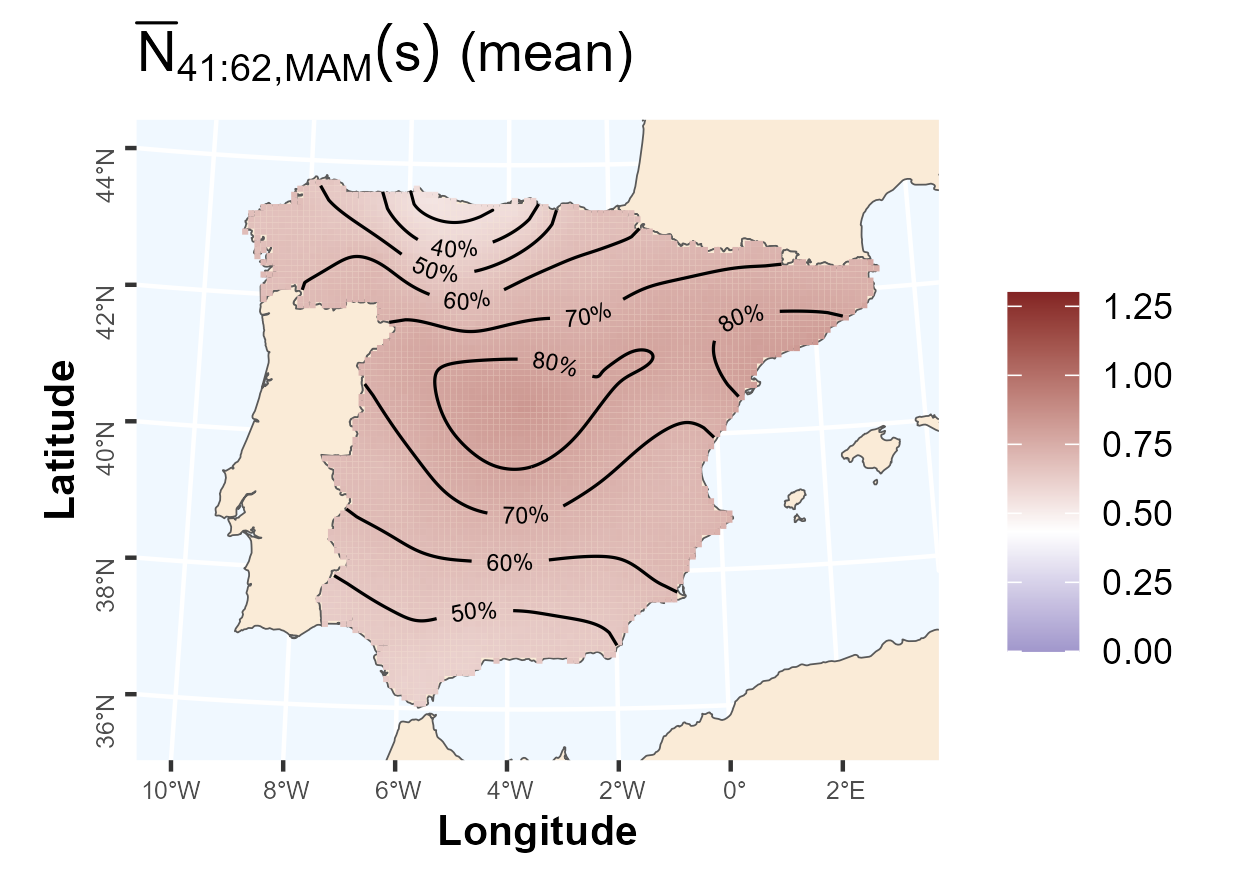}
    \includegraphics[width=.32\textwidth]{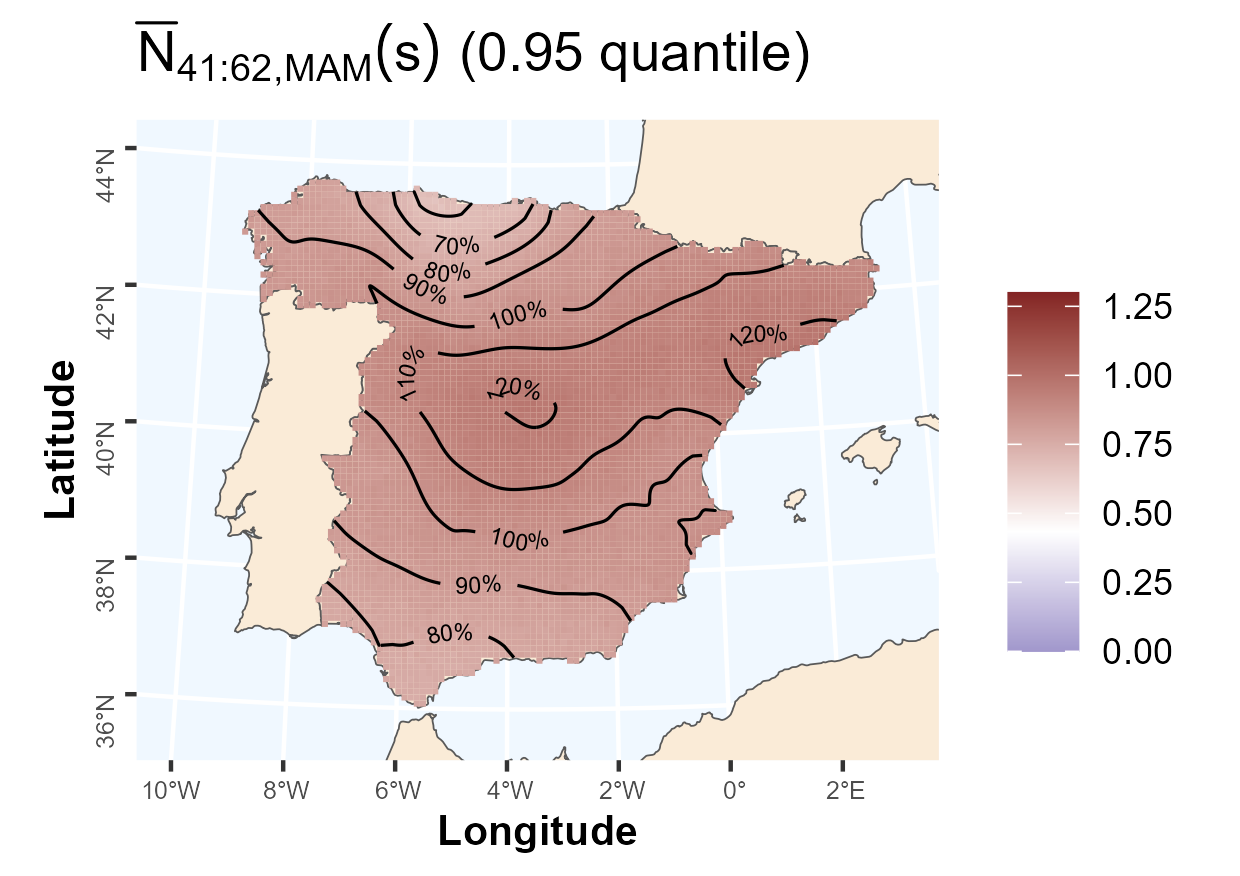}
    \includegraphics[width=.32\textwidth]{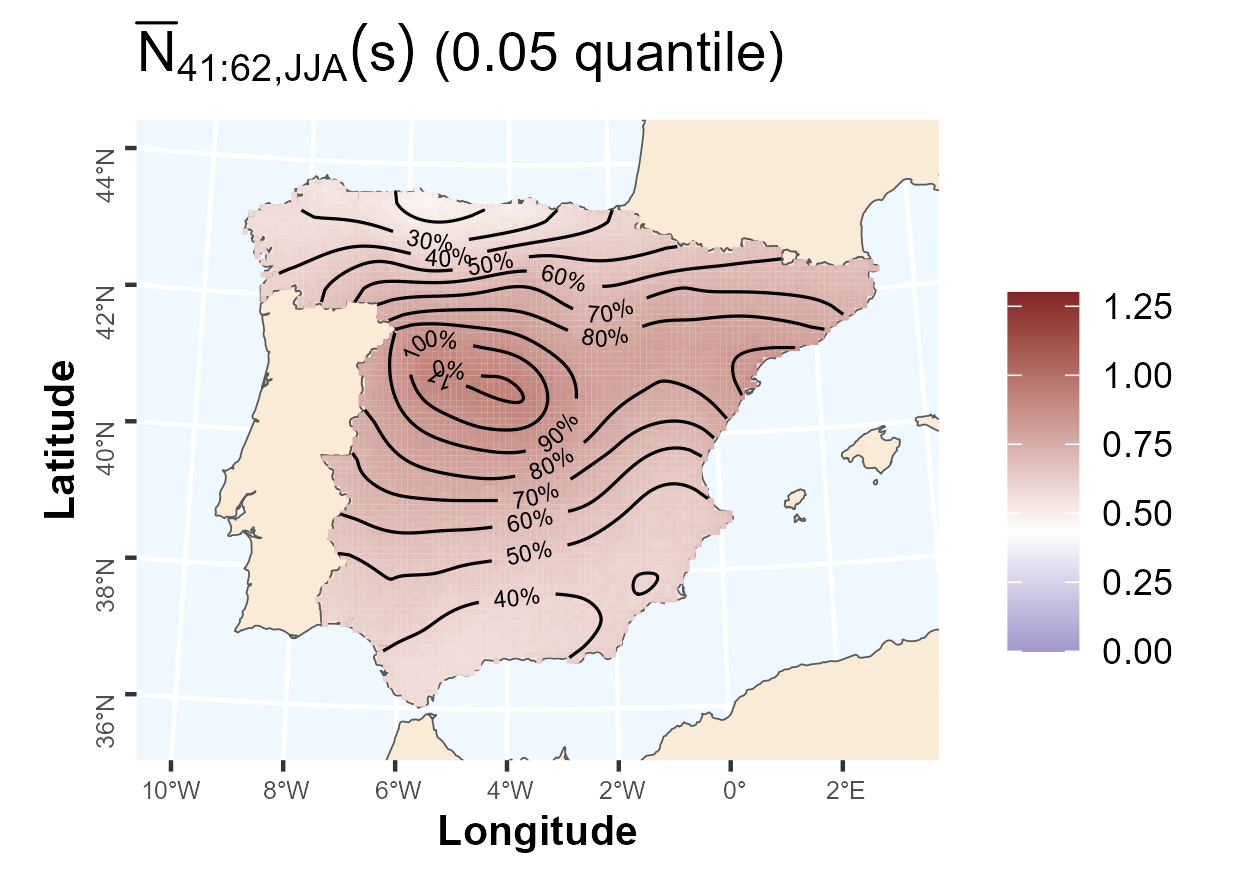}
    \includegraphics[width=.32\textwidth]{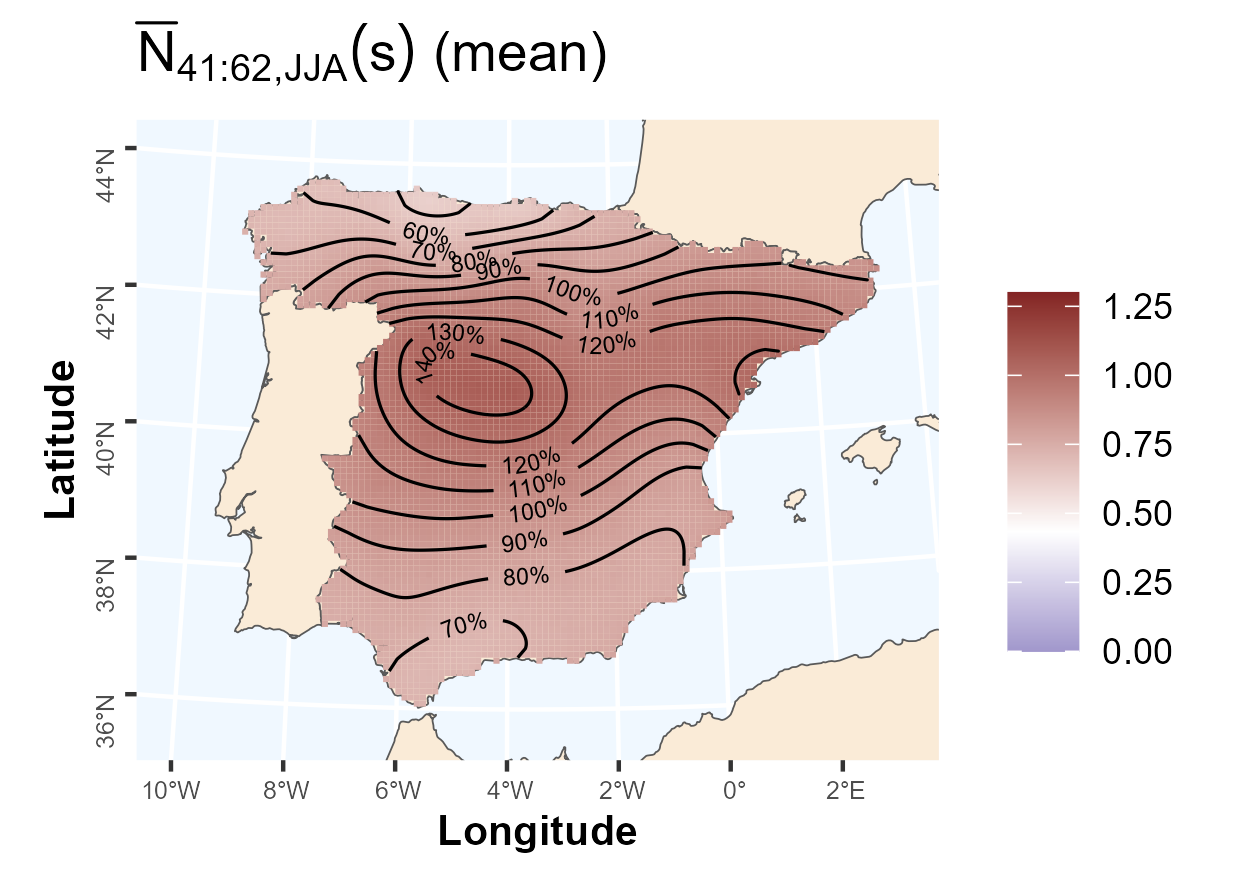}
    \includegraphics[width=.32\textwidth]{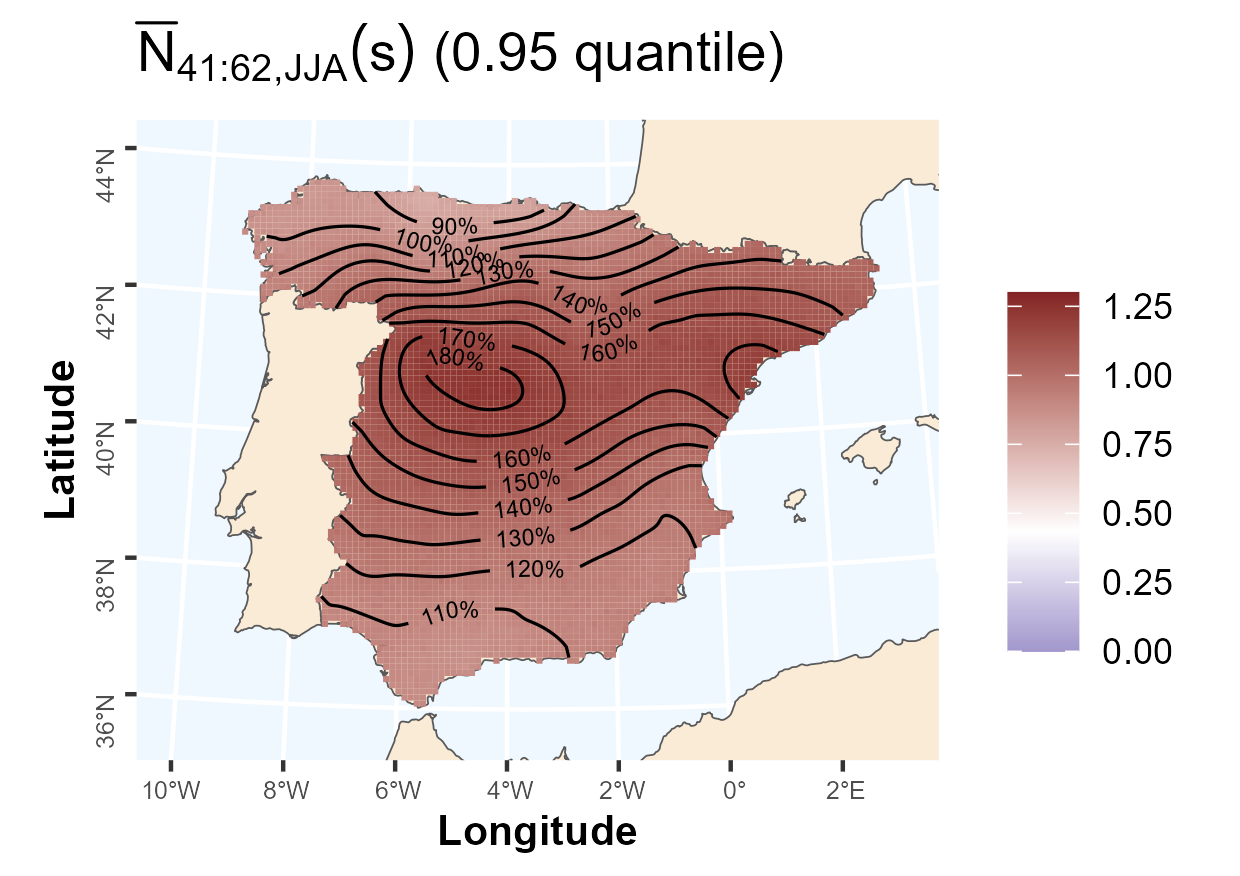}
    \includegraphics[width=.32\textwidth]{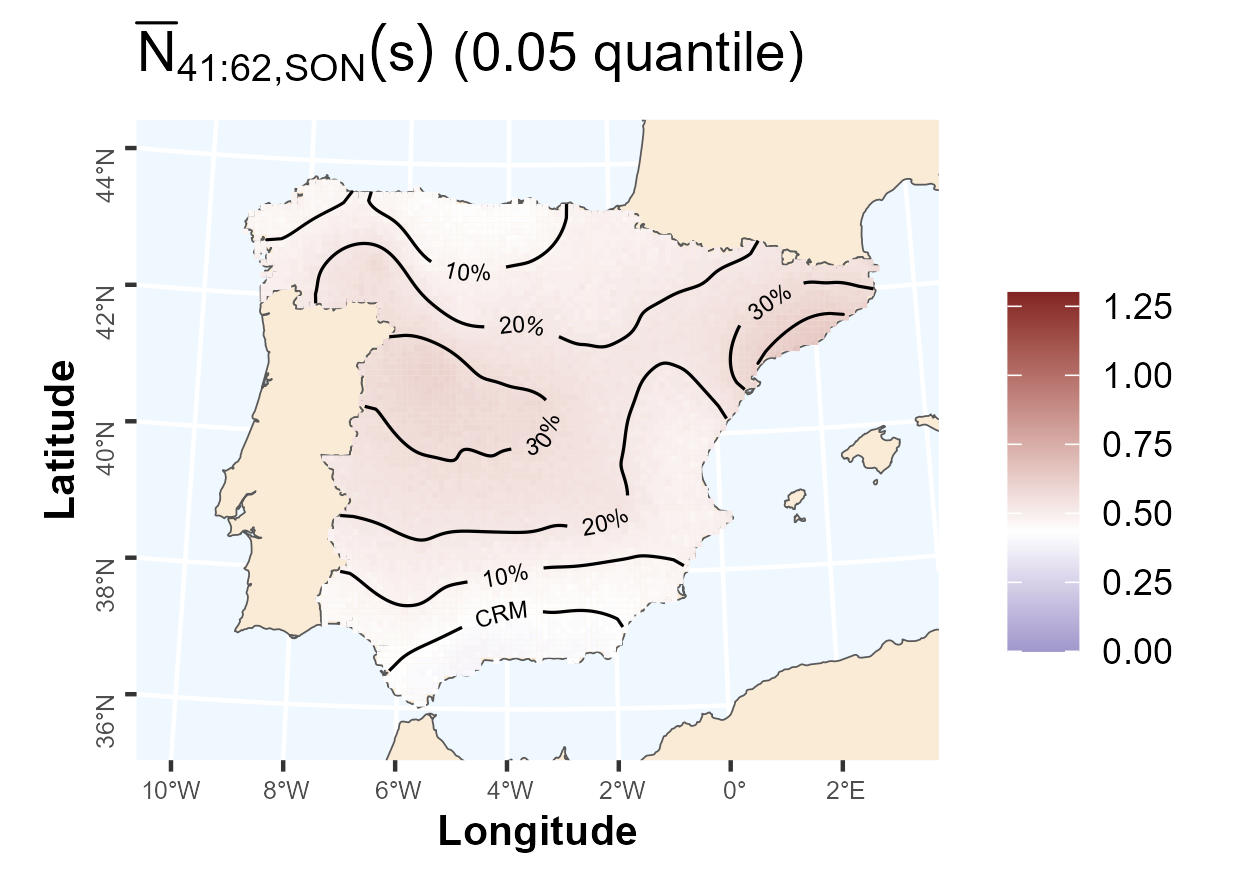}
    \includegraphics[width=.32\textwidth]{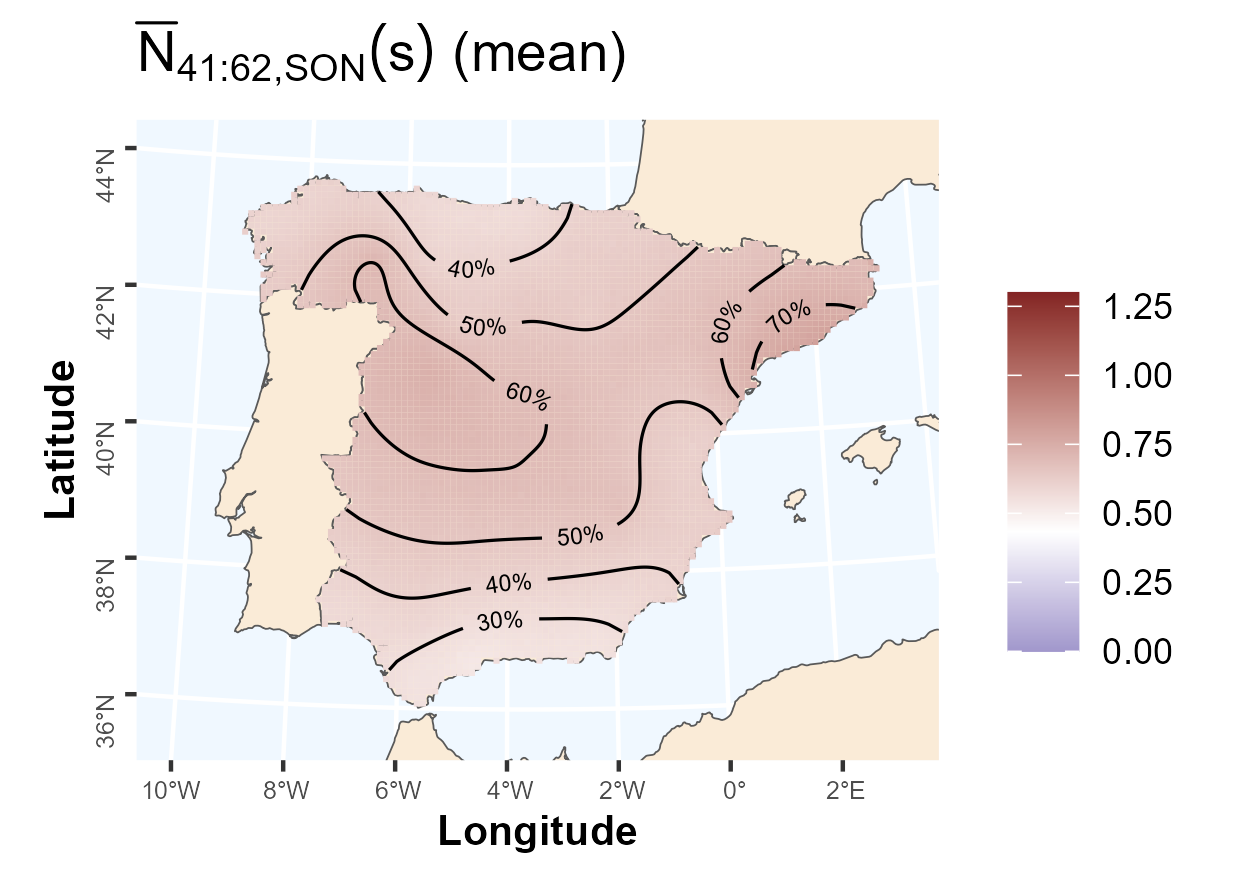}
    \includegraphics[width=.32\textwidth]{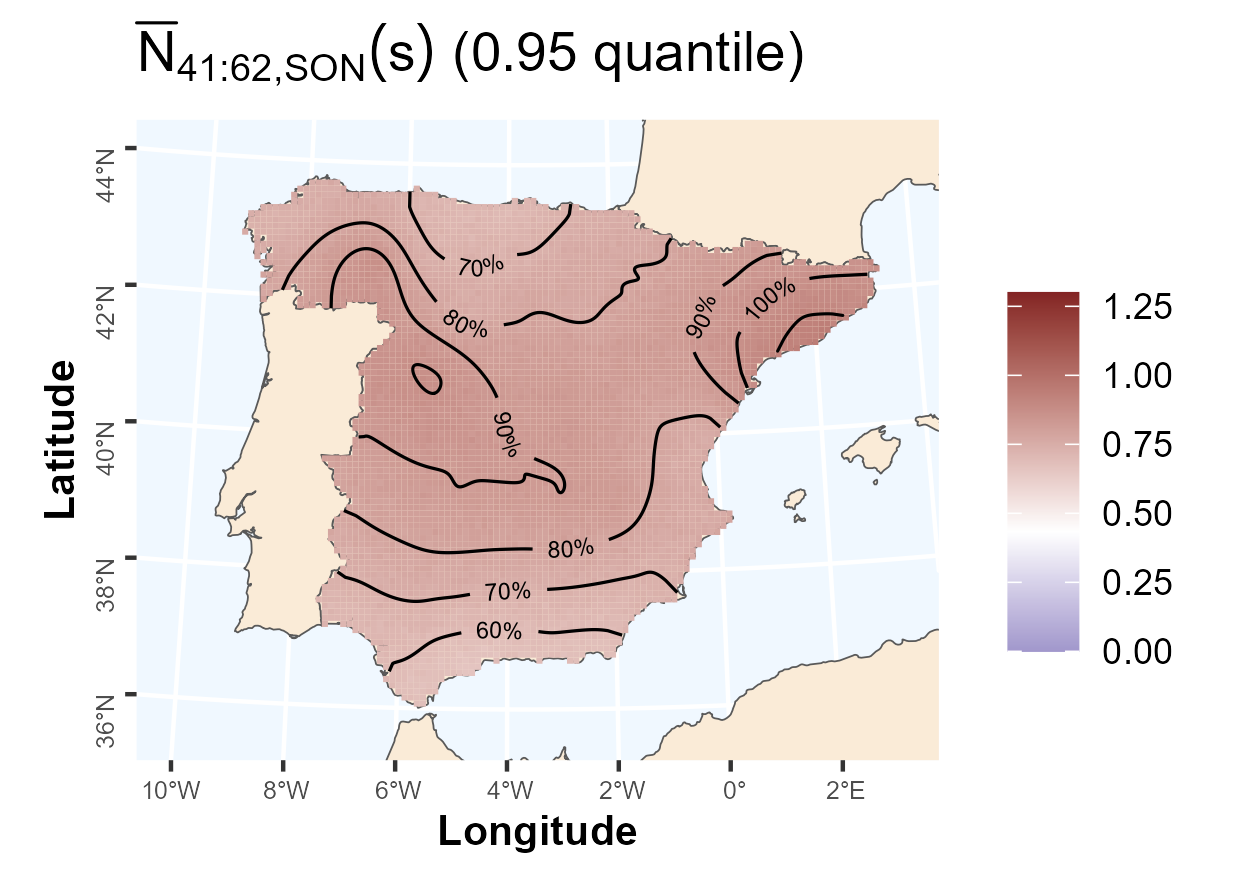}
    \caption{Maps of the posterior $0.05$ quantile (left), mean (center), and $0.95$ quantile (right) of $\bar{N}_{41:62,DJF}(\bs)$ (winter, first row), $\bar{N}_{41:62,MAM}(\bs)$ (spring, second row), $\bar{N}_{41:62,JJA}(\bs)$ (summer, third row), and $\bar{N}_{41:62,SON}(\bs)$ (autumn, fourth row). The stationary reference value $0.43$ is in white. Contour lines indicate deviations from stationarity in $10\%$ intervals.
    }
    \label{fig:N21stSeason}
\end{figure}

Figure~\ref{fig:N21stSeason} displays the posterior mean and the $90\%$ CI of $\bar{N}_{41:62,season}(\bs)$, the average number of records, now during the 21st century and again in each $season$. Comparing the results with the expected value under stationarity, we estimate that global warming has increased the average number of temperature records from $0.43$ to $0.72$ $(0.71,0.73)$. The increase is not homogeneous throughout the year, with summer being the most affected season, showing an increase of $0.86$ $(0.83,0.88)$. Spring follows with an increase of $0.71$ $(0.69,0.74)$, and both winter and autumn show an increase of $0.66$ $(0.64,0.68)$ and $0.65$ $(0.63,0.67)$, respectively. Furthermore, the increase is not homogeneous across space. For instance, the increase in spring and summer is not significant in some parts of the Atlantic coast while it is significant over the rest of Spain.

\begin{figure}[t]
    \centering
    \includegraphics[width=.45\textwidth]{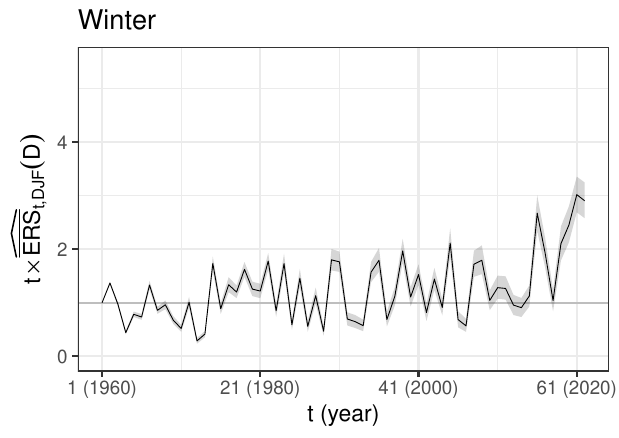}
    \includegraphics[width=.45\textwidth]{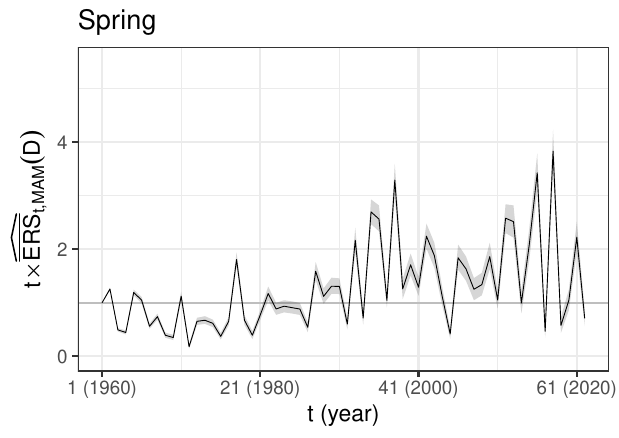}
    \includegraphics[width=.45\textwidth]{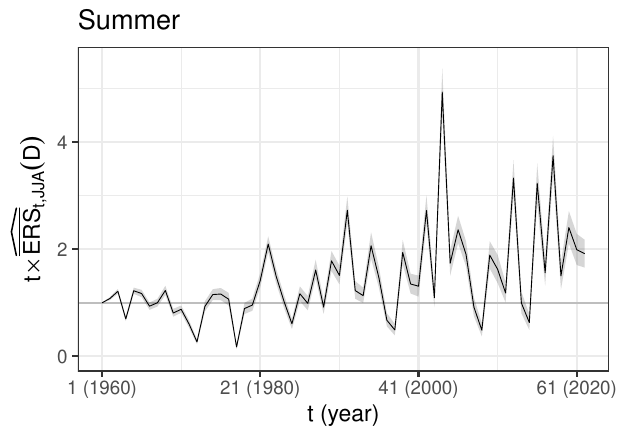}
    \includegraphics[width=.45\textwidth]{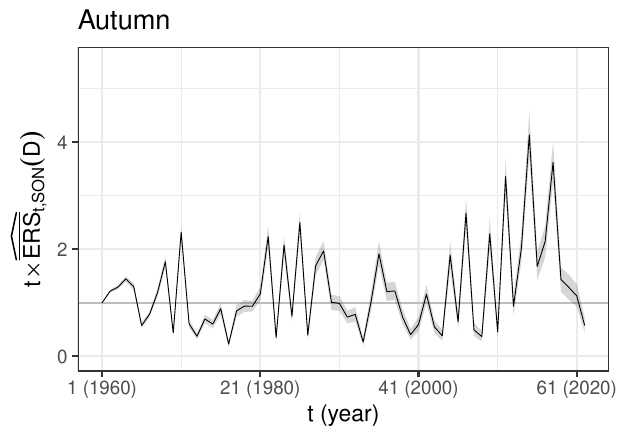}
    \caption{Posterior mean (black line) and $90\%$ CI (gray ribbon) of $t \times \widehat{\overline{\text{ERS}}}_{t,DJF}(D)$ (Winter, top--left), $t \times \widehat{\overline{\text{ERS}}}_{t,MAM}(D)$ (Spring, top--right), $t \times \widehat{\overline{\text{ERS}}}_{t,JJA}(D)$ (Summer, bottom--left), and $t \times \widehat{\overline{\text{ERS}}}_{t,SON}(D)$ (Autumn, bottom--right) against $t$. Expected value of $1$ under stationarity remarked for reference.}
    \label{fig:ERSseason}
\end{figure}

\paragraph{Extent of record surface.}
Figure~\ref{fig:ERSseason}, displays the posterior mean and $90\%$ CI of $t \times \widehat{\overline{\text{ERS}}}_{t,season}(D)$ for each $season$. A large variability between seasons and years is observed. Plots for spring, summer and autumn show an increasing values from the 80's. The highest deviation from stationarity observed across the seasons is the summer of 2003 which multiplies by $5$ the expected value under stationarity. Another notable fact is that winter did not present a particularly remarkable increase in the number of records until the last seven years of the study, during which, up to three of those years experienced around three times the number of records expected under stationarity. In contrast, in spring, the number of records is notably lower than the stationary case in some of those same years.


\clearpage

\bibliographystyle{agsm}

\bibliography{bibliography}